\documentclass[useAMS]{mn2e}

\usepackage[dvips]{graphicx}		
\usepackage{amsfonts}
\usepackage{amsmath}
\usepackage{amssymb}
\usepackage{cleveref}
\usepackage{longtable}
\usepackage{lscape}
\usepackage{times}
\usepackage{pifont}
\usepackage{booktabs}
\usepackage{xspace}		
\usepackage{tabularx}
\usepackage{rotating}
\usepackage{url}
\usepackage{stfloats}
\usepackage{mathrsfs}

\newcommand{\kms}{\mbox{km\,s$^{-1}$}}
\def\kms{km${\rm s}^{-1}$}
\def\arcdeg{\hbox{$^\circ$}}
\def\arcmin{\hbox{$^\prime$}}
\def\arcsec{\hbox{$^{\prime\prime}$}}
\def\ergcms{erg\,cm$^{-2}$s$^{-1}$}

\def\ha{H$\alpha$}
\def\hb{H$\beta$}

\def\NII{[N\,\textsc{ii}]}
\def\SII{[S\,\textsc{ii}]}

\def\OII{[O\,\textsc{ii}]}
\def\OIII{[O\,\textsc{iii}]}
\def\OIV{[O\,\textsc{iv}]}

\def\NeV{[Ne\,\textsc{v}]}

\def\ArIV{[Ar\,\textsc{iv}]}
\def\ArV{[Ar\,\textsc{v}]}
\def\HI{H\,\textsc{i}}
\def\HII{H\,\textsc{ii}}

\def\HeII{He\,\textsc{ii}}

\def\B-V{$B$--$V$}
\def\V-I{$V$--$I$}
\def\EBV{$E$($B$--$V$)}
\def\msun{$M_{\odot}$}
\def\SR{$S_{\rm H\alpha}$--$r$}
\def\sgen{$S$--$r$}
\def\p0{\phantom{0}}

\def\lessim{\raise-.5ex\hbox{$\buildrel<\over{\scriptstyle\mathtt{\sim}}$}}
\def\grtsim{\raise-.5ex\hbox{$\buildrel>\over{\scriptstyle\mathtt{\sim}}$}}
\DeclareMathAlphabet{\mathcalligra}{T1}{calligra}{m}{n}

\begin{document}

\title[The \ha\ surface brightness -- radius relation]{The \ha\ surface brightness -- radius relation: a robust statistical distance indicator for planetary nebulae}
\author[D.J. Frew, Q.A. Parker and I.S. Boji\v{c}i\'c]{David J. Frew$^{1,2}$\thanks{E-mail: djfrew@hku.hk}, Q.A. Parker$^{1,2,3}$ and I.S. Boji\v{c}i\'c$^{1,2,3}$
\\
$^{1}$Department of Physics, The University of Hong Kong, Pokfulam Road, Hong Kong, China\\
$^{2}$Department of Physics and Astronomy, Macquarie University, NSW 2109, Australia\\
$^{3}$Australian Astronomical Observatory, P.O. Box 296, Epping, NSW 1710, Australia}
\date{Accepted ; Received ; in original form }

\pagerange{\pageref{firstpage}--\pageref{lastpage}} \pubyear{2002}
\maketitle
\label{firstpage}

\begin{abstract}
Measuring the distances to Galactic planetary nebulae (PNe) has been an intractable problem for many decades.  We have now established a robust optical statistical distance indicator, the \ha\ surface brightness -- radius or \SR\ relation, which addresses this problem.   We developed this relation from a critically evaluated sample of primary calibrating PNe. 
The robust nature of the method results from our revised calibrating distances with significantly reduced systematic uncertainties, and the recent availability of high-quality data, including updated nebular diameters and integrated \ha\ fluxes.   The \SR\ technique is simple in its application, requiring only an angular size, an integrated \ha\ flux, and the reddening to the PN.  From these quantities, an intrinsic radius is calculated, which when combined with the angular size, yields the distance directly.  Furthermore, we have found that optically thick PNe tend to populate the upper bound of the trend, while optically-thin PNe fall along the lower boundary in the \SR\ plane.  This enables sub-trends to be developed which offer even better precision in the determination of distances, as good as 18 per cent in the case of optically-thin, high-excitation PNe.  This is significantly better than any previous statistical indicator.  We use this technique to create a catalogue of statistical distances for over 1100 Galactic PNe, the largest such compilation in the literature to date.  Finally, in an appendix, we investigate both a set of transitional PNe and a range of PN mimics in the \SR\ plane, to demonstrate its use as a diagnostic tool.  Interestingly, stellar ejecta around massive stars plot on a tight locus in \SR\ space with the potential to act as a separate distance indicator for these objects.
\end{abstract}


\section{Introduction}

One of the greatest difficulties still facing the study of planetary nebulae (PNe) in our own Galaxy has been the problem of determining accurate distances to them.  Due to the wide range of effective temperatures and bolometric luminosities seen in their ionising stars, they are not suitable as standard candles\footnote{However the well-known PN luminosity function (PNLF) works as an effective distance indicator for an ensemble of luminous PNe (see Ciardullo 2012, for a recent review).}, nor can their expanding PNe be used as standard rulers.  Indeed, the most reliable distances are for PNe located in external galaxies, such as M\,31 and the Large and Small Magellanic Clouds (e.g. Jacoby \& De Marco 2002; Reid \& Parker 2006).  This problem has led to the application of a range of secondary distance methods for Galactic PNe, which we will evaluate as part of this work.  For reviews of the older Galactic distance scales, the reader is referred to the works of Minkowski (1965), Gurzadyan (1970), Smith (1971) and Liller (1978). 
The PN distance-scale  problem was nicely summarised by Ciardullo et al. (1999, hereafter CB99) who stated that ``\emph{it is unfortunately less obvious $\ldots$ how one could devise a new `grand unification' calibration that simultaneously handles both the lower surface brightness objects that prevail among the nearby nebulae and the brighter PNe that dominate samples like those in the Galactic bulge and extragalactic systems.  We leave this daunting task to future workers.}''

So far accurate primary distances (with uncertainties $<$10\%) are known for less than one per cent of the more than 3400 Galactic PNe that have so far been catalogued (Parker et al. 2015, in preparation), of which the most accurate come from trigonometric parallaxes of their central stars (CSPNe; Benedict et al. 2003, 2009; Harris et al. 2007).  Generally speaking, distance estimates to the bulk of PNe are statistical in nature and rely on quantities which have a large observed dispersion (e.g. Cahn, Kaler \& Stanghellini 1992, hereafter CKS; Stanghellini, Shaw \& Villaver 2008, hereafter SSV).  Uncertainties in the Galactic PN distance scale have been significant, up to factors of three or more (e.g. Zhang 1995, hereafter Z95; Van de Steene \& Zijlstra 1995; CB99; Napiwotzki 2001; Phillips 2002; SSV).   This uncertainty severely hampers attempts to derive meaningful physical quantities for most Galactic PNe.  Almost every quantity of interest, including nebular radii, masses, luminosities and dynamical ages, and the luminosities and masses of their CSPNe, depends on accurate knowledge of their distances, as do all statistical determinations of the PN scale height, space density, and formation rate (Ishida \& Weinberger 1987).   

In this paper we develop and calibrate a new optical statistical distance indicator, the \ha\ surface brightness -- radius relation (\SR\ relation hereafter).  Here we address the problem posed by CB99, and our results show that the controversy surrounding the long-running PN distance scale problem has finally been put to rest.   Our technique is relatively simple in its application, requiring an angular size, an integrated \ha\ flux, and the reddening of the PN.  From these quantities, an intrinsic radius is calculated, which when combined with the angular size, yields the distance directly. We have chosen \ha\ as the most optimum emission-line, firstly as it best represents the nebular ionized mass, and secondly because a number of narrowband \ha\ imaging surveys have recently become available, from which large numbers of accurate integrated fluxes, diameters, and surface brightnesses can be determined.  These include the SuperCOSMOS H-alpha Survey (SHS; Parker et al. 2005; Frew et al. 2014a), the INT Photometric H-Alpha Survey (IPHAS; Drew et al. 2005),  
the VST Photometric H-Alpha Survey (VPHAS+; Drew et al. 2014), and the lower-resolution Southern H-Alpha Sky Survey Atlas (SHASSA; Gaustad et al. 2001) and Virginia-Tech Sky Survey (VTSS; Dennison, Simonetti \& Topasna 1998).

Our paper is arranged as follows: in \S2 we review the various distance methods that have been used in the literature, while we compile a sample of critically-assessed primary distances in \S3, which underpins our new relation. In \S4 we describe the \SR\ relation in detail, and discuss the increase in accuracy obtained by using specialised sub-trends. We also examine the theoretical basis for the relation in this section.  We present  our catalogue of \SR\ distances in \S5 (presented in full as an online supplement), and in \S6 we investigate the dispersion of the relation, before comparing our final mean distance scale with previous work in \S7.  This work refines the distance scales presented by Frew (2008; hereafter F08), and the earlier preliminary results given by Pierce et al. (2004), Frew \& Parker (2006, 2007), and Frew, Parker \& Russeil (2006).  We present our conclusions and suggestions for future work in \S8, including a discussion of the data expected from the recently launched GAIA astrometric satellite, and how our \SR\ relation will remain complimentary to that well into the future.  Finally, in an appendix, we investigate both a set of transitional PNe and a range of PN mimics in the \SR\ plane, to test its use as a diagnostic tool.  Preliminary results show it to have great promise.

\section{Previous Statistical Methods}\label{sec:statistical}

The last few decades have seen a wide range of techniques used to measure PN distances, both primary methods which generally have the highest accuracy, and statistical (secondary) methods, which can have considerable uncertainties (of factors of two or more), even if appropriately calibrated.   In this section we briefly review the standard statistical techniques previously used in the literature.  The reader is referred to the review of Smith (2015) for a fuller discussion of the limitations and biases of each distance technique. 

The classical Shklovsky method was the first statistical method to be applied that had any claim to veracity.  It assumed a constant ionised mass (typically 0.2\,M$_{\odot}$) for the PN shell and was first applied by Minkowski \& Aller (1954) and Shklovsky (1956).  Osterbrock (1960) applied this method to NGC\,3587 and O'Dell (1962) used newly-determined H$\beta$ fluxes to derive an early distance scale, based on emission theory and the assumption of constant ionised mass; several calibrating nebulae were used to determine the mean ionised mass for PNe.
This was followed by the work of Abell (1966), using `photored' fluxes for over 90 evolved PNe, before being further developed by Cahn \& Kaler (1971).  This distance scale was later utilised by Kaler (1983), Shaw \& Kaler (1989), and Kaler, Shaw \& Kwitter (1990).   Other Shklovsky scales have used the observed proper motions of the central stars, in combination with assumptions regarding their space motions (e.g. O'Dell 1962) to fix the zero point.  Cudworth (1974) undertook a statistical calibration of the PN distance scale using a large set of uniformly obtained proper motions, obtaining one of the longest scales to date.  However, as these are constant-mass scales, distances to the youngest compact PNe and the largest evolved PNe were in general overestimated and underestimated respectively.

In the simplest terms, and assuming a constant ionised mass, the nebular radius ($r$) increases as the PN evolves, and the mean electron density ($n_{e}$) falls in sympathy.  
If the mean electron density can be determined from measurements of \OII\ or \SII\ doublet intensities, the intrinsic nebular radius can be calculated.  Comparing this to the angular size of the PN leads directly to a distance via simple trigonometry.  Variations on this technique, by assuming an ionised mass derived from a set of calibration objects at known distance and using the observable electron density and \hb\ flux to infer a distance, have been utilised by Kingsburgh \& Barlow (1992) and Kingsburgh \& English (1992).  A more novel method has been utilised by Meatheringham, Wood \& Faulkner (1988), who found that Magellanic Cloud (MC) PNe fall on fairly tight plane in dynamical age -- density -- excitation-class space.  For a sample of Galactic PNe the dynamical age was estimated from the observed electron density and excitation class, and once the expansion velocity is measured, the intrinsic radius can be inferred.  Comparing this number with the angular size leads directly to a distance.  

An equally common approach in the literature is a variable-mass derivation of the Shklovsky method, as it is now known that PNe have a range of ionised masses, and the standard method can be inaccurate for highly evolved PNe with more massive shells (e.g. Buckley, Schneider \& van Blerkom 1993).  
An initial method was developed by Daub (1982), who empirically related the ionised mass to an optical thickness parameter, derived from the observed 5\,GHz (6\,cm) radio flux density ($F_{\rm \tiny 5}$), and the angular radius, $\theta$ (in arcsec).  The thickness parameter $\mathscr{T}$, is defined as:
\begin{equation} 	
\label{eq:daub}
\mathscr{T} = {\rm log}\,({\theta^2}/{F_{\rm 5}})
\end{equation} 

A value of $\mathscr{T}$ = 3.65 (corresponding to $r$ = 0.12 pc) was found to separate optically-thick from optically-thin PNe, which were assumed to have a constant mass at large radii.  This approach was re-calibrated by CKS, based on a refined set of nebulae with primary distance estimates, setting the thick-thin transition at $\mathscr{T}$ = 3.13 (corresponding to $r$ = 0.09 pc).  The ionized mass was determined using: 
\begin{equation} 	
\label{eq:CKS92}
{\rm log}\,M = 
\begin{Bmatrix}
\mathscr{T} - 4~~~~{\rm for}~ \mathscr{T} < 3.13   \\
-0.87~~~~{\rm for}~ \mathscr{T} > 3.13   
\end{Bmatrix}
\end{equation} 


The intrinsic radius (in pc) was then calculated from the following expression 
\begin{equation} 	
\label{eq:CKS_radius}
{\rm log}\,r = 0.4\,{\rm log}\,M + 0.2\,\mathscr{T} - 1.306
\end{equation} 

Finally the distance, $D$ (in pc), was determined from the well-known formula: 
\begin{equation} 	
\label{eq:small_angle}
D = \frac{206,265~r}{\theta}
\end{equation} 

Recently SSV re-calibrated the CKS scale using updated Galactic distances as well as data for a large set of LMC and SMC PNe, where the thick-thin transition was now determined to be at $\mathscr{T}$ = 2.1,  or a smaller radius, $r$ = 0.06 pc.  The SSV scale has been commonly used to date.  We will compare our distance results with their work in \S\ref{sec:SB_comparison}.  

Other statistical approaches use an ionised mass that is a continuous function of linear radius, as estimated from the surface brightness (e.g. Maciel \& Pottasch 1980;  Pottasch 1980, 1984).     
In general terms the ionised mass--radius relation can be expressed as:
\begin{equation} 	
\label{eq:mass-radius}
M_{\rm ion} \propto r^{\beta} 
\end{equation} 
where $\beta$ is a power-law index determined through observation.    While Maciel \& Pottasch (1980) found $\beta$ = 1, other authors derived significantly different values for $\beta$ (see Milne 1982; Pottasch 1984; Kwok 1985; Zhang 1995), to be further discussed in \S\ref{sec:SBr_theory}.  For more detailed discussions of this point, the reader is referred to Kwok (1985, 1993) and Samland et al. (1993).

A natural variant of the $M_{\rm ion}$--$r$ relationship is the brightness temperature--radius ($T_b$--$r$) relationship.   Again the primary observables are the 5\,GHz radio flux, or an equivalent radio or optical Balmer-line flux, and the angular radius, from which a surface brightness can be calculated.   
Various versions in the radio domain have been proposed by Amnuel et al. (1984), Van de Steene \& Zijlstra (1994, 1995), Buckley \& Schneider (1995),  Zhang (1995), Bensby \& Lundstr\"om (2001), Phillips (2002, 2004b) Uro\u{s}evi\'c et al. (2009) and Vukoti\'c et al. (2009), amongst others.  
The 5\,GHz brightness temperature, $T_{\rm b}$ (in K), is defined as:
\begin{equation}
\label{eq:temperature}
T_{\rm b} = \frac{c^2}{2 \pi k \nu^2} \frac{F_{\rm \tiny 5}}{\theta^2}~~~ \simeq 18.3 \frac{F_{\rm \tiny 5}}{\theta^2}
\end{equation}

Based on a set of calibrating nebulae with known distances, an expression for the distance can then be derived, of the form:
\begin{equation}
\label{eq:radio-dist}
{\rm log}\,D = a - b\,{\rm log}\,\theta - c\,{\rm log}\,F_{\rm \tiny 5}
\end{equation}

where $a$, $b$ and $c$ are empirically determined constants.  Relations of this form were used by Zhang (1995), Van de Steene \& Zijlstra (1995), and Bensby \& Lundstrom (2001), with relatively small ($<$10\%) differences in the proportionality constants derived in each study.  Schneider \& Buckley (1996) took an alternative approach, since they considered a single power-law inadequate to handle both young and old PNe. They fit a second-order polynomial to their calibration sample.  However, with the exception of the youngest PNe, a single power law is a reasonable fit to the range of surface brightness seen in PNe, from compact nebulae down to the very faintest objects dissolving into the interstellar medium (ISM).  Also, in an attempt to develop a novel robust approach for distance scale calibration, Vukotic et al. (2014) utilized the calibrating sample from SSV. Instead of using the usual fitting procedure they calculated the density of the data points in the fitting plane which resulted in probability distributions of diameter values for selected values of surface brightness.  A comparison of some of these radio-based distance scales with our  \SR\ distance scale is given in \S\ref{sec:SB_comparison}.

Another potential distance technique is based on the subset of PNe which have central stars still evolving left along the constant luminosity track in the theoretical Hertzsprung-Russell (HR) diagram.  If a canonical central star mass of 0.6\,M$_{\odot}$ (or a similar value) is assumed, and a temperature of the CSPN can be determined, then an absolute visual magnitude can be predicted using an appropriate bolometric correction (e.g. Vacca, Garmany \& Shull 1996).  
If accurate reddening-corrected photometry is available, then a distance directly follows.  Note that the resultant distance scale depends on the adopted mean CSPN mass.  Mal'kov (1997, 1998) seems to be the first to mention such a technique, but did not apply it, and it was first utilised (using bolometric magnitudes) by Phillips (2005b).  A related approach is to assume a constant absolute magnitude (i.e. a standard candle) for a homogenous sub-sample of CSPNe.  Phillips (2005a) took this approach for a set of CSPNe on the cooling track in evolved PNe but there appears to be a significant spread ($\sim$2 mag) in the absolute magnitudes of the CSPNe in old PNe (see F08), meaning the technique needs to be applied with caution.

Other statistical methods assume a standard ruler technique such as the angular size of the waists in Type\,I bipolar PNe, assuming these all have a similar intrinsic diameter (Phillips 2004a), but this approach was criticised by Frew et al. (2006).   Similarly, Gurzadyan (1970) used the angular diameter of the \HeII\ Str\"omgren zone at the centre of optically-thick PNe to estimate a distance.  However the systematics are not well quantified, and the method saw little application owing to the wide variety of intrinsic diameters, structural parameters, and excitation classes seen in PNe.
Finally there are also methods based on mid-infrared (MIR) fluxes, obtained either from IRAS (Tajitsu \& Tamura 1998) or MSX data (Ortiz et al. 2011).   These generally utilized an assumed dust mass, scaling the distances according to the observed MIR fluxes.

\section{Calibration of a new statistical distance scale}\label{sec:scale_calibration}

CB99 stressed the importance of deriving a statistical calibration that simultaneously handles both luminous PNe and the demographically common evolved, faint PNe.  These represent a population that are usually avoided as calibrating objects, and this may be the reason for the systematic offsets that have plagued the various statistical distance scales in the past (e.g.  Pottasch 1996).   
Previously, Stanghellini et al. (2002) found a relationship between \ha\ surface brightness and radius for a sample of LMC PNe, and Jacoby et al. (2002) outlined the potential of using an \SR\ relation as a distance indicator.  Such a relation is analogous to the radio $T_b$--$r$ relationships that have been the basis of many previous statistical distance scales (see \S\ref{sec:statistical}).  

Independently, we came to the same conclusion regarding the benefits of using an \SR\ relation as a distance indicator, based on a sample of \emph{ Galactic PNe} (see Pierce et al. 2004).  Our new relation also has the added benefit of including the most extreme PNe at the very bottom of the PNLF, which have traditionally been selected against in the radio regime (Zhang \& Kwok 1993; CB99).  
We chose to use the \ha\ emission-line (rather than the radio continuum) owing to the recent increase in reliable \ha\ fluxes available for Galactic PNe.  In particular, Frew et al. (2013, hereafter FBP13) and Frew et al. (2014a) have presented accurate \ha\ fluxes for about 1300 PNe in total, a significant fraction for the first time. However, a disadvantage in using the brighter \ha\ flux instead of \hb\ is that a correction for the \NII contribution is often required, though if done correctly the derived \ha\ integrated flux is accurate (see the discussion in FBP13).   Drawing on our new database of fundamental parameters for PNe (Parker et al., in prep.), which includes fluxes, extinctions, emission-line ratios and angular diameters, the \SR\ relation has been calibrated across the full range of surface brightness seen in PNe, from young, high-density, luminous objects like NGC\,7027 through to some of the faintest known PNe such as TK\,1 (Ton\,320).  

It is crucially important that the sample be as free from systematic bias as possible.  Earlier authors have diluted the precision of their calibrating sample by including PNe with poorly known distances, or by not weighting the individual distance estimates to the PN calibrators with appropriate uncertainties (cf. Bensby \& Lundstr\"om 2001; Phillips 2002, 2004b).  Furthermore, more than one study has inadvertently included \HII\ regions, symbiotic outflows, and other mimics as `PN calibrators', which add significant noise to the derived relationship.  We have used a range of diagnostic tools to remove these contaminants (Frew \& Parker 2010), so our approach does not suffer from the same issues.

\subsection{A Critical Evaluation of Primary Methods and Distances}\label{sec:evaluation}

Unfortunately, published primary distances are of widely varying quality, but a number of primary methods have been used with varying degrees of success; for earlier reviews, see Acker (1978) and Sabbadin (1986).   These techniques include direct trigonometric parallaxes of the CSPN (Harris et al. 2007; Benedict et al. 2009), or a photometric or spectroscopic parallax determined for a physical companion  to the CSPN (Bond \& Ciardullo 1999; CB99).  The analysis of eclipsing binary CSPNe (e.g. Bell, Pollacco \& Hilditch 1994) is potentially one of the most accurate to constrain PN distances, and the membership of a PN in a star cluster of known distance is also a highly promising technique, especially for the future (see Parker et al. 2011).  

A description of the primary distance methods used to define the Galactic calibrating sample are briefly described in the following subsections.  Individual PN distances are tabulated in each section, and a critical assessment of their associated uncertainties also follows.   These literature distances have been carefully examined, and in many cases revised with better systematics, and we also include several new kinematic and extinction distance determinations derived as part of this work.    
We then present a final set of calibrating distances in  \S\ref{sec:cal_sample}, which has allowed an \SR\ relationship to be defined over six decades in log \ha\ surface brightness.   It should also be emphasised that no statistical distances from other studies have been used as calibrators for our \SR\ relation (cf. Bensby \& Lundstr\"om 2001; Ortiz et al. 2011).

\subsubsection{Trigonometric distances}\label{sec:trig_parallax}

Direct trigonometric parallaxes have been measured for more than a dozen nearby CSPNe, either from the ground (e.g.  Harris et al. 1997, 2007), the Hipparcos satellite (Acker et al. 1998; van Leeuwen 2007), or the Hubble Space Telescope (HST; Benedict et al. 2003, 2009).  
The ground-based US Naval Observatory (USNO) CCD parallaxes of Harris et al. (2007) form an homogenous sample of accurate distances for several nearby PNe, and Smith (2015) has shown that they form a reliable, internally consistent dataset.  Owing to uncertain systematics, we have not used the ground-based data from Guti\'errez-Moreno et al. (1999).  
The Hipparcos parallaxes (van Leeuwen 2007) have also been shown to be problematic (e.g. Smith 2015), especially for compact PNe where subtle surface brightness variations across the PN may have had an undue influence on the astrometric reductions, therefore the Hipparcos parallaxes have not been used as calibrating data (cf. F08).  
Finally, we also adopt the distance to the young, compact nebula K\,3-35 (Tafoya et al. 2011), determined using VLBI Exploration of Radio Astrometry (VERA) array observations of a bright water maser in the nebula\footnote{Maser trigonometric distances for several pre-PNe are discussed in Vickers et al. (2015).}.   

Note that the trigonometric method is susceptible to the so-called Lutz-Kelker (L-K) bias (Lutz \& Kelker 1973; Smith 2003, 2006; Francis 2014) which causes measured parallaxes to  be systematically greater than their actual values in a statistical sense, and is broadly related to the Trumpler-Weaver bias (Trumpler \& Weaver 1953).  As emphasised by van Leeuwen (2007) and Francis (2014), the L-K bias is a sample statistical correction, and has not been applied to individual distances.   
In the future, the number of trigonometric parallaxes for CSPNe will be revolutionised with the results from the Gaia satellite (Perryman et al. 2001; Bailer-Jones 2002).  This point will be further discussed in our conclusions.  Table~\ref{trig_distances} summarises the determinations taken from the literature.

\begin{table}
\begin{center}
\caption{Trigonometric distances for planetary nebulae from the literature used as calibrating objects. Note that the Hipparcos parallaxes have been excluded from this table.}
\label{trig_distances}
\begin{tabular}{llc}
\hline
Name           					&     ~~~$D$ (pc)  					& Reference	\\
\hline
Abell 7     					& 	 676$^{+267}_{-150}$  			&  H07 	 	\\
Abell 21     					& 	 541$^{+205}_{-117}$  			&  H07  		\\
Abell 24     					& 	 521$^{+112}_{-79}$     			&  H07 		\\
Abell 31     					& 	 621$^{+91}_{-70}$  				&  B09 		\\
Abell 74     					& 	 752$^{+676}_{-242}$    			&  H07 		\\
Bode 1						&     477$^{+28}_{-25}$    				&  H13		\\   	
K 3-35        					&     3900$^{+700}_{-500}$~~~~~~~ 	&  T11 		\\
NGC 6720		 			&      704$^{+445}_{-196}$       			&  H07 		\\
NGC 6853~~~~~~~~~~~  		&      405$^{+28}_{-25}$				&  B09 		\\
NGC 7293$^{\ast}$  			&      216$^{+14}_{-12}$				&  B09 		\\
PuWe 1     					&     365$^{+47}_{-37}$     				&  H07 		\\
Sh 2-216$^{\ast}$    			&     129$^{+6}_{-5}$   				&  H07 		\\
TK 1    						&     532$^{+113}_{-80}$     			&  H07 		\\
\hline
\end{tabular}
\end{center}
{\scriptsize
\begin{flushleft}
References:  B09 -- Benedict et al. (2009); H07 -- Harris et al. (2007); H13 -- Harrison et al. (2013); T11 -- Tafoya et al. (2011). 
\end{flushleft}
}
\end{table}

\subsubsection{Photometric distances}\label{sec:phot_distances}

This method estimates a spectroscopic or photometric parallax for a companion star of normal spectral type.  The advantage of using this method was noted early on by Minkowski \& Baum (1960) and  Cudworth (1973, 1977).   The archetype is the well-studied, high-excitation PN, NGC\,246 (Bond \& Ciardullo 1999) and the method has been applied to a number of more distant PNe with wide binary companions, mostly by CB99.   Still other binary systems are dominated by the companion star, usually a B or A main-sequence star, or a cooler giant or subgiant, and for these a spectroscopic parallax is also feasible (e.g. Longmore \& Tritton 1980).  Absolute magnitudes have been taken from De Marco et al. (2013) for main sequence stars, and Schmidt-Kaler (1982) or Jaschek \& G\'omez (1998) for the evolved stars.     

\begin{table}
\begin{center}
\caption{Photometric / spectroscopic distances for resolved companions taken from the literature or derived as part of this study. Spectral types inferred from colours are given in italics.}
\label{photometric_distances}
\begin{tabular}{llll}
\hline
Name      			&   SpT (comp)	  		& ~~$D$   (kpc)					& Reference 			\\
\hline
Abell 14 				&    B7 V      				& 	  $5.6^{+1.0}_{-0.9}$     			&  D14; t.w.     			\\	
Abell 33      			&    \textit{K3 V}     			& 	  $1.17^{+0.18}_{-0.16}$~~~   	&  CB99, t.w.			\\   	
Abell 34		      		&    \textit{G0 V}     			& 	  $1.22^{+0.18}_{-0.16}$ 		&  t.w.					\\     
Abell 79        			&    F0 V           			&      $3.0^{+0.8}_{-0.6}$  			&  RC01, DP13,  t.w.		\\  	
HaTr 5     			&    G8 IV		   		&      $2.10^{+0.40}_{-0.35}$     		&  D14, t.w.      			\\   	
Hen 2-36				&    A2 II-III     				& 	 $1.5^{+1.3}_{-0.8}$     			&  M78, t.w.				\\   	
Hen 2-39				&    C-R3 III		     		&      $7.6^{+1.5}_{-1.3}$  		   	&  MB13, t.w. 			\\   	
H 3-75        			&    G8 III		     			&      $3.3^{+0.8}_{-0.5}$     			&  CB99; BP02, t.w.       	\\   	
K 1-14        			&    \textit{K2 V}	  		&      $3.14^{+0.52}_{-0.44}$ 	      		&  CB99,  t.w.			\\   	
K 1-22        			&    \textit{K2 V}	 		&      $1.34^{+0.22}_{-0.19}$ 	      		&  CB99,  t.w.			\\
LoTr 1         			&    K1 IIIe		       		&    	 $2.4^{+0.4}_{-0.3}$      			&  WG11, TJ13, t.w.		\\    	
LoTr 5         			&    G5 III       				& 	 $0.58^{+0.15}_{-0.14}$         	&  LT80, SH97, t.w.		\\    	
Me 1-1         			&    K2-K3 II		    		&      $6.0^{+1.9}_{-1.4}$         		&  SL04, PM08, t.w.       	\\    	
MPA J1824-1126~~ 	&    K2-K5 III	  			&      $11.8\pm4.1$    				&  FNC14				\\    	
Mz 2           			&    \textit{F3 V}	  		&      $2.33^{+0.58}_{-0.46}$       		&  CB99, t.w.			\\
NGC 246     			&    K0 V	 				&      $0.495^{+0.145 }_{-0.100}$ 		&  WW93, BC99 			\\
NGC 1514       		&    A0-A1 III       			& 	 $0.55^{+0.19}_{-0.15}$         	&  G72, RC10, t.w.		\\   	
NGC 1535       		&    \textit{K0 V}     			& 	 $2.19^{+0.40}_{-0.34}$       		&  CB99,  t.w. 			\\   	
NGC 2346       		&    A5 V 		      			& 	 $0.65^{+0.25}_{-0.20}$          	&  M78, t.w.				\\   	
NGC 3132      			&    A2 IV-V      			& 	 $0.70^{+0.29}_{-0.20}$       		&  M78, CB99, t.w.		\\     
NGC 6818			&    \textit{K1: V} 	  		&      $1.75^{+0.56}_{-0.42}$   		&  BC03, t.w. 			\\     
NGC 6853      			&    \textit{M5 V}     			&      $0.43\pm0.06$			       	&  CB99,  t.w.			\\   	
NGC 7008      			&    \textit{G8 IV}    			& 	 $0.97^{+0.17}_{-0.15}$       		&  CB99, SK92, t.w.		\\   	
Sp 3               			&    \textit{G0 V}       		& 	 $2.22^{+0.61}_{-0.48}$       		&  CB99, t.w. 			\\	
We 3-1          			&    \textit{F7 V}     			&     $1.55^{+0.30}_{-0.25}$  			&  t.w.					\\
WeBo 1         			&    K0 II-III pe	  			&     $3.0^{+0.8}_{-0.7}$  			&  BP03, t.w.   			\\	
\hline
\end{tabular}
\end{center}
{\scriptsize
\begin{flushleft}
Reference:  ~BC99 -- Bond \& Ciardullo (1999); BC03 -- Benetti et al. (2003); BP02 -- Bond \& Pollacco (2002); BP03 -- Bond et al. (2003); CB99 -- Ciardullo et al. (1999); D14 -- Douchin (2014);  FK83 -- Feibelman \& Kaler (1983); FNC14 -- Flagey et al. (2014); G72 -- Greenstein (1972);  LT80 -- Longmore \& Tritton (1980); M78 -- M\'endez (1978); MB13 -- Miszalski et al. (2013); PM08 -- Pereira et al. (2008); RC01 -- Rodr\'iguez et al. (2001); RC10 -- Ressler et al. (2010);  SH97 -- Strassmeier et al. (1997); SL04 -- Shen et al. (2004);  TJ13 -- Tyndall et al. (2013); WG11 -- Weidmann \& Gamen (2011b);  WW93 -- Walsh et al. (1993); t.w. -- this work. 
\end{flushleft}
}
\end{table}

The binary associations evaluated by CB99 have been re-evaluated here using better estimates for the reddening, derived both from unpublished spectroscopic data and from all available CSPN photometry (see F08; De Marco et al. 2013).  Furthermore, none of CB99's {\it possible} or {\it doubtful} associations have been considered (cf. F08), and of the probable associations, the distance of K\,1-27 has been rejected.  This distance, based on the companion being a white dwarf (WD) which was fit to the cooling sequence, is only a quarter of a newly calculated gravity distance (see Table\,\ref{grav_dist}), derived from the  data presented by Reindl et al. (2014).   If the companion to K 1-27 is in turn an unresolved dM / WD pair, the true colour of the WD would be bluer and hence the luminosity larger.  Alternatively, though with low probability, the companion is a background quasar.  This object is further noted in \S\ref{sec:subluminous_PNe}.  

We have revised the luminosity class of the companion of NGC\,7008 to IV (from V as assumed by CB99).  Using $M_V$ = $+3.1$   
for the G8 star (Schmidt-Kaler 1982), the distance is $\sim$900\,pc adopting the C99 reddening, or 970\,pc using our revised value.  Now the ionising star is no longer underluminous as it was using the original distance.  We also determine a revised distance of 3.0\,kpc to the barium K-type giant in WeBo\,1, based on the same arguments as Bond et al. (2003). However, we adopt a larger stellar mass of 4\,$M_{\odot}$, based on the nebula's type\,I chemistry as inferred from the observed \NII/\SII\ ratio (see Fig.\,10 of Smith, Bally \& Walawender 2007).  Table~\ref{photometric_distances} summarises the distance determinations.

Some other companion-dominated systems are not used as calibrators owing to the uncertain luminosity class of the cool star; examples include Abell\,70 (Miszalski et al. 2012), Abell\,82 (CB99), Hen\,3-1312 (Pereira 2004), K\,1-6 (Frew et al. 2011), and IC\,972 (Douchin et al. 2014).   
In other cases, the identification of the central star is in doubt (e.g. RCW\,21; Rauch et al. 1999), or the object is unlikely to be a true PN, such as the nearby bowshock nebula Abell\,35 (F08; Ziegler et al. 2012b).    

\subsubsection{Eclipsing / Irradiated Binaries}\label{sec:eclipsing_binaries}
Eclipsing binaries are fundamental astrophysical yardsticks, but the analysis of the small sample of eclipsing binary CSPNe has led to few distance determinations to date (Pollacco \& Bell 1993, 1994; Bell et al. 1994).  Close binary CSPNe showing a large irradiation (reflection) effect can also be used, such as DS\,1  (Drilling 1985), and LTNF\,1 around BE\,UMa (Liebert et al. 1995; Ferguson et al. 1999).  These methods are partly model dependent however, but offer great promise if the systematics are well understood.  Unfortunately, eclipsing CSPNe are rather uncommon, but a recent very accurate distance for the double-lined binary in SuWt\,2 has been obtained by Exter et al. (2010).  Table\,\ref{table:eclipsing} summarises the adopted calibrating distances.

\begin{table}
\begin{center}
\caption{PN distances derived from modelling close binary central stars.}
\label{table:eclipsing}
\begin{tabular}{lll}
\hline
Name  				&   ~~$D$ (kpc)  						& 	Reference    	 	\\
\hline
Abell 46       			&     1.70 $\pm$ 0.60         				&    PB94          		\\
Abell 63			      	&     2.40 $\pm$ 0.40      				&    BP94         		\\  	
DS 1    				&     0.70 $\pm$ 0.10~~~~~~~~ 			&    RB11       			\\   	
Hen 2-11~~~~~~~    	&     0.70 $\pm$ 0.18$^{\dagger}$			&    JB14       			\\   	
HFG 1                     	&     0.63 $\pm$ 0.32					&    EP05				\\	
LTNF 1     			&     2.0 $\pm$ 0.5$^{\dagger}$			&    F99   			\\   	
SuWt 2			       	&     2.30 $\pm$ 0.20 					&    EB11	 			\\    	
TS 1   				&     21.0 $\pm$ 4.0					&    SM10, TY10		\\
\hline
\end{tabular}
\end{center}
{\scriptsize
\begin{flushleft}
Note:~$^{\dagger}$Assumed uncertainty.\\  References: ~BP94 -- Bell et al. (1994);  EB11 -- Exter et al. (2011); EP05 -- Exter  et al. (2005); F99 -- Ferguson et al. (1999);  JB14 -- Jones et al. (2014);  PB94 -- Pollacco \& Bell (1994); RB11 -- Ribeiro \& Baptista (2011); SM10 -- Stasi\'nska et al. (2010); TY10 -- Tovmassian et al. (2010).     
\end{flushleft}
}
\end{table}

We also note the bipolar object Hen\,2-428, which has recently been suggested to contain a super-Chandrasekhar, double-degenerate nucleus (Santander-Garc\'ia et al. 2015), indicating this is a potential Type Ia supernova progenitor.  However, this interpretation has been questioned by Garcia-Berro, Soker \& Althaus (2015).  We note that a short model distance of 1.4\,kpc is derived from the analysis of Santander-Garc\'ia et al. (2015), making the surrounding nebula very underluminous as well as the central star's luminosity discrepant with standard post-AGB evolutionary tracks (Garcia-Berro, Soker \& Althaus 2015).  Our mean \SR\ distance of 2.7\,kpc suggests the evolutionary interpretation of Garcia-Berro et al. may be more likely.

\subsubsection{Cluster Distances}\label{sec:cluster-distances}  
Physical membership of a PN in an open or globular star cluster provides an accurate distance, and is an important key that can help to unlock many of the problems facing PN research (Parker et al. 2011).   At present the number of Galactic PNe thought to be genuine members of clusters is small; a few at best in open clusters, with four Galactic globular clusters currently thought to contain PNe  (Jacoby et al. 1997).
Historically, NGC\,2438 was assumed to be a member of the young open cluster M~46 (NGC\,2437), but Kiss et al. (2008) showed that they were unrelated.\footnote{Vickers et al. (2015) summarised the evidence showing that the bipolar, symbiotic-like outflow OH\,231.8+4.2 is a bona fide member of this cluster.} 
Additionally, NGC 2818 was thought to be physically associated with the open cluster of the same name (e.g. Tifft, Connolly \& Webb 1972), but Mermilliod et al. (2001) claimed the objects were unrelated.  However, recent work by V\'azquez (2012), has shown that the PN velocity is consistent with membership.   In the meantime, PHR\,J1315-6555 was shown by Parker et al. (2011) to be a physical member of the intermediate-age open cluster ESO\,96-SC04. The compact object NGC 6741 has been suggested to be a possible member of Berkeley\,81 (Sabbadin et al. 2005), but while the distances are comparable, the radial velocity of the cluster is 8\kms\ greater (Sabbadin et al. 2005; Magrini et al. 2015), suggesting non-membership of the PN.  
It should be noted that the recent increases in numbers of both Galactic PNe and open clusters (Dias et al. 2002) have increased the probability of positional coincidences between these two classes of object.   Lists of coincidences between clusters and PNe have been given by Kohoutek (2001) and Majaess, Turner \& Lane (2007), and recently two more possible associations (Abell\,8 and Hen\,2-86) have been presented in the literature (Turner et al. 2011; Moni Bidin et al. 2014).  The currently suggested associations are discussed individually below.\\

\noindent{\it NGC 2818}: ~~Tifft et al. (1972) argued that NGC 2818 was a member of the open cluster of the same name, and this became accepted as a valid association.  Dufour (1984) and Pedreros (1989) also assumed a physical association, but gave conflicting distances to the cluster.   However, Mermilliod et al. (2001) obtained accurate velocities for 12 cluster red giants to obtain a mean velocity of  $V_{\rm hel}$ = $+20.7 \pm 0.3$ kms$^{-1}$, significantly different to the PN velocity of $-1\pm3$ kms$^{-1}$ (Meatheringham et al. 1988), suggesting a line-of-sight coincidence. More recently, V\'azquez (2012) reanalysed the complex kinematics of the nebula, finding a systemic heliocentric velocity ($\sim$27 \kms) in closer agreement with the open cluster, suggesting  membership, with which we now concur.  The cluster distance of 3.0\,kpc is derived from the reddening and distance modulus given by Mermilliod et al. (2001), in turn based on a deep colour-magnitude diagram from Stetson (2000).\\

\noindent{\it PHR J1315-6555}:~~Parker et al. (2011) undertook a detailed study of the physical association between this bipolar PN and the intermediate-age open cluster ESO\,96-SC04 (AL 1).  Majaess et al. (2014) refined the distance to the cluster to $10.0\pm0.4$\,kpc, which we have adopted herein.\\

\noindent{\it BMP J1613-5406}:~~This evolved bipolar PN is a likely member of the Cepheid-hosting open cluster NGC 6067, based on positional coincidence and close agreement in radial velocities.  A full account of this very interesting association will be published separately (Frew et al., in preparation). \\ 

\noindent{\it Abell~8}:~~Bonatto, Bica \& Santos (2008) have identified a new intermediate-age open cluster in the field of this faint, round PN, giving a reddening of \EBV\ = 0.29 $\pm$ 0.03 and a distance, $D$ = 1.7 $\pm$ 0.1 kpc.  Based on their similar radial velocities, Turner et al. (2011) argued that this is a real association.    However, there are difficulties with this assessment.  Using the integrated flux of $F$(\ha) = $-11.90 \pm 0.10$ (FBP13), an average reddening, \EBV\ = 0.51 $\pm$ 0.09 (Kaler 1983; Ali 1999;  Phillips, Cuesta \& Kemp 2005), and a diameter of 60\arcsec\ (Abell 1966), the PN plots well below other optically thick PNe of similar surface brightness in  \SR\ space.  We conclude that the PN is either a cluster non-member or that the cluster distance is significantly in error.  Owing to these uncertainties, we have not used Abell\,8 as a primary calibrator.\\ 

\noindent{\it Hen 2-86}:~~Moni Bidin et al. (2014) suggested this compact PN was a likely member of NGC 4463, primarily based on the similarities in their radial velocities.  However the reddening to the PN, \EBV\ = 1.3 -- 1.5, is much greater than the cluster value, \EBV\ = 0.42.  Those authors suggested the PN shows high internal reddening, but the amount would be greater than any other PN reliably measured to date (see Phillips 1998).  Owing to this discrepancy, we prefer not to use Hen\,2-86 as a primary calibrator.\\   

\noindent{\it Globular cluster PNe}:~~Both Pease\,1, also known as K\,648  (Buell et al. 1997; Alves, Bond \& Livio 2000) and the peculiar H-deficient nebula GJJC\,1 (Cohen \& Gillett 1989; Borkowski \& Harrington 1991) are bona fide members of their respective globular clusters, M\,15 (NGC\,7078) and M\,22 (NGC\,6656).  ~Pease\,1 has been imaged with HST and has good estimates of its angular size (Alves et al. 2000) and integrated flux which qualify it to be a primary  calibrator.    
Jacoby et al. (1997) conducted an extensive search for PN candidates in Galactic globular clusters, finding two new examples, JaFu\,1 in Palomar\,6 and JaFu\,2 in the luminous cluster NGC\,6441.  JaFu\,2 is a certain member of NGC 6441, but JaFu\,1 was a less convincing candidate, owing to its large separation from the core of Pal\,6 (though still within the tidal radius), and its radial velocity being only marginally consistent with membership.  However, a new cluster velocity,  $V_{\rm hel}$ = +181 $\pm$ 3\,kms$^{-1}$ (Lee, Carney \& Balachandran 2004), greatly increases the membership probability.  JaFu\,1, JaFu\,2 and Ps\,1 are all adopted as primary calibrators.

\begin{table}
\begin{center}
\caption{Adopted PN calibrators from cluster associations, separated into young and intermediate-age clusters (top) and old globular clusters (bottom).}
\label{cluster_distances}
\begin{tabular}{llcl}
\hline
PN   						&    Cluster   			&  $D_{\rm clust}$ (kpc)  	&  References     			\\
\hline
Abell 8        					&    Bica 6   			&  1.60 $\pm$ 0.11 		&   TR11          			\\	
BMP J1613-5406   				&    NGC 6067   		&  1.70 $\pm$ 0.10  		&   F15		    			\\	
Hen 2-86		  				&    NGC 4463   		&  1.55 $\pm$ 0.10  		&   MB14     				\\	
NGC 2818 					&    NGC 2818           	&  3.0 $\pm$ 0.8   			&   MC01, V12    			\\   	
PHR J1315-6555 				&    ESO 96-SC4   		&  10.0 $\pm$ 0.4			&   PF11, T14     			\\	
\hline
GJJC 1  						&    NGC 6656      		&   3.2 $\pm$ 0.3			&   H96        				\\  	
JaFu 1     					&    Palomar 6      		&   7.2 $\pm$ 0.7   			&   H96, J97, LC04  		\\
JaFu 2     					&    NGC 6441      		&  13.6 $\pm$ 1.4			&   H96, J97, D08       		\\     %
Pease 1      					&    NGC 7078        	&  10.3 $\pm$ 0.9  			&   vB06    	 			\\  	
\hline
\end{tabular}
\end{center}
{\scriptsize
\begin{flushleft}
References:  D08 -- Dallora et al. (2008); F15  -- Frew et al. (2015, in prep.); H96 -- Harris (1996);  J97 -- Jacoby et al. (1997);   LC04 -- Lee et al. (2004);  MB14 -- Moni Bidin et al. (2014);  MC01 -- Mermilliod et al. (2001);  PF11 -- Parker et al. (2011);  TR11 -- Turner et al. (2011);  V12 -- V\'azquez (2012); vB06 -- van den Bosch et al. (2006).   
\end{flushleft}  	
}
\end{table}

\subsubsection{Model Atmosphere (Gravity) Distances}\label{sec:gravity}

This is potentially a powerful method to determine spectroscopic distances directly for the CSPN (cf. Heap 1977).  It aims to determine the stellar effective temperature and the surface gravity based on an NLTE model atmosphere analysis (e.g. M\'endez et al. 1988; Napiwotzki 2001).  In principle, it is an elegant method, albeit partly model dependent. It appears most published distances have systematic errors, with the greatest observational uncertainty being the determination of the surface gravity, expressed as log\,$g$ (e.g. Pottasch 1996; Rauch et al. 2007).   
The other observables are the visual magnitude and reddening.  From these data, the surface flux, mass and intrinsic radius of the star can be inferred, and using the reddening-corrected magnitude, a distance can be directly determined.   The distance is derived using the following equation (M\'endez et al. 1988):
\begin{equation}
\label{eq:grav}
D^{2} = 3.82 \times 10^{-9} \ {M_{c} F_{\star} \over g} \ 10^{0.4V_{0}}
\end{equation}

where $D$ is the distance in kpc, $M_{c}$ is the stellar (core) mass in solar units, $F_{\star}$ is the monochromatic Eddington flux in units of \ergcms\,\AA$^{-1}$ at $\lambda$5480\,\AA\ (Heber et al. 1984), $g$ is the surface gravity in cm\,s$^{-1}$ and $V_{0}$ is the extinction-corrected visual magnitude.
In turn the Eddington flux can be suitably approximated by the following linear equation if the effective stellar temperature, $T_{\star}$ (in K), is known (Cazetta \& Maciel 2000):
\begin{equation}
\label{eq:eddington}
F_{\star} = 1.85 \times 10^{4} \ T_{\star} \ - \ 9.97 \times 10^{7}
\end{equation}

Nonetheless there are caveats to this approach, and a number of criteria have been employed to minimise any bias in the adopted distance scale.  Because it is often difficult to simultaneously fit a model atmosphere to all the Balmer lines in the optical spectrum of a hot WD (the Balmer line problem) due to the incomplete treatment of metal opacities in the models (e.g. Werner 1996), there can be  significant errors in the effective temperature and the surface gravity, though modern analyses consider more detailed treatments of the metal lines (e.g. Gianninas et al. 2010).  The problem had also been noted by Pottasch (1996) who found that the log\,$g$ values derived from the often-used H$\gamma$ line profile are often systematically too low  (see also Rauch et al. 2007).  

Indeed, several independent lines of evidence point to problems with some of the published determinations, especially some of the older ones (see Pottasch 1996; Smith 2015).  More specifically, the log\,$g$  values are often underestimated, especially at low to moderate surface gravities.  This is illustrated in Fig.\,2 of Napiwotzki (1999), where the mean mass of CSPNe with log\,$g$ $<$6.0 is considerably less than the mean mass of the higher gravity objects, indicating a systematic underestimation of the gravities.  As a further example, the gravity distances derived from the Lyman-line data of Good et al. (2004) are in better agreement with the USNO trigonometric distances, than the Balmer-line determinations, and in turn, the older Balmer determinations of Napiwotzki (1999, 2001).  As another consistency check, the mean mass of an ensemble of DAO WDs (see table\,5 of Good et al. 2005) using the Lyman method agrees better than the Balmer method with the canonical WD average mass of 0.60\,$M_{\odot}$ (e.g. Tremblay, Bergeron \& Gianninas 2011; Kleinman et al. 2013).   Yet despite recent advances in NLTE modelling, systematic errors in the determination of the surface gravity persist.  Traulsen et al. (2005) give a surface gravity for the CSPN of the Helix nebula, as log\,$g$ = 6.3 (in cgs units).  The resulting distance of 780\,pc is way outside the error bar of the recent trigonometric distance of 216$^{+14}_{-12}$\,pc (Benedict et al. (2009).   Even for the well-studied star LS\,V\,+46~21, the CSPN of Sh\,2-216, there remains an unexplained discrepancy between the recent spectroscopic distance of Rauch et al. (2007) and the well-determined parallax distance from Harris et al. (2007).    

Pauldrach, Hoffmann  \& M\'endez (2004) have taken a different approach, also based on model atmospheres.  The mass and radius of the CSPN are calculated from the mass loss rate, $\dot{M}$ and the terminal wind velocity $v_{\infty}$, as estimated from a fit to the spectral lines.  However, very high masses were determined for some CSPNe, near the Chandrasekhar limit, and the resulting very large distances have not been supported by other methods (see the discussion of Napiwotzki 2006).   They have not been considered further.  

In order to derive appropriately weighted mean gravity distances (in cases where two or more NLTE analyses exists in the literature), all suitable $T_{\rm eff}$ and log\,$g$ determinations have been compiled to be used in conjunction with updated reddening values and visual magnitudes (e.g. F08; De Marco et al. 2013) to calculate a new, internally consistent data-set.   Preference has been given to the most recent  analyses.    
Table~\ref{grav_dist} gives the various PN central stars and the resulting gravity distances derived using equations~\ref{eq:grav} and \ref{eq:eddington} above.  The stellar mass (needed for the equation~\ref{eq:grav}) has been determined from the log\,$g$ -- $T_{\rm eff}$ diagram (not shown) from a comparison with the evolutionary tracks of Bl\"ocker (1995) and Vassiliadis \& Wood (1994), interpolating linearly if necessary.    Our new distances may differ somewhat from values published prior, due to slight differences between our adopted magnitudes, reddenings, and temperatures, and individual determinations found in the literature.

\begin{table*}
{\footnotesize	
\begin{center}
\caption{~Updated gravity distances using homogenised literature data. For CSPNe with multiple data, the adopted values are weighted means.}   
\label{grav_dist}
\begin{tabular}{lccccccl}
\hline
Name &~~$T_{\star}$ (kK)~~& ~~log\,$g$~~& ~$M_{\star}$/$M_{\odot}$~ &$V$ &   \EBV\ 	&   $D$ (kpc)~					&   ~References  					\\ 
\hline
Abell 7			&	  97	&  ~7.28~	&	0.59		&  ~15.50~	&	0.04	 		&	$0.53\pm0.18$			&	~N99, G04, GB10, Z12			\\
Abell 15			&	110	&	5.70	 	&	0.58		&	15.73	&	0.04	 		&	$4.0\pm1.2$				&	~MM97						\\
Abell 20			&	119	&	6.13 	&	0.57		&	16.47	&	0.10	 		&	$3.16\pm0.95$			&	~RK99						\\	
Abell 21			&	135	&	7.25	 	&	0.62		&	15.99	&	0.07	 		&	$0.82\pm0.34$			&	~N93, RK04, U				\\	
Abell 31			&	  91	&	7.15	 	&	0.58		&	15.54	&	0.04	 		&	$0.60\pm0.30$			&	~N99, G04, Z12				\\
Abell 36			&	111	&	5.75	 	&	0.57		&	11.55	&	0.04 		&	$0.53\pm0.17$			&	~TH05, Z12					\\
Abell 39			&	108	&	6.41		&	0.57		&	15.62	&	0.05 		&	$1.57\pm0.57$			&	~MM97, N99, G04, Z12			\\
Abell 43			&	107	&	5.54	 	&	0.60		&	14.74	&	0.17 		&	$2.47\pm0.30$			&	~N99, ZR09, RF11				\\
Abell 52			&	110	&	6.00	 	&	0.57		&	17.66	&	0.40 		&	$3.95\pm1.20$			&	~RK04, U					\\     
Abell 61			&	  95	&	7.06	 	&	0.58		&	17.41	&	0.05 		&	$1.60\pm0.30$			&	~N99, U						\\     
Abell 74			&	108	&	6.82	 	&	0.56		&	17.05	&	0.08 		&	$1.9\pm0.9$				&	~N99						\\     
Abell 78			&	113	&	5.25	 	&	0.64		&	13.26	&	0.14 		&	$1.92\pm0.62$			&	~WK92, RW98				\\     
AMU 1			&	  80	&	5.30 	&	0.55		&	13.67	&	0.09			&	$1.8\pm0.5$				&	~DM15						\\   	%
DS 2			&	  85	&	5.10 	&	0.58		&	12.37	&	0.20	 		&	$1.10\pm0.35$			&	~M88						\\     
EGB 6			&	101	&	7.38	 	&	0.59		&	16.00	&	0.04	 		&	$0.61\pm0.18$			&	~LB05, LG13					\\	
HaTr 7			&	100	&	6.00	 	&	0.56		&	15.11	&	0.09			&	$1.80\pm0.70$			&	~SW97						\\
HaWe 4			&	108	&	7.04	 	&	0.56		&	17.19	&	0.24			&	$1.15\pm0.70$			&	~N99, GB10					\\
HaWe 13			&	  68	&	6.38	 	&	0.40		&	16.90	&	0.44			&	$1.1\pm0.5$				&	~N99 						\\	
HbDs 1			&	111	&	5.70		&	0.59		&	12.53	&	0.14			&	$0.78\pm0.06$	 		&	~TH05, HB11, Z12				\\     
IC 2448			&	  95	&	5.40		&	0.58		&	14.26	&	0.07			&~~~$2.40\pm0.73$~~~		&	~HB11						\\
IC 2149			&	  39 	&     3.80		&	0.56 	&	11.34	&	0.20			&	$1.95\pm0.64$			&	~HM90, FH94					\\	
IC 4593			&	  41	&	3.70	 	&	0.62		&	11.33	&	0.05			&	$3.0\pm1.0$				&	~TL02, HB11, M12				\\
IsWe 1			&	100	&	7.00	 	&	0.56		&	16.56	&	0.22			&	$0.72 \pm0.23$			&	~NS95, WH06					\\
Jacoby 1			&	150	&	7.25	 	&	0.63		&	15.52	&	0.00			&	$0.70\pm0.30$			&	~W95, DH98, WD06			\\    	
Jn 1				&	145	&	6.75	 	&	0.56		&	16.17	&	0.07			&	$1.55\pm0.50$			&	~N93, W95, WH06				\\	
JnEr 1			&	130	&	7.00	 	&	0.60		&	17.14	&	0.02			&	$1.9\pm0.8$				&	~RW95, WR05				\\	
K 1-16			&	160	&	6.10	 	&	0.58		&	15.08	&	0.04			&	$2.20\pm0.88$			&	~HB95, W95, KW98, WR07		\\
K 1-27			&	135 	&	6.40		&	0.57		&	16.11	&	0.06			& 	$2.20\pm0.90$			&	~RR14b						\\
Lo 1				&	110	&	6.85	 	&	0.58		&	15.21	&	0.00			&	$0.85\pm0.26$	 		&	~HB04, Z12					\\
Lo 4				&	170	&	6.00 	&	0.62		&	16.60	&	0.14			&	$4.6\pm1.4$	 			&	~WR07						\\	
Lo 8				&	  90	&	5.10	 	&	0.58		&	12.97	&	0.05			&	$1.9\pm0.7$		 		&	~HM90						\\	
LoTr 4			&	120	&	5.80	 	&	0.60		&	16.65	&	0.17			&	$4.7\pm1.3$		 		&	~RR14						\\     
M 2-29			&	  50	&	4.00 	&	0.65		&	15.50	&	0.65			&      $7.1\pm2.1$		  		&	~MM11, U					\\     
MeWe 1-3		&	100 &	5.50	 	&	0.59		&	17.10	&	0.37			&	$5.5\pm1.6$				&	~SW97						\\     
MWP 1			&	163	&	6.61	 	&	0.565	&	13.13	&	0.03			&	$0.51\pm0.06$			&	~CA07						\\     	
NGC 246			&	150	&	5.97 	&	0.59		&	11.84	&	0.02			&	$0.58\pm0.35$			&	~HB95, W95, DW98, WR07		\\     
NGC 650-1		&	138	&	7.31		&	0.60		&	17.53	&	0.14			&	$1.38\pm0.40$ 			&	~KP98, CA06					\\    	
NGC 1360		&	105	&	5.80	 	&	0.56		&	11.34	&	0.01			&	$0.46\pm0.08$			&	~HD96, HB11, Z12				\\    	
NGC 1501		&	136	&	5.80	 	&	0.57		&	14.38	&	0.67			&	$0.82\pm0.24$			&	~KH97, W06, CA09, U			\\   	
NGC 1535		&	  71	&	4.60	 	&	0.63		&	12.09	&	0.02			&	$2.18\pm0.40$			&	~BH95, HB11					\\    	
NGC 2371/2		&	150	&	6.00 	&	0.59		&	14.85	&	0.04			&	$2.15\pm0.50$			&	~QF07, WR07					\\  	
NGC 2392		&	  44	&	3.83 	&	0.64		&	10.60	&	0.09			&	$1.70\pm0.50$			&	~HB11, MU12					\\  	
NGC 2438		&      114	&	6.62 	&	0.57		&	16.82	&	0.17			&	$1.88\pm0.57$			&	~RK99, OK14					\\  	
NGC 2867		&	141	&	6.00 	&	0.60		&	16.03	&	0.32			&	$2.44\pm0.60$			&	~QF07						\\  	
NGC 3587		&	  94	&	6.97 	&	0.57		&	15.74	&	0.01			&	$0.87\pm0.26$			&	~N99, Z12					\\
NGC 4361		&	126	&	6.00	 	&	0.58		&	13.26	&	0.02			&	$0.93\pm0.28$			&	~TH05, Z12					\\
NGC 5189		&	135	&	6.00	 	&	0.60		&	14.53	&	0.34			&	$1.13\pm0.40$			&	~QF07						\\
NGC 6720		&	112	&	6.93	 	&	0.58		&	15.29	&	0.04			&	$0.92\pm0.28$			&	~N99, Z12					\\
NGC 6853~~~~~	&      114  &	6.82	 	&	0.60		&	13.99	&	0.05			&	$0.49\pm0.20$			&	~HB95, N99, TH05, GB10, Z12	\\   	
NGC 6905		&	141	&	6.00	 	&	0.60		&	14.58	&	0.14			&	$1.62\pm0.48$			&	~QF07						\\
NGC 7094		&	110	&	5.56		&	0.59		&	13.61	&	0.12			&	$1.75\pm0.36$			&	~KW98, N99, ZR09			\\   	
NGC 7293		&	107	&	7.10		&	0.60		&	13.48	&	0.02			&	$0.29\pm0.13$			&	~WD97, N99, GB10, Z12, U		\\
Pa 5				&      145  &	6.70 	&	0.56		&	15.69	&	0.10			&	$1.35\pm0.3$				&	~DM15, U					\\   	%
Ps 1				&	  38	&	3.95 	&	0.60		&	14.73	&	0.10			&	$9.3\pm1.1$				&	~BB01, RH02					\\   	
PuWe 1			&	100	&	7.25 	&	0.58		&	15.55	&	0.10			&	$0.50\pm0.16$			&	~MM97, N99, G04, GB10, Z12	\\
RWT 152			&	  45	&	4.50	 	&	0.55		&	13.02	&	0.12			&	$2.4\pm0.9$				&	~EB82						\\   	%
Sh 2-78			&	120 	&	7.50	 	&	0.70		&	17.66	&	0.32			&	$0.91\pm0.27$			&	~D99						\\
Sh 2-188			&	  95	&	7.41		&	0.58		&	17.45	&	0.33			&      $0.73\pm0.24$			&	~N99, GB10					\\
Sh 2-216			&	  91	&	7.07		&	0.56		&	12.67	&	0.04			&	$0.17\pm0.05$			&	~RZ07, GB10					\\
TK 1				&	  86	&	7.48	 	&	0.58		&	15.70	&	0.02			&	$0.45\pm0.25$ 			&	~G04, GB10					\\	
WeDe 1			&	127	&	7.55	 	&	0.68		&	17.24	&	0.09			&	$0.99\pm0.29$			&	~LB94, N99, U				\\						
\hline
\end{tabular}
\end{center}
}
{\scriptsize
\begin{flushleft}
References:~~
BB01 -- Bianchi et al. (2001);  BH95 -- Bauer \& Husfeld (1995);  CA06 -- C\'orsico \& Althaus (2006); CA07 -- C\'orsico et al. (2007); CA09 -- C\'orsico et al. (2009); D99 -- Dreizler (1999); DH98 -- Dreizler \& Heber (1998);  DM15 -- De Marco et al. (2015); EB82 -- Ebbets \& Savage (1982); FH94 -- Feibelman, Hyung \& Aller  (1994);  G04 -- Good et al. (2004); GB10 -- Gianninas et al. (2010);  HB95 -- Hoare et al. (1995);  HB04 -- Herald  \& Bianchi (2004); HB11 --  Herald \& Bianchi (2011);  HD96 -- Hoare et al. (1996); HM90 -- Herrero, Manchado \& M\'endez (1990); KH97 -- Koesterke \& Hamann (1997); KP98 -- Koorneef \& Pottasch (1998); KW98 -- Kruk \& Werner (1998); LB94 -- Liebert et al. (1994); LB05 --  Liebert et al. (2005); LG13 -- Liebert et al. (2013); M88 -- M\'endez et al. (1988); M12 -- M\'endez et al. (2012); MK92 -- M\'endez et al. (1992);  MM97 -- McCarthy et al. (1997); MM11 -- Miszalski et al. (2011); MM15 -- Manick et al. (2015); N99 -- Napiwotzki (1999); NS93 -- Napiwotzki \& Sch\"onberner (1993);  NS95 -- Napiwotzki \& Sch\"onberner (1995); OK14 -- \"Ottl, Kimeswenger \& Zijlstra (2014); QF07 -- Quirion et al. (2007); RF11 -- Ringat et al. (2011);  RH02 -- Rauch et al. (2002);  RK99 -- Rauch et al. (1999); RK04 -- Rauch et al. (2004); RR14 -- Reindl et al. (2014b);  RZ07 -- Rauch et al. (2007);  SW97 -- Saurer et al. (1997);  TH05 -- Traulsen et al. (2005); TL02 -- Tinkler \& Lamers (2002); U -- unpublished data;  W95 -- Werner (1995);  WD97 -- Werner et al. (1997);  WD07 -- Werner et al. (2007a); WH06 -- Werner \& Herwig (2006); WK92 -- Werner \& Koesterke (1992); WR07 -- Werner et al. (2007b); ZR09 -- Ziegler et al. (2009);  Z12 -- Ziegler et al. (2012a).    
\end{flushleft}
}
\end{table*}

\subsubsection{Expansion parallaxes}\label{sec:expansion}

A potentially powerful technique is the expansion parallax method, where the PN's angular expansion in the plane of the sky over a suitably long time period is compared to the shell's radial velocity, based on either optical or radio data; ~Terzian (1997) and Hajian (2006) provide reviews of the technique.    
We have decided that the expansion parallaxes based on older, ground-based, optical photographs (e.g. Chudovicheva 1964; Liller 1965; Liller et al. 1966) are not of sufficient quality to be useful.  
Several PNe have distance estimates based on multi-epoch Very Large Array (VLA) 6\,cm radio observations (Masson 1986, Hajian, Terzian \& Bignell 1993, 1995; Hajian \& Terzian 1996; Kawamura \& Masson 1996), and are potentially far more accurate than the older optical determinations.  Other distance determinations are given by Christianto \& Seaquist (1998), Guzm\'an, G\'omez \& Rodr\'iguez (2006), Guzm\'an-Ramirez et al. (2009), and Guzm\'an et al. (2011).  Precise HST optical parallaxes, also based on multi-epoch nebular images, have become available in the last decade (Reed et al. 1999; Palen et al. 2002;  Li, Harrington \& Borkowski 2002; Hajian 2006) which promise to have a significant impact on the local PN distance scale.  Furthermore, Meaburn et al. (2008) and Boumis \& Meaburn (2013) have used the proper motions of fast-moving outer optical knots (assuming ballistic motion) to derive distances for NGC\,6302 of 1170 $\pm$ 140 pc, and KjPn~8 of 1800 $\pm$ 300 pc, respectively, though the extended nebula of KjPn\,8 does only as a rough integrated \ha\ flux available, so has been excluded as a calibrator (but see \S\ref{sec:background}).

While expansion parallaxes were thought to be a relatively simple, yet powerful method, it has become apparent that there are serious sources of systematic error in the technique which need to be considered before reliable distances can be determined.  Firstly, the majority of PNe are aspherical, so various corrections for prolate ellipsoidal geometries have been applied (e.g. Li et al. 2002), and secondly, the angular expansion rate on the sky  (a  pattern velocity) was assumed to be equal to the spectroscopically measured gas velocity.  However, these are usually not identical in nature.  Mellema (2004) modelled the jump conditions for both shocks and ionization fronts, and found that the pattern velocity is typically $\sim$30\% larger than the matter velocity, hence the calculated distances are too short by this amount.  Sch\"onberner, Jacob \& Steffen (2005b), using 1-D hydrodynamical modelling, also found that the pattern velocity is always larger than the material velocity.  These authors found that the necessary correction factor ranged between 1.3 and 3.0, depending on the evolutionary state of the CSPN.
That such biases in expansion distances do exist is provided by the study of the symbiotic nebula Hen\,2-147 (Santander-Garc\'ia et al. 2007).  These authors found that the expansion parallax method gave a distance of 1.5 $\pm$ 0.4 kpc, a factor of two lower than the distance of 3.0 $\pm$ 0.4 kpc obtained from the period-luminosity (P-L) relationship for the central Mira variable.  Correcting for the jump condition described earlier, these authors find $D$ = 2.7 $\pm$ 0.5 kpc, in much better agreement with the P-L distance.

Following Mellema (2004), the exact value of the correction factor depends upon the shock's Mach number\footnote{The Mach number is defined as $\mathcal{M}$ = $v$/$v_{s}$, where $v$ is the velocity of the object relative to the ambient gas and $v_{s}$ is the sound velocity in the gas.} ($\mathcal{M}$), given by:  
\begin{equation} 
\label{eq:mach}
\mathcal{M} = \frac{(\gamma + 1)(u_{1} - u_{0}) + [(\gamma + 1)^{2}(u_{0} - u_{1})^{2} +16a_{0}^{2}]^{1/2}}{4a_{0}}
\end{equation} 

where $\gamma$ is the adiabatic index (for isothermal shocks\footnote{Mellema (2004) shows that the isothermal case is justified as most PNe (at least the ones which have had expansion parallaxes determined), have relatively high densities and slow shocks.}, $\gamma$ = 1),  $u_{0}$ is the pre-shock velocity of the gas (taken to be $\sim$13 \kms, noting that the correction factor is only weakly dependent on the exact value),  $u_{1}$ is the spectroscopically derived expansion velocity, and $a_{0}$ is the pre-shock sound speed ($a_{0}$ = 11.7 \kms\ for nebular gas at 10$^{4}$\,K, following Mellema 2004).  The correction factor $\mathcal{R}$, is then found from equation (4) of Mellema (2004), viz:
\begin{equation} 
\label{eq:mellema}
\mathcal{R} = \frac{(\gamma + 1)\mathcal{M}u_{0} + (\gamma + 1)\mathcal{M}^{2}a_{0}}{(\gamma + 1)\mathcal{M}u_{0} + 2(\mathcal{M}^{2} -1)a_{0}}
\end{equation} 

The ratio tends to unity for high values of $\mathcal{M}$, that is high spectroscopic expansion velocities.  
Several PNe with optical expansion parallaxes have bright rims with attached shells, and so the rim can be considered to be shock bounded (Mellema 2004), and not indicative of an ionization front.  However, the very youngest PNe (e.g. Vy~2-2) need to be modelled as expanding (D-type) ionization fronts surrounded by neutral material (see also Sch\"onberner et al. 2005b).   In this case the correction factor is more difficult to evaluate (Mellema 2004) but has been applied to BD+30\arcdeg3639.  He obtains $D$ = 1.3 $\pm$ 0.2 kpc, in agreement with the distance from Sch\"onberner et al. (2005b).   The most recent distance for this PN comes from the detailed analysis by Akras \& Steffen (2012), who give $D$ = $1.52\pm0.21$ kpc, which we adopt here.

We have applied a numerical correction to all expansion distances taken from the literature to account for the jump condition, unless it had been specifically taken into account, or the distance is based on the ballistic motion of high-proper motion features.  In addition, unpublished HST expansion parallaxes were kindly provided by A. Hajian (2006, pers. comm.; see also Hajian 2006), that were also utilised by F08 and Smith (2015).  An example is given here (the southern PN, NGC 5882) to show how the correction factor ($\mathcal{R}$) is calculated.  For this object, the new (uncorrected) expansion distance is $D$ = 1.32 $\pm$ 0.2 kpc, with the additional note that the [N~{\sc ii}] and [O~{\sc iii}] images give the same distance.  The [N~{\sc ii}] and [O~{\sc iii}] expansion velocities (Hajian et al. 2007) are also similar, with a mean of 25 \kms.  Correcting for the jump condition, and assuming an isothermal shock ($\gamma$ = 1)  following Mellema (2004), equations~\ref{eq:mach} and \ref{eq:mellema} can be used to estimate a correction factor, $\mathcal{R}$ = 1.3 $\pm$ 0.1.  The corrected distance is $D$ = 1.72 kpc, and a distance uncertainty of 25\% has been assumed.  
Table~\ref{exp_parallax} provides expansion distances compiled from the literature, including the unpublished data from Hajian (2006), except for the kinematically complex objects NGC\,6326 and NGC\,7026 (e.g. Clark et al. 2013).

\begin{table}
\begin{center}
\caption{Expansion distances for 29 planetary nebulae.  For PNe with more than one determination, the adopted values are weighted means.}
\label{exp_parallax}
{\footnotesize
\begin{tabular}{lll}
\hline
Name           			& ~~$D$  (kpc) 				&   Reference  						\\
\hline
Abell 58      			&     $4.60\pm0.60$    			&  C13  							 	\\
BD+30 3639 ~~		&     1.52 $\pm$ 0.21       		&  LH02, AS12 						\\
DPV 1				&     2.9 $\pm$ 0.8       			&  HJ14 								\\
Hu 1-2  				& 	$>$2.7					&  MB12 								\\    	
IC 418   				& 	1.3 $\pm$ 0.4 				&  GL09  								\\    	
IC 2448      			&     2.2 $\pm$ 0.5      			&  PB02, M04, SJ05, H06$^{\ddag}$		\\
J 900	     			& 	4.8 $\pm$ 1.0          		&  H06$^{\ddag}$						\\   	
KjPn 8       			&     1.8 $\pm$ 0.3     			&  BM13  							\\    	
M 2-43      			&     6.9 $\pm$  1.5     			&  GG06   							\\
NGC 2392  			& 	 1.3 $\pm$ 0.3				&  GD15$^{\ddag}$		 		 		\\   	
NGC 3132  			& 	1.2 $\pm$ 0.4				&  H06								\\   	
NGC 3242 			&     0.78 $\pm$ 0.23 			&  HT95, M06, RG06   					\\
NGC 3918  			& 	1.45 $\pm$ 0.30			&  H06$^{\ddag}$		 		 		\\   	
NGC 5882  			& 	1.72 $\pm$ 0.43			&  H06$^{\ddag}$		 		 		\\   	
NGC 5979  			& 	2.0$\pm$ 0.5  				&  H06$^{\ddag}$		 		 		\\   	
NGC 6210   			&     2.1 $\pm$  0.5 			&  HT95, M04							\\
NGC 6302  			&    1.17 $\pm$ 0.14			&  ML08   							\\   	
NGC 6543   			&    1.55 $\pm$ 0.44      			&  RB99, M04   						\\    	
NGC 6572    			&     2.0 $\pm$ 0.5    			&  HT95, KM96, M04					\\    	
NGC 6578    			&     2.90 $\pm$ 0.78~~~~		&  PB04, M04							\\    	
NGC 6720  			&     0.72 $\pm$ 0.22			&  OD09, OD13  						\\
NGC 6741  			&     $>$1.5					&  SB05								\\
NGC 6826  			& 	 2.1 $\pm$ 0.5 			&  SJ05, H06$^{\ddag}$		 		 	\\   	
NGC 6881 			&     1.6 $\pm$ 0.6      			&  GR11 								\\   	
NGC 6884   			&     3.30 $\pm$ 1.24 			&  PB02, M04							\\	
NGC 6891  			& 	 2.9 $\pm$ 0.6				&  PB02, H06$^{\ddag}$				\\  	
NGC 7009   			&     1.45 $\pm$ 0.5      			&  S04   								\\   	
NGC 7027   			&     0.92 $\pm$ 0.10       		&  Z08	   							\\  	
NGC 7662  			&     1.19 $\pm$ 1.15       		&  HT96, M04 							\\  	
Vy 2-2        			&     4.68 $\pm$ 1.20      		&  CS98, M04     						\\  	
\hline
\end{tabular}
}
\end{center}
{\scriptsize
\begin{flushleft}
Notes:~$^{\dagger}$Assumed uncertainty; $^{\ddag}$corrected according to the precepts discussed in the text.\\
References:   AS12 -- Akras \& Steffen (2012);  BM13 -- Boumis \& Meaburn (2013); C13 -- Clayton et al. (2013); CS98 --  Christianto \& Seaquist (1998); GD15 -- Garc\'ia-D\'iaz et al. (2015); GG06 -- Guzm\'an et al.  (2006); GL09 -- Guzm\'an et al. (2009);  GR11 -- Guzm\'an-Ram\'irez et al. (2011);  H06 -- Hajian (2006);  HJ14 -- Hinkle \& Joyce (2014); HT95 -- Hajian et al. (1995);  HT96 --  Hajian \& Terzian (1996);  KM96 -- Kawamura \& Masson (1996);   LH02 --  Li et al. (2002);  M04 -- Mellema (2004); MB12 -- Miranda et al. (2012); ML08 -- Meaburn et al. (2008); OD09 -- O'Dell et al. (2009);  OD13 -- O'Dell et al. (2013);  PB02 --  Palen et al. (2002);  RB99 -- Reed et al. (1999);  RG06 -- Ruiz et al. (2006); S04 -- Sabbadin et al. (2004); SB05 -- Sabbadin et al. (2005); SJ05 -- Sch\"onberner et al. (2005b);  Z08 -- Zijlstra et al. (2008, and references therein).
\end{flushleft}
}
\end{table}

\subsubsection{Distances from Photoionization Modelling}\label{sec:phot_model}

Relatively accurate distance determinations using a self consistent treatment of spatiokinematic and photoionization modelling is a comparatively recent development.  The development of powerful 2-D and 3-D photoionization codes (e.g. Ercolano et al. 2003) allows the self-consistent determination of the PN structure, central star characteristics, and distance, once accurate spectrophotometric line mapping, narrowband imaging, and kinematic data are  available.  This technique as applied to individual PNe (e.g. Monteiro et al. 2004;  Schwarz \& Monteiro 2006; Monteiro et al. 2011) holds promise, with all recent determinations summarised in Table~\ref{phot_model_distances}.   However, we have not utilised the distance for Mz\,1 (Monteiro et al. 2005), owing to the lack of a reliable CSPN magnitude needed for modelling.  Additionally, Bohigas (2008) presented photoionization models for 19 PNe, deriving two distances per object by comparing the model output with the observed \ha\ flux and the angular size respectively.  We only used PNe which had the model distances consistent to better than $\pm$25\%, with the additional requirement that the input parameters agreed with those in our database (Parker et al., in prep.).  Only two PNe matched these requirements: JnEr\,1 and K\,3-72.  
~In Table\,\ref{phot_model_distances}, we present the photoionization model distances for 16 calibrating PNe.

\begin{table}
\begin{center}
\caption{PN distances from photoionization modelling.}
\label{phot_model_distances}
\begin{tabular}{lll}
\hline
Name					& ~~$D$ (kpc)  					& Reference    	 	\\
\hline
Abell 15        				&  4.01 $\pm$ 1.0$^{\dagger}$~~~~~~	& ER05		 		\\    	
Abell 20               			&  2.35 $\pm$ 0.60$^{\dagger}$       	& ER05 				\\	
Hb 5      					&  1.4 $\pm$ 0.3$^{\ddag}$      		& RSM04				\\	
IC 418          				&  1.25 $\pm$ 0.10$^{\dagger}$      	& MG09     			\\
JnEr 1        				&  1.1 $\pm$ 0.2   					& B08     				\\	
K 3-72    					&  $5.0 \pm 0.6$					& B08				\\     
MeWe 1-3         			&  3.95 $\pm$ 1.0$^{\dagger}$  		& EF04 				\\
NGC 40      				&  1.15 $\pm$ 0.12          			& M11  				\\	
NGC 2610~~~~~~~~~		&  2.5 $\pm$ 0.5       	  	 		& H06, U		  	      	\\
NGC 3132			 	&  0.93 $\pm$ 0.25$^{\dagger}$		& M00, SM06	    		\\	
NGC 3918      				&  $\geq$1.5      					& C87	 	        	\\	
NGC 6026				&  2.0 $\pm$ 0.5					& D13 				\\  	
NGC 6369       			&  1.55 $\pm$ 0.30$^{\dagger}$ 		& M04	          		\\	
NGC 6781       			&  0.95 $\pm$ 0.14~~~~~~	  		& SM06          			\\	
\hline
\end{tabular}
\end{center}
{\scriptsize 
\begin{flushleft}
Notes: ~$^{\dagger}$Estimated uncertainty; $^{\ddag}$distance given half-weight.\\ 
References:  ~B08 -- Bohigas (2008);  C87 -- Clegg et al. (1987);  D13 -- Danehkar et al. (2013); ER05 --  Emprechtinger, Rauch \& Kimeswenger (2005);  H06 -- Harrington (2006);  M00 -- Monteiro et al. (2000); M04 -- Monteiro et al. (2004);  M11 -- Monteiro et al. (2011);  MG09 -- Morisset  \& Georgiev (2009); RSM04 -- Rice et al. (2004);  SM06 -- Schwarz \& Monteiro (2006); U -- unpublished data.   
\end{flushleft}
}  
\end{table}

\subsubsection{Kinematic distances}

Kinematic distances can be determined for a restricted sample of PNe, namely those with little or no peculiar motion with respect to the local standard of rest.  In other words, the PN partakes of nearly circular orbital motion around the Galaxy.  The technique uses the position on the sky and the measured radial velocity of the PN to infer a distance (e.g. Corradi \& Schwarz 1993; Corradi et al. 1997; Phillips 2001), assuming a model for the Galactic rotation curve. The approach can also be used for any neutral hydrogen in the foreground of the PN which causes an absorption line at 21\,cm in the radio spectrum.  Thus the distance for the absorbing cloud can be determined, which is a lower limit to the distance of the PN (e.g. Pottasch et al. 1982; Gathier, Pottasch \& Goss 1986; Maciel 1995). This limit in some cases constrains the distance quite well.   
In this work an updated Galactic rotation curve slightly different to the IAU standard has been utilised:  the adopted values are $v_{\odot}$ = 240\,\kms, and $R_{\odot}$ = 8.3\,kpc  (Brunthaler et al. 2011).  A flat rotation curve in the range of  $4\leq R\leq 14$\,kpc has also been assumed.   For the cases where there is a kinematic ambiguity, the overall interstellar extinction proved useful in determining that the near distance was the only solution in each case.

Only a few kinematic determinations have been adopted as calibrating data.  Type\,I PNe (Peimbert 1978; Kingsburgh \& Barlow 1994), which are produced from higher-mass progenitor stars, are in general the only objects for which this approach is valid, where we assume these objects have a low peculiar velocity relative to its local ISM.   Their peculiar velocity is assumed to be equal to the velocity dispersion of main sequence stars of spectral types B3--A0,  $\sigma_{u}$ = 15 \kms\ (Cox 2000), as such stars, with main sequence masses of $>$3--4\,$M_{\odot}$ are the plausible progenitors for Type\,I PNe (cf. Karakas et al. 2009).  This uncertainty dominates the error budget for each distance determination, especially as most have accurate systemic velocities.  Table~\ref{kinematic_distances} summarises the best currently available distances (or limits) utilising this technique.  The radial velocities were taken from the references given in the table, and were all converted to the LSR frame.   Two distance determinations for non-Type\,I PNe are described in more detail below.\\

\noindent{\it HFG\,2 (PHR\,J0742-3247)}.  ~This high-excitation, optically-thin nebula was discovered by Fesen, Gull \& Heckathorn (1983), and later confirmed by Parker et al. (2006).  The 17th-mag central star is ionizing part of an extended \HII\ region of dimensions 7\arcmin $\times$ 4\arcmin.  That the source of ionization is the CSPN is shown by spectroscopically detectable [O~{\sc iii}]  emission in the nebulosity immediately closest to the PN (F08).   We adopt a revised  \ha\ flux from Frew et al. (2014a) to calculate the surface brightness.   A CO detection to the \HII\ region is reported by Brand et al. (1987), and the measured LSR velocity, +23.5\,\kms\ leads to a  distance for the PN of 2.1 $\pm$ 0.6 kpc.  \\

\noindent{\it NGC 6751}.  ~This is another example of an ambient \HII\ region ionised by a hot CSPN (Chu et al. 1991), in this case an early [WO] type.  A revised kinematic distance of $2.7\pm0.7$\,kpc has been determined from the radial velocity data presented by Clark et al. (2010).  See that reference for further details.

\begin{table}
{\footnotesize	
\begin{center}
\caption{Kinematic distances for PNe mostly of Peimbert's Type \,I.}
\label{kinematic_distances}
\begin{tabular}{lccl}
\hline
PN    		  		& ~~$v_{\rm LSR}$	&~~~~~$D$ (kpc)~~~~~ 		&  References~~~~~~     	\\
\hline
Abell 79              		&   $-44\pm8$     		&    $4.4\pm1.0$			&    RC01   	 			\\    
BV 5-1	         		&   $-73\pm1$          	&    $5.5\pm1.2$		  	&    JB00    				\\    
CVMP 1       			&   $-28\pm5$     		&    $1.9\pm0.7$	   		&    CV97     			 	\\
IPHAS-PN 1  			&   $-71 \pm 2$ 		&    $7.0^{+4.5}_{-3.0}$    	&    M06         				\\    
HaTr 10    			&   $+63 \pm 5$    		&    $4.0\pm1.0$	   		&    L12	 				\\    
Hen 2-111      			&   $-28 \pm 5$      		&     $1.9\pm0.6$			&   MW89	         		\\
HFG 2		 		&   $+23.5 \pm 1$    	&     $2.1\pm0.5$	   		&    B87 			          	\\
K 1-10       			&   $+52\pm 5$      		&     $5.0\pm1.3$      	 	&    L12   					\\	
K 3-72          			&   $+28\pm10$ 	      	&     $3.8^{+2.0}_{-1.6}$		&    CS93, L12    			\\	
M 2-53     			&   $-61 \pm 2$         	&    $6.0\pm1.0$			&   	HB05				\\ 	
M 3-3             			&   $+55\pm 2$     		&    $5.5^{+1.8}_{-1.3}$ 		&     H96   	 			\\  	
M 3-28        			&   $+32\pm 3$         	&	$2.5^{+1.1}_{-1.3}$		&     HB05         			\\     
M 4-14      			&   $+49 \pm 3$         	&  	$3.8\pm1.1$ 		   	&   	D08       				\\
Mz 3              			&   $-53\pm3$  		&    3.4 $\pm$ 0.8 			&  	R00			         	\\      
NGC 5189 			&   $-13.3\pm1$ 		&  	$1.0^{+0.7}_{-0.6}$ 	&	SV12				\\
NGC 6751			&   $+42\pm1$     		&    $2.7\pm0.7$ 	   		&    CM91, CG10      		\\	%
SuWt 2     			&   $+29 \pm 5$        	&     $2.3\pm0.6$         		&    JL10		  			\\	
We 1-4       			&   $+28 \pm 5$      	&     $4.8\pm1.5$       		&    L12     				\\	
We 2-5       			&   $-27 \pm 5$      		&     $2.3\pm0.6$       		&    L12 			 		\\	
WeSb 4       			&   $+69 \pm 3$      	&     $4.7\pm1.0$     		&    L12	  		 		\\	
\hline
\end{tabular}
\end{center}
}
{\scriptsize 
\begin{flushleft}
References:  B87 -- Brand et al. (1987); CG10 -- Clark et al. (2010); CM91 -- Chu et al.  (1991); CS93 -- Corradi \& Schwarz (1993);  CV97 --  Corradi et al. (1997); D08 --  Dobrin\v{c}i\'c et al. (2008);  F08 -- Frew (2008); HB96 -- Huggins et al. (1996);  HB05 -- Huggins et al. (2005); JB00 -- Josselin et al. (2000); JL10 -- Jones et al. (2010);  L12 -- L\'opez et al. (2012); M06 -- Mampaso et al. (2006);  MW89 --  Meaburn \& Walsh (1989); Ph01 --  Phillips (2001); PM02 -- Pena \& Medina (2002); R00 -- Redman et al. (2000); RC01 -- Rodr\'iguez et al. (2001);  SV12 -- Sabin et al. (2012); U -- unpublished data.\\
Note HFG\,2 and NGC\,6751 are non-Type\,I PNe ionising ambient interstellar gas. 
\end{flushleft}
}
\end{table}

\subsubsection{Extinction Distances}.  
Individual extinction distances can be determined for PNe by comparing their observed extinctions with stars in the immediate vicinity of the PN at a range of distances that bracket the PN's distance (Lutz 1973; Kaler \& Lutz 1985; Gathier et al. 1986).  While the method has the advantage of making no assumptions about the PN, it has proved difficult to calibrate in practice (Saurer 1995; Giammanco et al. 2011).   The  extinction is usually determined from the observed Balmer decrement of the nebular shell (e.g. Kimeswenger \& Kerber 1998; Giammanco et al. 2011; Navarro, Corradi \& Mampaso 2012), or by measuring the apparent colours of the CSPN, and assuming an intrinsic value for the colour index (see De Marco et al. 2013) to get the reddening directly (Weston, Napiwotzki \& Sale 2009).  
In general, extinction distances have been taken from the literature only if the PN is within 4\arcdeg\ of the Galactic plane (cf. Phillips 2006), which as an example, excludes all the distances from Martin (1994).  At greater latitudes, the extinction distances for more remote PNe can be greatly underestimated as it is effectively outside the main dust layer of the disk (see the discussion by Phillips 2006).  Furthermore, distance determinations based on {\it average} extinction-distance diagrams or their equivalents (e.g. Acker 1978; Pottasch 1984; Napiwotzki 2001) have been excluded as calibrating data owing to the potentially low precision of the method.

The distance uncertainties for the various literature determinations are rather inconsistent, with some being little more than rough estimates.  If the nominal uncertainty on an individual extinction distance is less than 25\%, it has been reset to that value here.  While individual distances have rather large errors, the method as a whole is not expected to be biased to a short or long scale, provided that a substantial number of PNe are used as calibrators and no high-galactic latitude PNe are included.  However, extinction distances to compact PNe might be overestimated if internal dust is significant (e.g. Ciardullo \& Jacoby 1999; Giammanco et al. 2011), and the effect has been seen in young PNe like NGC\,7027 (Navarro et al. 2012).    
Nevertheless, most PNe seem to show little or no internal extinction due to intrinsic dust (F08), verified from the observed blue colours of the CSPNe in evolved objects at high latitudes, such as NGC\,246 and NGC\,7293 (see Bond \& Ciardullo 1999; Landolt \& Uomoto 2007; F08; De Marco et al. 2013).   Table~\ref{extinction_distances} gives a summary of the adopted extinction distances, taken from the references listed following the table.

\begin{table}
\begin{center}
\caption{Extinction distances for planetary nebulae. PNe with $|b| \geq$ 5\arcdeg\ have been excluded from this table.  Weighted averages are quoted for PNe with more than one independent distance determination.}
\label{extinction_distances}
{\footnotesize
\begin{tabular}{lcl}
\hline
Name           		& ~~~~~~~$D$~~(kpc)~~~~~~~~~&    References	  			\\
\hline
Abell 14	            		& 	$5.4\pm0.8$       	     	&     GC11   					\\    
BV 5-1    				&     $3.0\pm0.4$        		&	 GC11					\\
CBSS 1				&     $4.0\pm1.0$        		&	 CB94					\\
CBSS 2				&     $4.8\pm1.5$        		&	 CB94 					\\
CBSS 3				&     $4.8\pm1.5$        		&	 CB94 					\\
CVMP 1    			&     $2.0\pm0.5$        		&	 CV97 					\\
Hen\,2-111    			&     $2.2\pm0.5$   			&      F08  					\\   
IC 1747  				&     $2.8\pm0.3$			&      A78, P84, KL85			\\   
IPHAS-PN1			&     $5.9\pm1.5$  			&      M06, F08    				\\ 	
J 900	    			&     $4.30\pm0.65$          	&      GC11   					\\
M 1-4   				&     $3.30\pm0.35$	 	&      GC11					\\
M 1-71	  			&     $2.9\pm0.4$   			&      GC11     					\\    
M 1-77   				&     $2.5\pm0.1$   			&      HW88       				\\ 	
Mz 2	 			&    	$2.0\pm0.5$    		&      F08  					\\    
NGC 2346  			& 	$1.06\pm0.15$	      	&      GP86     					\\    
NGC 2440  			&     $1.77\pm0.45$    		&      F08    					\\
NGC 2452  			& 	$3.70\pm0.36$      		&      A78, P84, GP86			\\
NGC 5189  			&     $1.50\pm0.30$    		&      F08 		 			\\   
NGC 6537  			&  	$2.81\pm0.45$           	&      NC12 					\\  
NGC 6565  			& 	$2.0\pm0.5$            	&      TC02  					\\   
NGC 6567  			& 	$1.68\pm0.17$            	&      GP86  					\\   
NGC 6741  			& 	$2.60\pm0.55$            	&      KL85, SB05, GC11	 		\\   
NGC 6781  			& 	$0.83\pm0.24$            	&      NC12  					\\
NGC 6842  			& 	$2.39\pm0.28$	   	&      HW88, GC11				\\   
NGC 6894  			& 	$1.15\pm0.25$          	&      P84, KL85, GC11 			\\   
NGC 7026  			& 	$1.70\pm0.35$           	&    P84, SW84, KL85, GC11 	\\    
NGC 7048  			& 	$1.80\pm0.50$         	&      A78, HW88, GC11			\\    %
NGC 7354  			& 	$1.1\pm0.5$	          	&      KL85, GC11$^{\star}$   		\\    
PHR J1327-6032  		&    $2.2\pm0.6$     		& 	 F08   					\\
SaWe 3				&    $2.1\pm0.3$   			& 	 F08    					\\    %
Sh 1-89				&    $2.2\pm0.3$   			& 	 HW88, F08, GC11   	 	\\    %
Vy 2-2 				&    $2.30\pm0.17$    	  	&   	 GC11      				\\    %
\hline
\end{tabular}
}
\end{center}
{\scriptsize
\begin{flushleft}
Notes:~$^{\star}$disparate values; object given half weight.\\
References:  ~A78 -- Acker (1978); CB94 -- Cappellaro et al. (1994);  CV97 --  Corradi et al. (1997);  F08 -- Frew (2008); GC11 -- Giammanco et al. (2011);  GP86 --  Gathier et al. (1986); HW88 -- Huemer \& Weinberger (1988); KL85 -- Kaler \& Lutz (1985); M06 -- Mampaso et al. (2006);  NC12 -- Navarro et al. (2012);  P84 -- Pottasch (1984); SB05 --Sabbadin et al. (2005); SW84 -- Solf \& Weinberger (1984).  \end{flushleft}
}
\end{table}

\subsubsection{Miscellaneous Distance Methods}\label{sec:miscell}
This section includes a small but varied set of distances obtained using methods other than those described above, as summarised in Table\,\ref{table:miscell}.   
For the historically observed final-flash CSPNe, we have assumed for visual maximum a luminosity of 5000\,$L_{\odot}$ and a bolometric correction of zero (i.e. $M_V$ = $M_{\rm Bol}$).  The peak visual brightness for V605\,Aql (Abell\,58),  FG\,Sge (Hen\,1-5), and V4334\,Sgr (Sakurai's object = DPV\,1) has been taken from Duerbeck et al. (2002), van Genderen \& Gautschy (1995), and Duerbeck et al. (2000), respectively.   An independent distance to FG\,Sge based on pulsation theory has been obtained by Mayor \& Acker (1980).  In addition, the classical nova V458 Vul is located inside a faint planetary nebula, which was flash-ionized by the nova outburst.  Wesson et al. (2006) has described the various distance determinations to this object, which all agree within the uncertainties.

As a further example, Wareing et al. (2006) modelled the morphology of the strongly asymmetric object Sh\,2-188 to determine the relative velocity in the plane of sky that best reproduces the observed PN/ISM interaction.  Combining this transverse velocity with a measured proper motion of the CSPN leads directly to a distance.  Lastly, Eggen (1984) has determined a convergent parallax to NGC\,7293 based on its assumed membership of the Hyades moving group.  While this distance is consistent with the trigonometric distance from Table\,\ref{trig_distances}, we have used the latter owing to its much smaller uncertainty.

\begin{table}
\begin{center}
\caption{Miscellaneous distance estimates for six PNe.}
\label{table:miscell}
{\footnotesize
\begin{tabular}{llll}
\hline
Name           			&     ~~$D$ (kpc)  					&     Method  							&    Reference  	\\
\hline
Abell 58				&	$5.0\pm1.5$					&	outburst brightness$^{\ddag}$		&	This work		\\   
DPV 1				&	$3.8 \pm1.1$					&	outburst brightness$^{\ddag}$		&	This work		\\   
Hen 1-5				&	$2.8\pm0.8$					&	outburst brightness$^{\ddag}$		&	This work		\\   
Hen 1-5				&	$2.5\pm0.5$					&	pulsation theory					&	MA80		\\   
NGC 7293			&	$0.18\pm0.03^{\dagger}$		&	convergent parallax				&	E84			\\   
Sh 2-188				&	$0.85^{+0.50}_{-0.42}$			&	proper motion 						&	WOZ06		\\   
V458 Vul				&	$13.4\pm2.0$					&	light travel-time					&	WB06 		\\   
V458 Vul				&	$11.6\pm3.0^{\dagger}$			&	nova decline						&	WB06 		\\   
\hline
\end{tabular}
}
\end{center}
{\scriptsize
\begin{flushleft}
Notes: ~$^{\dagger}$Assumed uncertainty;  $^{\ddag}$assumed luminosity of 5000\msun\ for the central stars of Abell\,58 (V605 Aql), DPV\,1 (V4334 Sgr) and Hen\,1-5 (FG Sge) at maximum brightness.    \\
References:  ~E84 -- Eggen (1984); MA80 -- Mayor \& Acker (1980); WB06 -- Wesson et al. (2006); WOZ06 -- Wareing et al. (2006).  
\end{flushleft}  
}
\end{table}

\subsection{The Bulge Sample}\label{sec:bulge_comparison}

We also use a restricted set of Galactic bulge objects as an adjunct to our calibration process (included in Table\,\ref{table:SBr_calibrators}).   To constrain Bulge membership and exclude foreground disk objects, we applied constraints on the flux and diameter as is usual.  We further constrained the sample using the observed radial velocities, taken primarily from the compilation of Durand et al. (1998).   We further assumed that Bulge PNe had $|V_{\rm hel}| > 125$~\kms.   While this approach excludes many bona fide Bulge PNe, it has the benefit of excluding the vast majority of foreground disk interlopers, which would add noise to the relation.   Integrated fluxes were taken from the sources discussed previously, and angular dimensions were mostly taken from Tylenda et al. (2003), Ruffle et al. (2004) and Kovacevic et al. (2011), and we have adopted the distance to the Galactic centre of $8.30\pm0.23$\,kpc from Brunthaler et al. (2011).  However, owing to the substantial line of sight distance through the Bulge, and the fact that the Bulge sample may not be symmetrically located around the Galactic centre, we have only given half-weight to these PNe in our final calibration.

\subsection{The Extragalactic Sample}\label{sec:Cloud_comparison}

PNe in the nearest satellite galaxies of the Milky Way are resolved with HST, and have the advantage of an accurately known distance.   F08 showed that the \SR\ relation for the Galactic sample is consistent within the uncertainties with the \SR\ relation seen for MC PNe.    In contrast to F08, we have now used these PNe in our analysis, enlarging our calibrating sample by a factor of two.  We adopt distances of $50.0\pm0.2$\,kpc ($\mu_{0}$ = 18.49) for the LMC (Pietrzy\'nski et al. 2013)
and $61.7\pm2.0$\,kpc ($\mu_{0}$ = 18.95) for the SMC (Graczyk et al. 2014), adopting line-of-sight depths of 1.0\,kpc and 2.0\,kpc respectively.   Similarly, we use three PNe belonging to the Sagittarius dSph galaxy (e.g. Zijlstra et al. 2006) as calibrating nebulae.   We adopt a distance to this system of $26\pm2$\,kpc, for consistency with the complementary analysis of Vickers et al. (2015).\footnote{There is a moderately bright PN in the Fornax dSph galaxy at a distance of $137\pm7$\,kpc (Kniazev et al. 2007), but no HST imagery is available for it, and a second peculiar H-deficient PN in its globular cluster Hodge\,5 (Larsen 2006), but like GJJC\,1 in M\,22, this exhibits no \ha\ emission.}   We should note that there is a significant line-of-sight depth to the SMC (Haschke, Grebel \& Duffau 2012, and references therein), but evidently a much smaller depth for the main body of the SMC (Graczyk et al. 2014), which contains most of our calibrating PNe.   There is considerable potential for a depth effect to be found in the Sgr dSph system as well, since none of the three PNe are located near the centre of the galaxy.  Thus we have given half-weight in our final calibration to the PNe in the latter system.

\section{The \SR\ relation}

The \SR\ relation requires only an angular size, an integrated \ha\ flux, and the reddening to the PN.  From these quantities, an intrinsic radius is calculated, which when combined with the angular size, yields the distance.  Recall that  the \SR\ relation has better utility than the equivalent [O{\sc\,iii}] and [N{\sc\,ii}] relations (Shaw et al. 2001; F08), as it includes both bright objects and the most senile PNe over a broad range of excitation, and best reflects the underlying ionised mass. The [N{\sc\,ii}] relation, especially, is strongly influenced by abundance variations between objects, and furthermore, there is negligible  [N{\sc\,ii}] emission in the PNe of highest excitation.  The \ha\ relation is also preferred to the equivalent \hb\ relation, as at a minimum, \ha\ fluxes are a factor of approximately three brighter.  As mentioned above, a number of high-quality \ha\ imaging surveys have recently become available, which have also allowed the determination of accurate integrated \ha\ fluxes for a significant fraction of Galactic PNe.  

Overall, the inclusion of additional calibrating PNe and the use of refined input data (fluxes, extinctions, and angular dimensions) have led to a slight improvement of the distance scale with respect to F08; the present mean scale is about four per cent longer, and more in agreement with the independent theoretical tracks computed by Jacob, Sch\"onberner \& Steffen (2013).  
While some previous authors (e.g. Schneider \& Buckley 1996) have suggested that a single power-law is inadequate to handle both young and old PNe, we find that a linear \SR\ relation is applicable as a robust distance method, excluding only the very youngest optically-thick PNe and transitional objects.

\subsection{Fundamental observables}\label{sec:fundamental_observables}
\subsubsection{Angular Dimensions}

For the brighter Galactic calibrating PNe, the angular dimensions have been taken from Tylenda et al. (2003) and Ruffle et al. (2004) if available.  These works quote major and minor axes at the 10\% level of the peak surface brightness isophote, which is a standard adopted throughout this work where feasible.  Note that the adopted dimensions are for the main PN shell, which encloses the rim, or primary shock, but does not include any faint outer halo(s) if present (e.g. Corradi et al. 2003; Frew et al. 2012). 
Major and minor dimensions for most of the largest PNe have been determined here anew, based on available digital broadband red or \ha\ + {[\rm N \sc ii]} images at the same isophote level.  These were primarily taken from the SHS, SSS, and IPHAS surveys with some recent images from the POPIPLAN survey (Boffin et al. 2012) also utilized.  For compact Galactic PNe, we utilised HST images if available, either from the literature (e.g. Sahai et al. 2007; Gesicki et al. 2014; Hsia et al. 2014) or from the Hubble Legacy Archive.\footnote{see \url{http://hla.stsci.edu/}}  The dimensions of compact PNe derived from ground-based measurements were corrected using a PSF deconvolution if needed (e.g. Ruffle et al. 2004).   We then calculated geometric mean diameters and radii for each PN.   The uncertainties have been adopted directly from the relevant references if present, or calculated from inverse variances if more than one determination is available.  

For the LMC and SMC PNe we adopt the major and minor axial dimensions from Shaw et al. (2001), Stanghellini et al. (2002, 2003) and Shaw et al. (2006), based on HST imagery.  For consistency with the sample of Galactic calibrating objects, the angular dimensions at the 10\,per cent brightness contour have been used from these references, rather than the `photometric radii', encompassing 85\% of the total flux, defined by Stanghellini et al. (1999).   For the three calibrating PNe belonging to the Sagittarius dSph galaxy, we adopted the dimensions from Zijlstra et al. (2006).

\begin{figure*}
\begin{center}
\includegraphics[width=15.5cm]{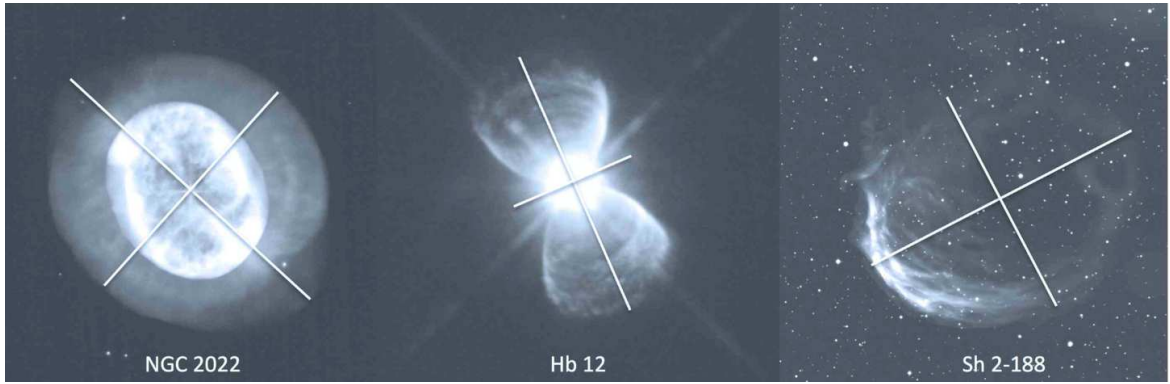}
\caption{Major and minor axes over-plotted on three PNe, to show how the dimensions are determined; the elliptical isophotes have been omitted for clarity.  The objects are (from left to right) the double-shell elliptical NGC 2022, the bipolar Hubble 12, and the strongly asymmetric Sh\,2-188 (image credits: Hubble Legacy Archive and INT Photometric \ha\ Survey of the Northern Galactic Plane).}  
\label{fig:dimensions}
\end{center}
\end{figure*}

The isophote method is best suited for elliptical and round PNe.  However, some highly evolved PNe strongly distorted by interaction by the ISM have been treated differently.  In these cases a strict application of the 10 per cent isophote rule may only give dimensions of the bright interacting rim, a typical example being Sh\,2-188 (Wareing et al. 2006).  In this case an isophote which includes the non-interacting part of the main shell is used to give the overall dimensions of the object.  Similarly, the dimensions for some evolved bipolar PNe are sometimes hard to define, and are dependent on the exact orientation of the `waist'.  In most cases these are relatively large PNe, so the subjective effect of choosing an appropriate contour has only a relatively small percentage change on the overall dimensions of the nebula.  Figure\,\ref{fig:dimensions} shows how the major and minor axes have been determined for three PNe of differing morphological types.    

\subsubsection{Integrated Fluxes}

For Galactic PNe, the integrated \ha\ fluxes and their uncertainties are mostly adopted from Kohoutek \& Martin (1981), Dopita \& Hua (1997), Wright, Corradi \& Perinotto (2005), and FBP13, for the brighter objects, or from F08 and Frew et al (2014a) for a few of the largest and most evolved PNe.   For the LMC and SMC PNe we adopt the \ha\ fluxes and associated uncertainties from Shaw et al. (2001), Stanghellini et al. (2002, 2003) and Shaw et al. (2006), supplemented with data from Reid \& Parker (2010b).  For the PNe belonging to the Sagittarius dSph galaxy, we average the integrated \ha\ fluxes from Ruffle et al. (2004), Zijlstra et al. (2006) and FBP13.

Note that the integrated fluxes for less-evolved PNe, especially those measured with photoelectric photometry through large apertures or from CCD surveys of limited resolution may include some or all of any faint surrounding AGB halo.  We expect this to be a minor effect, as the typical halo surface brightness is a factor $10^{-3}$ less than the main shell, while the surface area of the halo is an order of magnitude larger than the main shell (see Corradi et al. 2003).  This means that on average, only about one per cent of the total flux resides in a typical AGB halo.  Moreover, since the calibrating sample includes several PNe with surrounding haloes, there should be little error in the application of our method to other objects.

\subsubsection{Extinction Constants}

The logarithmic extinction constants,  $c_{\rm H\beta}$, for Galactic PNe are widely scattered in the literature.  Extensive data compilations include CKS, Tylenda et al. (1992), Condon \& Kaplan (1998), Ruffle et al. (2004), Giammanco et al. (2011), Kovacevic et al. (2011), and FBP13.   The extinction constants are usually determined from the Balmer decrement, as derived from optical spectroscopy, or by comparing the Balmer and radio continuum fluxes.  To be as homogenous as possible, we re-calculated the radio--\ha\ extinctions using the radio data and methods given in Boji\v{c}i\'c (2010) and Boji\v{c}i\'c et al. (2011a, 2011b, and references therein).

However, since the extinctions for many faint PNe were either previously unknown, or unreliable, new values were determined where applicable.  Similarly, extinctions for brighter PNe were re-derived from published \ha/\hb\ ratios adopting a Howarth (1983) reddening law.  For the \ha/\hb\ ratios we adopted an average of the data presented in Acker et al. (1992), the \emph{Catalog of Relative Emission Line Intensities Observed in Planetary Nebulae} (ELCAT) compiled by Kaler, Shaw \& Browning (1997), the extensive database of $>$2000 spectra taken as part of the MASH survey and related programmes, supplemented with data taken from more recent papers in the literature, including Torres-Peimbert \& Peimbert (1977), Kohoutek \& Martin (1981), Guti\'errez-Moreno, Cortes \& Moreno (1985), Shaw \& Kaler (1989), Dopita \& Hua (1997), Tsamis et al. (2003, 2008), Liu et al. (2004), Wright, Corradi \& Perinotto (2005), Zhang et al. (2005), and Wang \& Liu (2007). Other papers were outlined in Frew et al. (2013).  
For higher-latitude objects that still had poor-quality data, we utilised the reddening data from  Schlafly \& Finkbeiner (2011) as a cross-check.  Finally, for PNe with adequate central star data, \EBV\ values for the central stars have been calculated following F08, De Marco et al. (2013) and Douchin et al. (2014), using available $UBVR_cI_cJ$ photometry from the literature.   

An average of measurements from several independent sources should be fairly representative of the extinction for each PN.  The extinction uncertainties have been adopted directly from the relevant references if present, or calculated from inverse variances if more than two independent values are available.    We plan to publish the individual extinction determinations separately, in Version\,1 of our global MASPN Database (see Parker et al. 2015). 

For the extragalactic PNe we calculate the extinction constants from the flux data presented by Shaw et al. (2001, 2006), Stanghellini et al. (2002, 2003), Ruffle et al. (2004), Zijlstra et al. (2006), and Reid \& Parker (2010b), adopting a minimum value from Schlafly \& Finkbeiner (2011) if the calculated extinction is less than this.  
The \hb\ and \ha\ logarithmic extinction constants, $c_{\beta}$ and $c_{\alpha}$, are related to the reddening following the Howarth (1983) extinction law:
\begin{equation}
\label{eq:reddening_constant}
\begin{matrix}
{c_{\beta}   = 1.45~E(B-V)}\\
{c_{\alpha}  = 0.99~E(B-V)}
\end{matrix}
\end{equation}

The \ha\ extinction coefficient was added to the observed logarithmic \ha\ flux to get the reddening-corrected flux for each PN.  The intrinsic \ha\ surface brightness\footnote{To convert a log flux per steradian to a log flux per square arcsec, subtract 10.629 dex.}  in units of \ergcms\ sr$^{-1}$ was then calculated from the angular geometric radius ($\theta$) and reddening-corrected flux, using the formula:
\begin{equation}
\label{eq:SB_flux}
{S_{\rm H\alpha}} = \frac{F_{\rm H\alpha}}{4\pi \theta^2}
\end{equation}

\subsection{Final Calibrating Sample}\label{sec:cal_sample}

Nearly 30 Galactic PNe have distances based on more than one primary method.  For these PNe, a weighted average distance has been calculated based on the quoted uncertainties of each individual distance determination.  For consistency, individual distances  were combined within each method first (after removing outlying data points using a 2$\sigma$ cut).  These were then combined with distances from other primary methods weighted by inverse variances to determine the final weighted distance, using:  
\begin{equation}
\label{eq:weighted_mean}
    D_w = \frac{ \sum_{i=1}^n w_i D_i}{\sum_{i=1}^n w_i}
\end{equation}

where [$D_{1}, D_{2} \ldots \,D_{n}$] are the individual distance estimates, with associated weights [$w_{1}, w_{2} \ldots \,w_{n}$] determined from the inverse variances, $w_{i} = 1/\sigma_{i}^{2}$.
The uncertainty of the weighted mean distance was calculated (following FBP13) as:
\begin{equation}
\label{eq:weighted_error}
\sigma_{D_w}  =  \bigg( \frac {V_1} {V_1^2-V_2} \sum_{i=1}^n w_i \left(F_i - \bar{F}_{w} \right)^2 \bigg)^{0.5}
\end{equation}
   
where $V_1 = \sum_{i=1}^n w_i$  and $V_2 = \sum_{i=1}^n {w_i^2}$.
\smallskip

Finally, for each calibrator, the linear radius was determined from the angular radius and the adopted distance using Equation\,\ref{eq:small_angle}.

This approach is quite robust to any error in the angular dimensions, because this flows through to both the surface brightness and the radius. For example, a 20 per cent uncertainty in each angular dimension (40 per cent uncertainty in the calculated surface brightness) leads to only a $\sim$10 per cent uncertainty in the distance.  Similarly, owing to the form of the \SR\ relation, an uncertainty of 20 per cent in the \ha\ flux leads to only a 5 per cent error in the computed radius, i.e. the PN distance.  Hence, scatter introduced into the \SR\ relation due to observational uncertainties in the angular dimensions, fluxes or extinctions are generally minor compared to the uncertainties in the distances of the calibrating PNe, or the dispersion in the relation due to cosmic scatter (see below).  However, for highly reddened PNe, the uncertainty in the surface brightness is dominated by the extinction uncertainty which can reach 0.3\,dex in some cases, leading to an additional uncertainty of 20 per cent in the distance.. 

Table~\ref{table:SBr_calibrators} gives the relevant observational and derived data for the full calibrating sample of 322 PNe.  These range from the very nearest objects out to PNe at the distance of the SMC (0.13\,$\leq$\,$D$\,$\leq$\,60 kpc).   The columns in Table~\ref{table:SBr_calibrators} consecutively give the PN designation, common name, adopted distance (in kpc), the method of distance determination, a simplified morphological code (E = elliptical, B = bipolar, R = round, A = Asymmetric) after Parker et al. (2006), the major and minor dimensions in arcseconds, the adopted $E(B-V)$ value (in mag), the reddening-corrected \ha\ surface brightness (in cgs units per steradian), and the logarithm of the nebular radius (in pc).

\begin{table*}
\begin{center}
\caption{Final calibrating nebulae for the \SR\ relation.  The table is published in its entirety as an online supplement. A portion is shown here for guidance regarding its form and content.}
\label{table:SBr_calibrators}
\begin{tabular}{llccccccccc}
\hline
PN G		&	Name		&     $D$ (pc)				&Method	&    Trend	& Morph 	&~~$a$ (\arcsec)~~ &~~$b$ (\arcsec)~~   & \EBV\ 	&   $S_0$(H$\alpha$) 	&  log $r$ (pc) 	\\ 
\hline
002.1$+$01.7	&	JaFu~1		&	7200 $\pm$ 700		&	C	&	Inter		&	Eb	&	8.0		&	8.0		&	$1.93 \pm 0.21$	&	$-2.20 \pm 0.26$	&	$-0.86$		\\
002.4$+$05.8 	&	NGC~6369	&	1550 $\pm$ 300 		&	M 	&	Inter 	&	Eb	&	30.0		&	29.0		&	$1.31 \pm 0.16$	&	$-1.01 \pm 0.17$ 	&	$-0.96$		\\
003.5$-$04.6	&	NGC~6565	&	2000 $\pm$ 500		&	X	&	Inter		&	E	&	18.0		&	13.0		&	$0.31 \pm 0.10$ 	&	$-1.95 \pm 0.12$	&	$-1.13$		\\
004.0$-$03.0	&	M~2-29		&	7100 $\pm$ 2200		&	G	&	Thin		&	E	&	4.8		&	3.6		&	$0.72 \pm 0.14$	&	$-1.25 \pm 0.15$	&	$-1.16$		\\
010.4$+$04.4	&	DPV~1		&	3400 $\pm$ 500		&    M; Z	&	Thin		&	R	&	44.0		&	44.0		&	$0.71 \pm 0.08$	&	$-4.35 \pm 0.15$	&	$-0.51$		\\
010.8$-$01.8	&	NGC~6578	&	2900 $\pm$ 800		&	E	&	Inter		&	E	&	12.1		&	11.8		&	$0.93 \pm 0.10$	&	$-1.18 \pm 0.12$	&	$-1.08$		\\
011.7$-$00.6	&	NGC~6567	&	1680 $\pm$ 170		&    E; H	&	Thin		&	E	&	8.1		&	6.4		&	$0.48 \pm 0.10$	&	$-0.79 \pm 0.11$	&	$-1.52$		\\
013.8$-$02.8	&	SaWe~3		&	2100 $\pm$ 300		&	X	&	Thick	&	B	&  110.0		&	80.0		&	$0.72 \pm 0.27$	&	$-3.82 \pm 0.27$	&	$-0.32$		\\
019.6$+$00.7	& MPA J1824-1126	&	11800 $\pm$ 4100		&	P	&	Inter		&	E	&	13.0		&	13.0		&	$1.19 \pm 0.14$	&	$-3.30 \pm 0.20$	&	$-0.43$		\\
021.8$-$00.4	&	M~3-28		& $2500^{+1100}_{-1300}$	&	K	&	Thick	&	B	&	24.1		&	12.1		&	$1.34 \pm 0.21$	&	$-2.32 \pm 0.21$	&	$-0.99$		\\
~~~~~~~\vdots &~~~~~~\vdots	&	     \vdots				& \vdots 	&     \vdots	& \vdots	&   \vdots		&  \vdots		&	  \vdots			&	 \vdots			&	\vdots		\\
\hline
\end{tabular}
\end{center}
{\scriptsize
\begin{flushleft}
Method codes: ~B -- eclipsing binary CSPN; C -- cluster membership;  E -- expansion parallax; G -- gravity distance; H -- \HI\ absorption distance; K -- kinematic method;  M -- photoionization model distance;  P -- photometric parallax; T -- trigonometric parallax; X -- extinction distance; Z - other distance estimate.
\end{flushleft}  
}
\end{table*}

The general form of the relationship between surface brightness and radius is expected to be a power law, with constants $\gamma$ and $\delta$ describing the slope and zero point respectively, viz:
\begin{equation}
\label{eq:HASB_R}
{\rm log}\,{S_{\rm H\alpha}} = \gamma\,{\rm log}\,r - \delta
\end{equation}

We use an ordinary least-squares (OLS) bisector fit (Isobe et al. 1990) to represent the full calibrating sample, since observational errors are present in both the nebular fluxes and diameters (i.e. the surface brightness) which are independent of the errors on the distances, and hence the physical radii.  The justification for this approach was discussed by Isobe et al. (1990) and Feigelson \& Babu (1992).   Disk PNe with formal uncertainties in the distance of less than 10\% have been given double weight in the calculation of the coefficients.  All other PNe have been assigned unit weight, except for the Bulge objects, assigned half weight.  The best fit based on our full sample of 332 PNe is represented by the equation:
\begin{equation}
\label{eq:HASB_R}
{\rm log}\,{S_{\rm H\alpha}} = -3.63 (\pm 0.06)\,{\rm log}\,r - 5.34 (\pm 0.05)
\end{equation}

with a Pearson correlation coefficient, $R = -0.96$.   The slope is steeper than the r$^{-3}$ law previously found for LMC and SMC PNe by Shaw et al. (2001) and Stanghellini et al. (2002), primarily due to the different treatment of the PN dimensions by those authors.  

\begin{figure}
\begin{center}
\includegraphics[width=9.15cm]{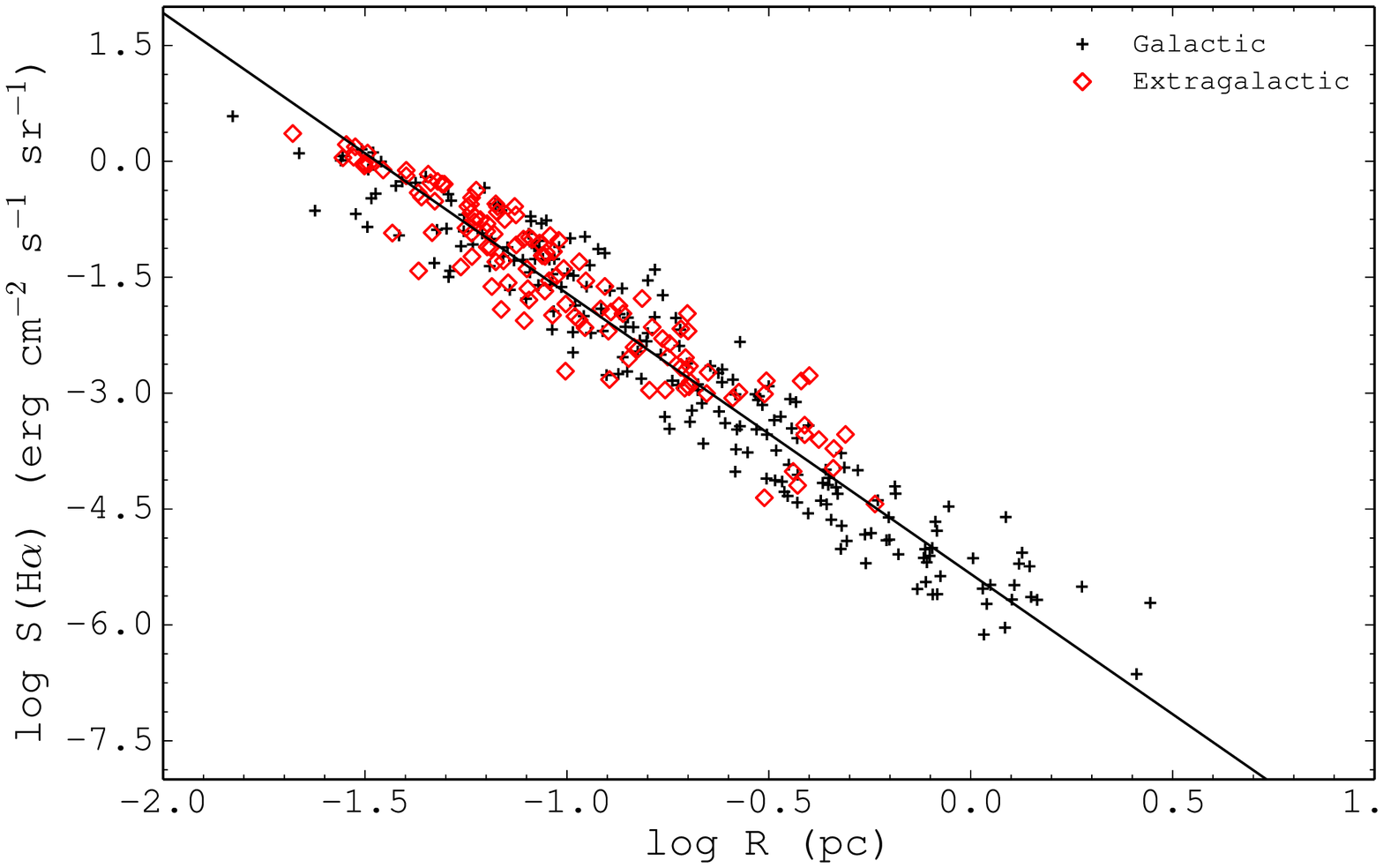}
\includegraphics[width=9.15cm]{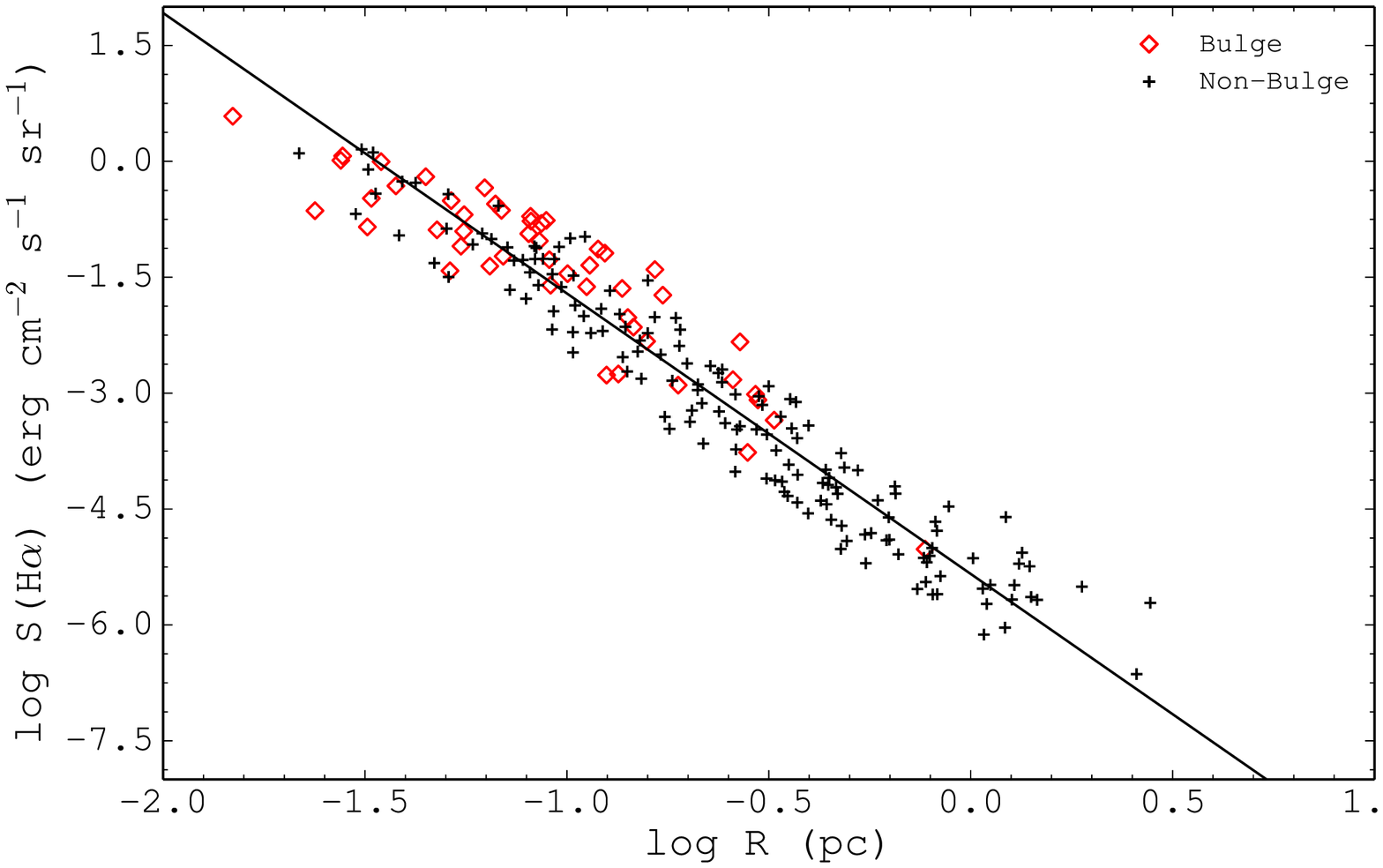}
\caption{Top panel: \SR\ relation plotting the Galactic calibrating sample of 206 PNe (crosses), as well as the the 126 extragalactic PNe from the LMC, SMC, and Sgr dwarf spheroidal galaxy (red diamonds), spanning $>$6.5 dex in surface brightness.  The line is a least-squares bisector fit to the entire calibrating sample.  The shallower gradient of the relation at small radii is compared with theoretical tracks in Fig.\,\ref{fig:jacob}.  Lower panel:  \SR\ relation comparing Galactic Bulge PNe with the remaining PNe.   
}
\label{fig:SBrel_all}
\end{center}
\end{figure}

The overall impression of the \SR\ relation (Fig.\,\ref{fig:SBrel_all}) is a well behaved linear trend, but with a shallower gradient at small radii, discussed further in \S\ref{sec:young_PNe}.  There may be a flattening of the slope at the very bottom of the locus, but this needs to be confirmed with more data.  The origin of the radius dependence at large radii may be due to more uncertain distances combined with lower quality \ha\ fluxes for the very largest PNe.  It is also possible that some of the very oldest PNe may be `re-brightened' by an interaction with the ISM (see Wareing 2010).  The most discrepant objects Sh\,2-188, Sh\,2-216, and WeDe\,1 are within $\sim$100\,pc of the Galactic mid-plane or less than the dust scale height (Spitzer 1978).  The surface brightness of these PNe might be enhanced by mass augmentation from the ISM, especially close to the Galactic plane, or alternatively by shock excitation for fast moving PNe (Wareing et al. 2006), which would lead to these objects lying above the power law derived from the total calibrating sample.  Indeed the optical spectrum of Sh\,2-188, shows extraordinarily strong \SII\ lines for a PN (Rosado \& Kwitter 1982), suggesting shock excitation is important in this object.

\subsection{PN subsamples in the \SR\ plane}\label{sec:subsamples}

We recommend applying the mean \SR\ trend (Equation\,\ref{eq:HASB_R}) for all PNe that have a spectroscopic signature that does not allow classification as definitively optically-thick (\S\ref{sec:thick_PNe})  or optically-thin (\S\ref{sec:thin_PNe}), and for other PNe for which the required optical spectroscopy is currently lacking.   
The calibrating PNe in the \SR\ relation represent the full range of properties manifested by PNe, such as morphological type, excitation class, ionised mass, metallicity, and central star luminosity, so we have hopefully circumvented the thorny problem of Malmquist bias\footnote{Malmquist bias is present when the intrinsic (cosmic) dispersion of a sample of objects is significant.  In other words, if a sample of objects (stars, PNe or galaxies, for example) is flux-limited, then only the most luminous objects are selected at large distances, so there is an observed increase in the average luminosity of a flux-limited sample as distance increases.} (Malmquist 1924).   Having new and revised data available for these calibrators also  provides the opportunity to investigate the presence of any sub-trends within the relation.   Table\,\ref{SB-r_revised_coeffs} provides a summary of the equation coefficients for the most important subsets of calibrating nebulae.  Excluding the very youngest PNe, the observed power-law slope of the \SR\ relation is between $-3.3$ and $-3.8$, depending on the subset used.  The small offset between the Galactic disk and extragalactic samples is due to one or more of Malmquist bias (the extragalactic sample is flux and surface brightness limited), systematic errors in measuring PN diameters (more difficult for extragalactic PNe), and possibly progenitor mass and metallicity differences (e.g. Jacob et al. 2013) between the different galaxies.

\begin{figure}
\begin{center}
\includegraphics[width=9.15cm]{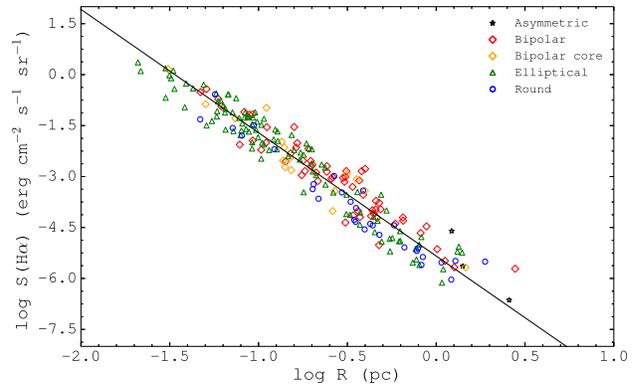}
\caption{
\SR\ relation for the calibrating sample (excluding the Bulge objects), with morphology indicated by different symbols (refer to the text for more details). A colour version of this figure is available in the online journal.}
\label{fig:SBrel_morph}
\end{center}
\end{figure}

Owing to the relative difficulty of morphologically classifying PNe from two-dimensional images (e.g. Kwok 2010; Chong et al. 2012), we do not formally calculate different sub-trends for the various morphological classes, but only provide a visual breakdown by class, seen in the right panel of Figure~\ref{fig:SBrel_morph}.  Canonical bipolar PNe and elliptical PNe with bipolar-cores tend to populate the upper part of the broad trend in the \SR\ plane.  Elliptical PNe without bipolar cores are more uniformly spread, while spherical PNe tend to plot beneath the mean trend-line, at moderate to large radii.   To help alleviate the problem of cosmic scatter, we now subdivide the full ensemble of PNe into different subsets {\sl based on spectroscopic criteria}, discussed in the following sections.

\begin{table}
\begin{center}
\caption{Summary of revised \SR\ relation best-fit constants for different PN subsets as defined in the text.  }
\label{SB-r_revised_coeffs}
{\footnotesize
\begin{tabular}{lrccc}
\hline
Subset								& ~~$n$~~ 	&        $\gamma$  			&		 $\delta$ 			&       $R$ 		\\
\hline
All Calibrators$^{\dagger}$				&    322		&	$-3.63\pm0.06$		&	$-5.32\pm0.05$		&  	$-0.96$		\\   	
Galactic Disk							&    147		&	$-3.58\pm0.06$		&	$-5.38\pm0.04$		&	$-0.96$		\\ 	
Galactic Bulge 						&      45		&	$-3.27\pm0.22$		&	$-4.85\pm0.25$		&	$-0.90$		\\   	
Extragalactic$^{\ddag}$					&    126 		&	$-3.50\pm0.11$		&	$-5.13\pm0.11$		&	$-0.94$		\\
\hline
Optically-thick 						&    126		&	$-3.32\pm0.12$		&	$-4.97\pm0.08$		&	$-0.95$		\\
Intermediate							&    106		&	$-3.59\pm0.09$		&	$-5.21\pm0.10$		&	$-0.97$		\\
Optically-thin 							&      90		&	$-3.75\pm0.11$		&	$-5.73\pm0.07$		&	$-0.97$		\\
Compact {\scriptsize ($r$$<$0.04\,pc)}  	&     34		&	$-2.74\pm0.51$		&	$-4.15\pm0.80$		&	$-0.68$		\\ 	
\hline
\end{tabular}
}
\end{center}
{\scriptsize
\begin{flushleft}
Note:~$^{\dagger}$Includes four Galactic halo PNe;~$^{\ddag}$LMC / SMC PNe, and 3 PNe from the Sgr dSph galaxy.
\end{flushleft}  
}
\end{table}

\subsubsection{Optically-thick PNe}\label{sec:thick_PNe}

These PNe have relatively strong low-excitation features such as the [N\,{\sc ii}], [O\,{\sc ii}] and [S\,{\sc ii}] lines.  We follow Kaler \& Jacoby (1989) and Jacoby \& Kaler (1989) in defining an optically-thick PN as having the reddening-corrected ratios $F(\lambda3727)$/$F$(\hb) $\geq$ 1.5 and/or $F(\lambda6584)$/$F$(\ha) $\geq$1.  Using only those calibrators that meet these spectroscopic criteria to define the relation, the optically-thick (or `long') trend is given by the equation:
\begin{equation}
\label{eq:HASB_typeI}
{\rm log}\,S_{\rm H\alpha} = -3.32 (\pm 0.12)\,{\rm log}\,r - 4.97 (\pm 0.08)
\end{equation}

Many optically-thick bipolar PNe also have Type~I chemistries, using the Kingsburgh \& Barlow (1994) definition.  A subset of 45 known Type\,I PNe was extracted from the overall calibration sample, all but one of which is morphologically bipolar, and the coefficients are given in Table\,\ref{SB-r_revised_coeffs}.    The resulting relation is statistically indistinguishable from the general optically-thick trend, which is preferred.

\subsubsection{Optically thin PNe}\label{sec:thin_PNe}

These PNe are the spectroscopic opposites of the optically-thick PNe, and are defined as PNe having {\sl very weak or absent low excitation lines} of [N\,{\sc ii}], [O\,{\sc ii}] and [S\,{\sc ii}] (cf. Kaler 1981; Frew et al. 2014c).  Formally we define optically-thin PNe as having the line ratio $F(\lambda6584)$/$F$(\ha) $\leq$ 0.1.   The [O\,{\sc ii}] and [S\,{\sc ii}] emission lines are similarly weak to absent.  A subset of high-excitation (HE) objects have the same [N\,{\sc ii}] criterion, but also have $F$(He{\sc ii}) $\geq$ 0.75\,$F$(H$\beta$), and relatively strong emission lines of other high-excitation species, such as \OIV, \ArIV, \ArV, and \NeV\ (cf. F08).   Representative examples of the latter group include NGC\,1360 (Goldman et al. 2004) and the more evolved object MWP\,1, which is invisible on deep \NII\ images (see Tweedy \& Kwitter 1996).   Note that the nebular excitation class (e.g. Dopita \& Meatheringham 1991; Reid \& Parker 2010a) does not map closely with our definition of optical depth, so has not been investigated further.

Most HE PNe appear to have CSPNe still on the nuclear burning track close to the turnaround point or `knee' in the HR diagram.  These PNe are optically thin to the \HI\ continuum and usually to the \HeII\ continuum as well, and consist essentially of a He$^{2+}$ Str\"omgren zone, i.e.  $T_{z}$(He\,{\sc ii}) $>$  $T_{z}$(H) (K\"oppen 1979; Torres-Peimbert et al. 1990).  The ionization parameter is high, and their spectroscopic uniformity reflects the systematically lower ionised masses of these nebulae.  Consequently, these PNe plot near the lower bound of the overall \SR\ locus.  However, their CSPNe are spectroscopically heterogeneous, with both H-rich and H-deficient nuclei, and at least two belong to the born-again class (e.g. Guerrero et al. 2012 and references therein).  This is suggestive that several evolutionary scenarios may produce low-mass PNe (Frew \& Parker 2007, 2010, 2012).
The optically-thin (or `short') trend should only be used for PNe that meet the spectroscopic criteria described above.  It is represented by the equation:
\begin{equation}
\label{eq:HASB_typeI}
{\rm log}\,S_{\rm H\alpha} = -3.75 (\pm 0.11)\,{\rm log}\,r - 5.73 (\pm 0.07)
\end{equation}

HE PNe typically have either round or elliptical morphologies, sometimes with amorphous filled centres, though some are strongly axisymmetric objects associated with post-common envelope nuclei (De Marco 2009; Corradi et al. 2011).  Indeed many post-common envelope PNe are optically-thin following our definition, from which Frew \& Parker (2007) and F08 suggested that these PNe have systematically lower ionised masses in the mean, typically only $\sim$0.1\,$M_{\odot}$.  Curiously, known close-binary PNe show a somewhat restricted range of H$\alpha$ surface brightness ($S_{\rm H \alpha}$ \lessim\ $-2.5$\,erg\,cm$^{-2}$s$^{-1}$sr$^{-1}$) compared to the full observed range for all PNe ($S_{\rm H \alpha} \simeq$ +0.2 to $-6.7$\,erg\,cm$^{-2}$s$^{-1}$sr$^{-1}$).  In other words, the PNe of highest surface brightness are rarely observed to host close-binary nuclei.  This has been traditionally interpreted as a selection effect (e.g. Bond \& Livio 1990), but may instead be pointing to a physical effect in that post-CE PNe are born ``old'', with moderate surface brightnesses at best, and with preferentially lower-mass CSPNe.  To address this problem, a more detailed statistical study of these PNe is planned for a future paper in this series.

\begin{figure}
\centering  
\includegraphics[width=9.15cm]{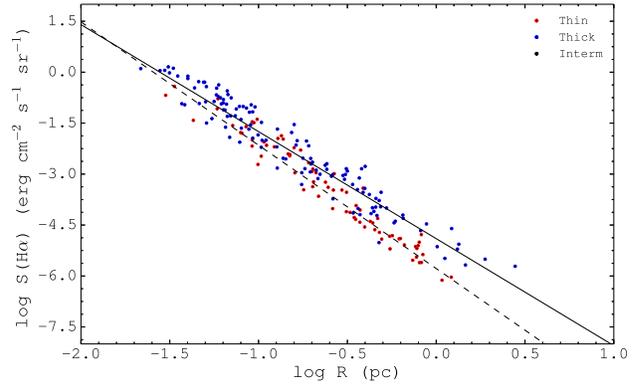}
\caption{\SR\ relations for optically-thick and optically thin PNe, plotted separately.  A colour version of this figure is available in the online journal.}
\label{fig:trends}
\end{figure}

\subsubsection{Compact high-SB PNe}\label{sec:young_PNe}

The overall impression of the \SR\ relation is that of a shallower gradient at small radii.  This was noted by F08, but is more apparent with the revised calibrating sample from this work, albeit demonstrated mostly by the Cloud and Bulge sub-samples.  To investigate this, we subdivided the calibrating sample into two groups on the basis of intrinsic radius, separated at log\,$r$ = $-1.40$ ($r$ = 0.04~pc).  A bisector fit to the compact PN sample ($n = 34$) is given by:    
\begin{equation}
\label{eq:SR_small}
{\rm log}\,S_{\rm H\alpha} = -2.74 (\pm 0.51)\,{\rm log}\,r - 4.15 (\pm 0.80)
\end{equation}

with a markedly lower correlation coefficient of $R = -0.68$.  This slope is shallower than the gradient we observe for the full calibrating sample.  However, as compact PNe in the Galaxy tend to have a lower surface brightness for a given radius than those observed to date in the Magellanic Clouds, likely due to selection effects, we recommend against using this relation at this point.  Alternatively, the youngest, dustiest PNe may be amenable to having distances calculated via the SED technique (e.g. Vickers et al. 2015), further described in \S\ref{sec:SED}.

\subsubsection{Subluminous PNe}\label{sec:subluminous_PNe}

We find evidence for a small heterogeneous group of peculiar, subluminous PNe that fall $>$3$\sigma$ below the main \SR\ locus, based on the primary distance estimates tabulated here.  These are RWT\,152, HbDs\,1, K\,1-27, HaWe\,13, Hen\,3-1357 (the Stingray nebula), and the central core of KjPn\,8 (discussed in \S\ref{sec:background}).  The first two appear to have low-mass H-normal stars and may represent a population of objects largely overlooked in current surveys, though there is some evidence that HbDs\,1 might be a wisp of ionized ISM (Frew et al., in preparation).  However RWT\,152 appears to have the typical morphology of a PN; its flux and diameter data have been taken from Pritchet (1984) and Aller et al. (2014) respectively. K\,1-27 has a rare O(He) CSPN (Reindl et al. 2014a) and its discrepant nature has been discussed previously by Frew \& Parker (2010).  HaWe\,13 has a ionizing star on either a post-RGB or post-EAGB track, based on the parameters given in Table\,\ref{grav_dist}, and its morphology appears to be consistent with it being produced via a common-envelope interaction (e.g. Hall et al. 2013).  A further object, Hen\,3-1357, has been argued to be the product of a post-EHB pathway (see Reindl et al. 2014b), being a young, compact nebula at an unusually short distance of $\sim$1.5\,kpc, calculated via the gravity method.  For these reasons, none have been used as calibrating nebulae, but are plotted in Fig.\,\ref{fig:mimics} for illustrative purposes.

\subsection{The physical basis for the \SR\ relation}\label{sec:SBr_theory}

Detailed photoionization modelling of the \SR\ relation and its relationship to central star evolutionary tracks (Kwok 1985; Van de Steene \& Zijlstra 1995; Jacob et al. 2013), is beyond the scope of this paper but making some simple assumptions from emission theory, we can relate the observed gradient of the \SR\ relation to other parameters such as the ionized mass and electron density.   For an uniform spherical nebula of radius $r$, the integrated flux $F_{\rm H\alpha}$ emitted by the \ha\ recombination line is given by:
\begin{equation} 
\label{eq:F_recomb}
{F_{\rm H\alpha}} = \left( \frac{r^3}{3D^2} \right)\, h\nu_{\rm H\alpha} n_{\rm e} n_{\rm p} \alpha^{\rm eff}_{{\rm H}\alpha}
\end{equation}

where $n_{\rm e}$ and $n_{\rm p}$ are the electron and proton densities respectively, and $D$ is the distance to the PN (see Hua \& Kwok 1999).  In practice, PNe are not homogenous, and a volume filling factor $\varepsilon$ is used to take this into account. Various values are presented in the literature, but a consensus value, $\varepsilon$ = 0.3, is often adopted (Boffi \& Stanghellini 1994;  cf. Gathier 1987; Pottasch 1996).   

In the absence of extinction, the nebular \ha\ surface brightness is given by:
\begin{equation}
\label{eq:surface}
S_{\rm H\alpha} = {\epsilon \over 3} \,n_{e}^{2}\,r\, h\,\nu_{{\rm H}\alpha}\, \alpha^{\rm eff}_{{\rm H}\alpha}
\end{equation}

while the nebular ionized mass, $M_{\rm ion}$ is calculated with the following expression:
\begin{equation}
\label{eq:M_i_basic}
M_{\rm ion} = \frac{4\pi}{3} n_{\rm p} \mu m_{\rm H} \varepsilon r^3   
\end{equation}

where $\mu$ is the mean atomic mass per hydrogen atom.  
Combining equations~\ref{eq:F_recomb} and \ref{eq:M_i_basic}, the ionized mass can be expressed in terms of the angular radius, $\theta$, and the \ha\ flux as:
\begin{equation}
\label{eq:M_i_prov}
M_{\rm ion} = \frac{4\pi\mu {\rm m_H}}{(3h\nu_{\rm H\alpha} x_e \alpha^{\rm eff}_{{\rm H}\alpha})^{1/2}}\, \varepsilon^{1/2} \theta^{3/2} D^{5/2} F_{\rm H\alpha}^{1/2}  
\end{equation}

where $x_{\rm e}=n_{\rm e}/n_{\rm p} \simeq$\,1.16 (Hua \& Kwok 1999).   
Simplifying, equation~\ref{eq:M_i_prov} can be finally expressed in terms of the distance:
\begin{equation}  
\label{eq:M_i}
M_{\rm ion}/ M_{\odot} = 0.035\, \varepsilon^{1/2} \theta^{3/2} D^{5/2} {F'}_{\rm H\alpha}^{1/2}  
\end{equation}

where now, $F'_{\rm H\alpha}$ is the nebular \ha\ flux in units of 10$^{-12}$ \ergcms, $\theta$ is in arcmin and $D$ in kpc.   
Since, from equation~\ref{eq:surface},  the surface brightness $S_{\rm H\alpha} \propto n_{e}^{2}\,r$, and since $n_{\rm e} \propto M\,r^{-3}$, we can simplify to:
\begin{equation}
\label{eq:surface-mass}
S_{\rm H\alpha} ~\propto~ M_{\rm ion}^{2} r^{-5}
\end{equation}

A natural consequence of the interacting stellar winds (ISW) model (Kwok, Purton \& FitzGerald 1978; Kwok 1982) is that the mass of a PN shell increases with age, due primarily to the expansion of the ionization front within the nebula, as well as the snow-plow effect when the PN becomes evolved (Villaver, Manchado \& Garc\'ia-Segura 2002).  Hence, unlike the Shklovsky method which assumes constant ionised mass, PNe manifest an observable mass-radius relation.  Recalling that $M_{\rm ion} = r^{\beta}$ from equation\,\ref{eq:mass-radius}, we can also write:   \begin{equation}
\label{eq:S_alpha}
S_{\rm H\alpha} \propto r^{2\beta-5}
\end{equation}

Now the Shklovsky constant-mass assumption ($\beta$ = 0) predicts a  $r^{-5}$ power law (Seaton 1968), which is much steeper than observed.  Using the set of calibrating PNe defined here, the observed mean $r^{-3.6}$ relation predicts a value for $\beta$ = 0.7, somewhat smaller than earlier determinations (e.g. Daub 1982; Milne 1982; Kwok 1985), which we attribute to this study including the most evolved PNe with very faint central stars.  
Since the temperature and luminosity of the ionizing star change markedly during the evolution of the PN, this has a direct influence on the index $\beta$ (Perinotto et al. 2004).  
Yet despite our simplifying assumptions, it is quite remarkable that a simple linear relationship essentially defines the full population of PNe in the \SR\ plane, excluding the very youngest objects.    

Both Kwok (1985, 1993) and Samland et al. (1993) showed that errors in statistical distances increase rapidly as $\beta$ $\longrightarrow$ 2.5, at which point the method becomes degenerate; i.e. there is no dependence of surface brightness on radius.  Since observationally, the value of $\beta$ is much less than this, we conclude that the various \sgen\ relations in the literature are valid if calibrated correctly, with the only disadvantage being the observed cosmic scatter.

\begin{figure*}
\begin{center}
\includegraphics[width=15cm]{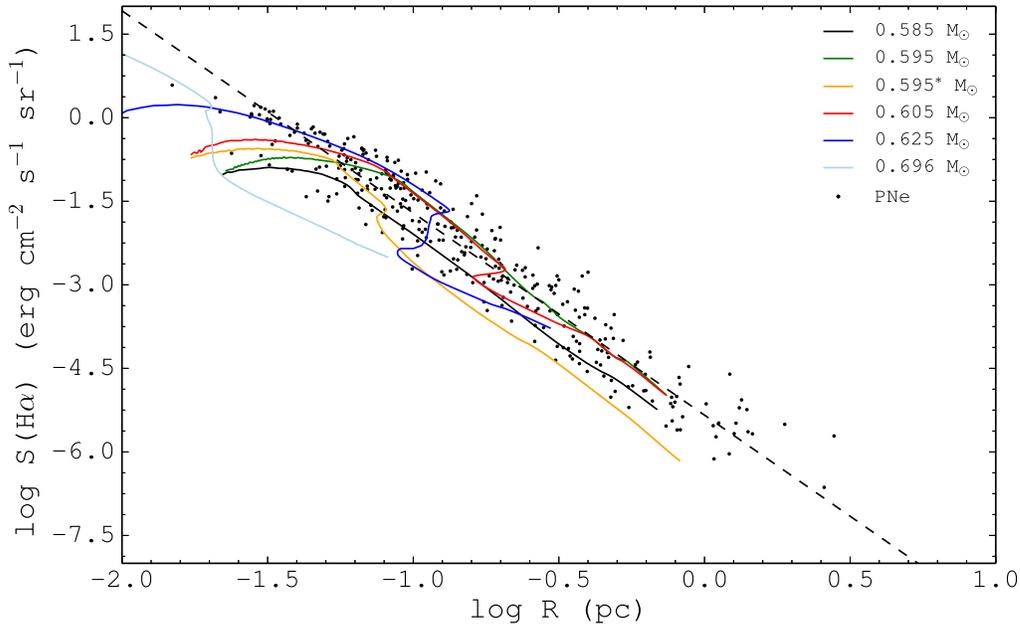}
\caption{The total calibrating sample, over-plotted with evolutionary tracks from Bl\"ocker (1995) and Sch\"onberner (1981) transformed to the \SR\ plane following Jacob et al. (2013).  The most evolved PNe are at bottom right.   A colour version of this figure is available in the online journal. }
\label{fig:jacob}
\end{center}
\end{figure*}

Our mean \SR\ scale is fully consistent with the theoretical evolutionary tracks of Jacob, Sch\"onberner \& Steffen (2013).  These tracks were generated from the hydrodynamical nebular models of Perinotto et al. (2004) and  Sch\"onberner et al. (2005a) along with the CSPN, using the evolutionary tracks for the latter from Bl\"ocker (1995) and Sch\"onberner (1981).  The nebular radius and surface brightness for PNe with a range of core masses were then over-plotted on the \SR\ plane in Fig.\,\ref{fig:jacob}.   The agreement is very good between these tracks and the observational data, with a slight offset owing to the slightly differing definition of angular size between the studies (see the discussion of Jacob et al. 2013).  Note that the evolutionary models do not extend to the lowest surface brightness owing to constraints in computational time.
Our results (see \S\ref{sec:catalogue}) show that with care, the mean-scale distances derived here have comparable accuracy to most direct methods currently in use, and significantly better than any other statistical distance indicator published in the literature to date (Jacob et al. 2013; Ali et al. 2015; Smith 2015).

We further note that Smith (2015) identified and discussed the scale error at large radii that affects the SSV distance scale. Recall that SSV used a constant mass assumption for all PNe larger than a radius of 0.06\,pc.   Following Van de Steene \& Zijlstra (1995), the optical thickness parameter of SSV is related to the brightness temperature by the following expression:
\begin{equation}
\label{eq:vdsz1}
{\rm log}\,T_{\rm b} = -\mathscr{T} + 4.86
\end{equation}

In the general $S$--$r$ plane, the two power laws have slopes of 1.64 (for thick PNe) and 5.0 for thin PNe; i.e. a constant mass trend.  However the observational data (Fig.\,\ref{fig:jacob}) appear to rule out a constant mass trend at large radii, at least for the PNe discovered and observed to date.   Similar scale errors afflicted the earlier studies of Daub (1982) and CKS, which had the optically thick/thin boundary at somewhat larger nebular radii of 0.12\,pc and 0.09\,pc respectively.


\section{The Distance Catalogue}\label{sec:catalogue}

Table~\ref{SB-r_distances} provides a catalogue of \SR\ distances for over 1100 Galactic PNe, published in its entirety as an online supplement.   The columns consecutively give the PN~G identifier, the usual name, the adopted geometric {\sl diameter} in arcsec, the adopted reddening and its uncertainty, the method used to determine the reddening, the logarithm of corrected \ha\ surface brightness and its uncertainty, the logarithm of the computed radius (in pc), and the resulting mean statistical distance in kiloparsecs (kpc) and its uncertainty.  The next two columns provide either a short (optically thin) or long (optically thick) distance if applicable.  The last column lists any notes, including if the PN is a calibrator for the relation.   The mean-trend distance is given for all objects, which can be conveniently used for future statistical comparisons with sets of primary distances or other secondary distance scales.  If an alternative distance is given, then this is the preferred distance to be used for studies of individual properties.

The inferred radius is derived from the adopted reddening-corrected surface brightness. Typical uncertainties in this parameter are calculated from the quadratic sum of the individual uncertainties in the angular size (actually the surface area), the integrated \ha\ flux and the extinction; typical logarithmic uncertainties are respectively 0.04\,dex, 0.02\,dex and 0.02\,dex  for a bright well observed PN (e.g. NGC\,2022 or NGC 3242), ranging to 0.10\,dex, 0.10\,dex and 0.04\,dex for a large asymmetric PN like Sh\,2-188 (recall Fig.~\ref{fig:dimensions}).  However, for highly reddened PNe, the uncertainty in the surface brightness is dominated by the extinction uncertainty, which can reach 0.3\,dex in the worst cases.  This is a contributing reason to our decision to give Galactic Bulge PNe reduced weight as calibrating objects.

\begin{table*}
\begin{center}
\caption{~A catalogue of \SR\ distances for Galactic PNe. The table is published in its entirety as an online supplement, and a portion is shown here for guidance regarding its form and content.  }
\label{SB-r_distances}
{\scriptsize
\begin{tabular}{llccccccccccc}
\hline
PN\,G & 
Name & 
$a$&
$b$ &
~~$E(B-V)$~~ &
method &
$\log$\,$S_0$(H$\alpha$) &
$\log$\,$r$  &
$D_{\rm mean}$ &
$D_{\rm thin}$&
$D_{\rm thick}$&
Notes 
\\
		&
		&
(\arcsec)	& 
(\arcsec)	& 
(mag)	&
		&
(cgs\,sr$^{-1}$)	&
(pc)		&
(kpc)		&
(kpc)		&
(kpc)		&
		&
\\
\hline
000.0$-$06.8&H~1-62&$       5.0$&$       4.0$&$0.49 \pm 0.29$&$1,3$&$-1.25 \pm 0.29$&$     -1.12$&$6.97 \pm 2.43$&$...$&$...$&...\\
000.1$+$17.2&PC~12&$       2.3$&$       2.2$&$0.54 \pm 0.31$&$1,3$&$-0.65 \pm 0.32$&$     -1.29$&$9.46 \pm 3.39$&$...$&$...$&...\\
000.1$-$01.7&PHR~J1752-2941&$      16.7$&$      12.2$&$0.99 \pm 0.31$&$1$&$-3.07 \pm 0.33$&$     -0.62$&$6.95 \pm 2.54$&$...$&$...$&...\\
000.1$-$02.3&Bl~3-10&$       7.2$&$       6.9$&$0.64 \pm 0.25$&$3$&$-2.41 \pm 0.30$&$     -0.80$&$9.24 \pm 3.24$&$7.62 \pm 2.12$&$...$&...\\
000.1$-$05.6&H~2-40&$      18.3$&$      16.9$&$0.50 \pm 0.22$&$1$&$-3.22 \pm 0.23$&$     -0.58$&$6.19 \pm 1.99$&$...$&$...$&...\\
000.2$+$01.7&JaSt~19&$       7.2$&$       6.4$&$1.59 \pm 0.07$&$1,3$&$-2.22 \pm 0.13$&$     -0.85$&$8.50 \pm 2.49$&$...$&$...$&...\\
000.2$+$06.1&Terz~N~67&$      16.0$&$      12.0$&$0.76 \pm 0.13$&$1,3$&$-3.57 \pm 0.26$&$     -0.48$&$9.79 \pm 3.27$&$...$&$...$&...\\
000.2$-$01.9&M~2-19&$       9.4$&$       8.5$&$0.83 \pm 0.21$&$1,3$&$-1.78 \pm 0.22$&$     -0.97$&$4.89 \pm 1.55$&$...$&$...$&...\\
000.3$+$12.2&IC~4634&$      20.5$&$       6.6$&$0.35 \pm 0.06$&$1,3$&$-1.31 \pm 0.08$&$     -1.10$&$2.79 \pm 0.79$&$2.35 \pm 0.44$&$...$&...\\
000.3$-$01.6&PHR~J1752-2930&$       8.6$&$       7.9$&$1.07 \pm 0.21$&$3$&$-2.90 \pm 0.23$&$     -0.67$&$10.79 \pm 3.48$&$...$&$...$&...\\
~~~~~~~~~~$\vdots$     		&	~~~~~~$\vdots$  	&  $\vdots$  		&   $\vdots$  		& $\vdots$  	&   $\vdots$			&   $\vdots$			&     $\vdots$		&  $\vdots$		&    $\vdots$ 	&    $\vdots$   &    $\vdots$ \\
\hline
\end{tabular}
}
\end{center}
{\scriptsize
\begin{flushleft}
Notes:~C -- calibrating object; ~P -- object has vetted primary distance but not used as a calibrator (see text).  Method code for reddening determination: 1- Nebular extinction from Balmer decrement or radio flux/Balmer comparison; 2 - CSPN photometry; 3 - Line-of sight galactic extinction.
\end{flushleft}
 }
\end{table*}

The distances given in Table~\ref{SB-r_distances} supersede any \SR\ distances previously published (Pierce et al. 2004; Frew et al. 2006b, 2011; F08; Viironen et al. 2009, 2011; Boji\v{c}i\'c et al. 2011b; Corradi et al. 2011) using earlier calibrations of the \SR\ relation, though in all cases the differences in distances are less than five per cent.

\section{Intrinsic Dispersion of the \SR\ relation}\label{sec:accuracy_SBr}

\begin{figure}
\centering  
\includegraphics[width=9.1cm]{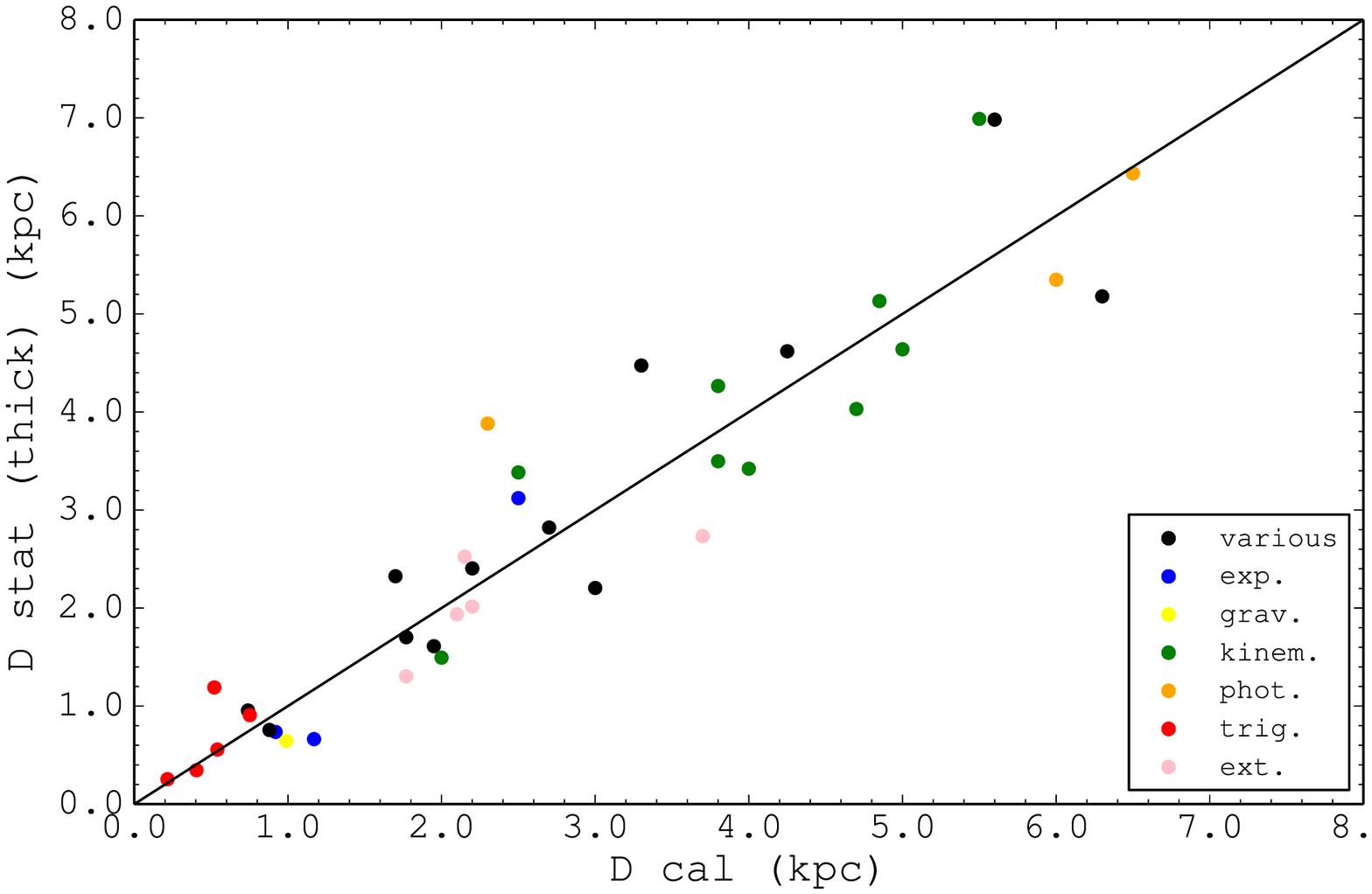}
\includegraphics[width=9.1cm]{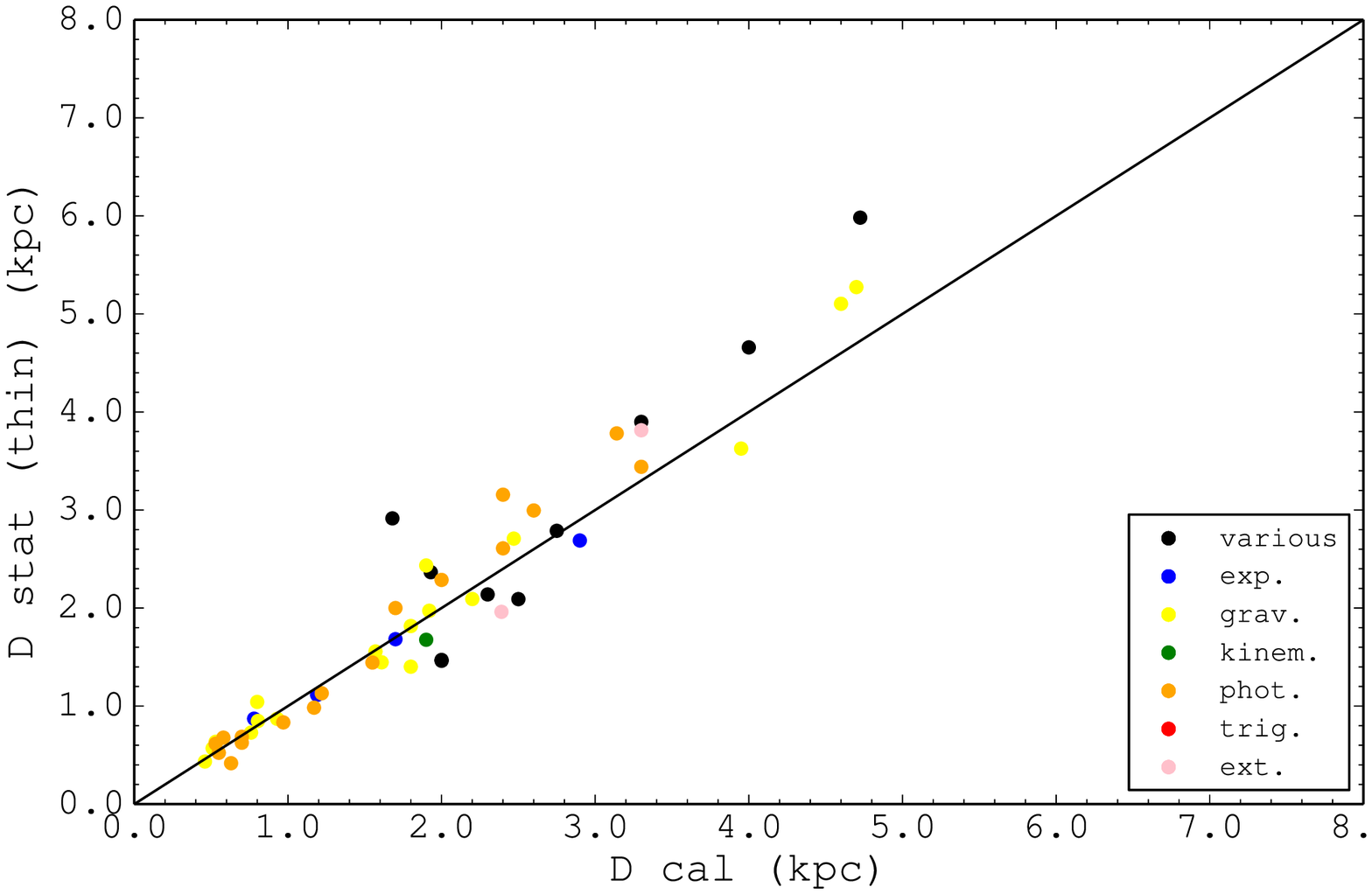}
\caption{Comparison of primary calibrating distances with our statistical distances for two subsets of Galactic calibrating PNe.  Individual distance techniques are colour-coded, as shown in the key, and error bars are omitted for clarity.  Three cluster distances are off-scale, and are not plotted.
The top panel plots the primary calibrating distance (abscissa) against the long-trend \SR\ distance (ordinate) for optically-thick PNe; the resulting dispersion is 28\%.  The lower panel plots the primary distance against the short-trend \SR\ distance for optically-thin PNe; the resulting dispersion is only 18\%.  The lines in each panel have a slope of unity.  A colour version of this figure is available in the online journal.}
\label{D_stat_calib_comparison2}
\end{figure}

The \SR\ relation is a robust statistical distance indicator for all PNe, and especially for those for which no primary distance technique is available.  In the fist instance, a measure of the dispersion of the technique can be evaluated by comparing the distances of the PNe in the calibrating sample with the distances derived for these PNe from the mean \SR\ relation.   The calculated distances have a        
dispersion of $\pm$\,28 per cent across the full range of intrinsic diameter.  In figure~\ref{D_stat_calib_comparison2} we refine this approach, by plotting individual PNe using the high- and low-trend statistical distances separately.  Using the relation for optically thick PNe only, a 28 per cent dispersion is similarly obtained.  In addition, using the `short' trend for optically-thin PNe gives a small resulting dispersion of only $\pm$\,18 per cent.  This 1$\sigma$ dispersion is considerably better than any previous statistical distance indicator, validating the use of sub-trends based on spectroscopic criteria.  The dispersion in the thick relation is higher than in the shorter thin relation, and close inspection shows that for a few bipolar PNe, the thick relation appears to be less accurate than the mean trend.  This may be due in part to the difficulty of accurately measuring the angular sizes of many bipolar PNe, but is also likely that the bipolar PNe are a heterogeneous group.  It appears likely that bipolar nebulae may be produced by both high-mass single progenitors as well as lower-mass close-binary stars (e.g. De Marco 2009).  SSV also find that their distance scale does not work well for bipolar PNe.

The observed dispersion includes a convolution of the uncertainties in both the calibrating distances and the statistical distances.  In order to gauge the uncertainties of each primary technique, the distances for individual PNe were compared with the adopted \SR\ distances.  Table~\ref{dist_method_dispersion} shows the results, 
which reveal that the gravity, kinematic, and extinction distance methods have the greatest uncertainties, unsurprisingly given the discussion in \S\,\ref{sec:evaluation}.  The problems with the gravity method have already been discussed.  The kinematic method was primarily applied to Type~I PNe, but it seems even these can sometimes have significant peculiar velocities, meaning that the technique should be used with caution.   The extinction method, while powerful in the sense that it can be applied to many PNe, is problematic, and care should be taken to avoid using PNe that are found in fields with significant differential extinction over small spatial scales (Giammanco et al. 2011).

\begin{table}
\begin{center}
\caption{Averaged distance ratios and nominal uncertainties of individual techniques using the mean \SR\ relation. The 1$\sigma$ uncertainties are a convolution of the uncertainties in the individual distances and the uncertainties in the adopted \SR\ distances (using sub-trends as described in the text).  }
\label{dist_method_dispersion}
{\footnotesize
\begin{tabular}{lccc}		
\hline
Distance technique		            		&~~$\kappa_{\rm mean}$~~	&~~$\kappa_{\rm adopt}$~~		&	~$n$~		\\
\hline
Trigonometric parallax				&	$0.93\pm0.29$		&	$1.03\pm0.24$			&	11			\\	
Photometric / spect. parallax 			&	$1.19\pm0.28$		&	$1.08\pm0.28$			&	31			\\
Cluster membership				&	$1.08\pm0.37$		&	$1.12\pm0.27$			&	6			\\
Gravity method					&     $1.17\pm0.58$		&	$1.03\pm0.39$			&	46			\\
Expansion parallax 				&	$1.02\pm0.24$		&	$0.96\pm0.20$			&	29			\\
Photoionization modelling   			&	$1.07\pm0.36$		&	$0.99\pm0.26$			&	13			\\
Kinematic method					&	$1.02\pm0.25$		&	$1.06\pm0.26$			&	25			\\    
Extinction distances				&	$1.05\pm0.38$		&	$1.08\pm0.37$			&	31			\\
Extragalactic PNe  					&	$0.99\pm0.27$		&	$0.95\pm0.26$			&	119			\\
Bulge PNe  						&	$0.98\pm0.36$		&	$0.96\pm0.35$			&	49			\\
\hline
\end{tabular}
}
\end{center}
\end{table}

\section{Comparison with Other Distance Scales}\label{sec:SB_comparison}

\begin{figure*}
\centering  
\includegraphics[width=8.82cm]{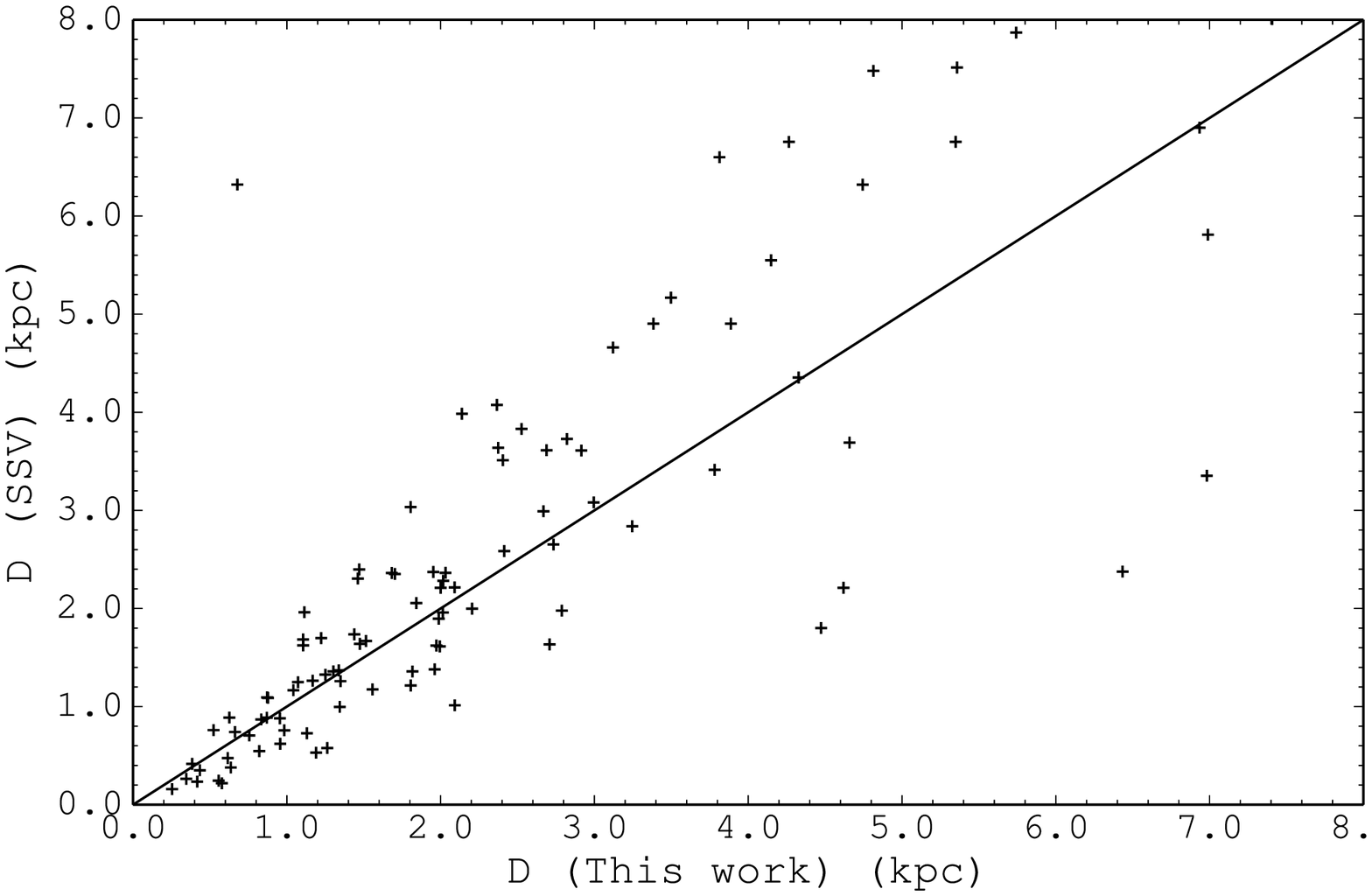}
\includegraphics[width=8.82cm]{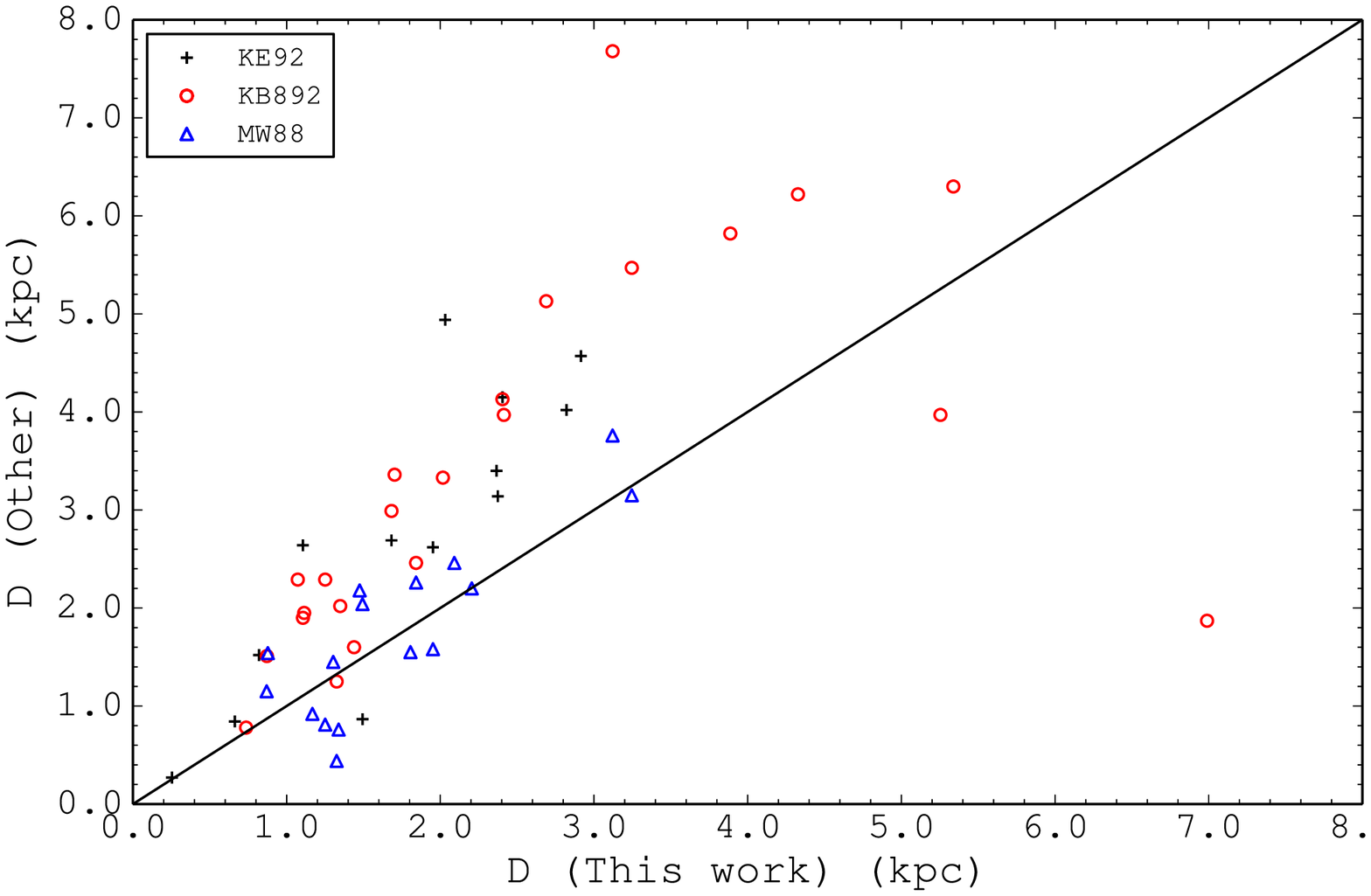}
\includegraphics[width=8.82cm]{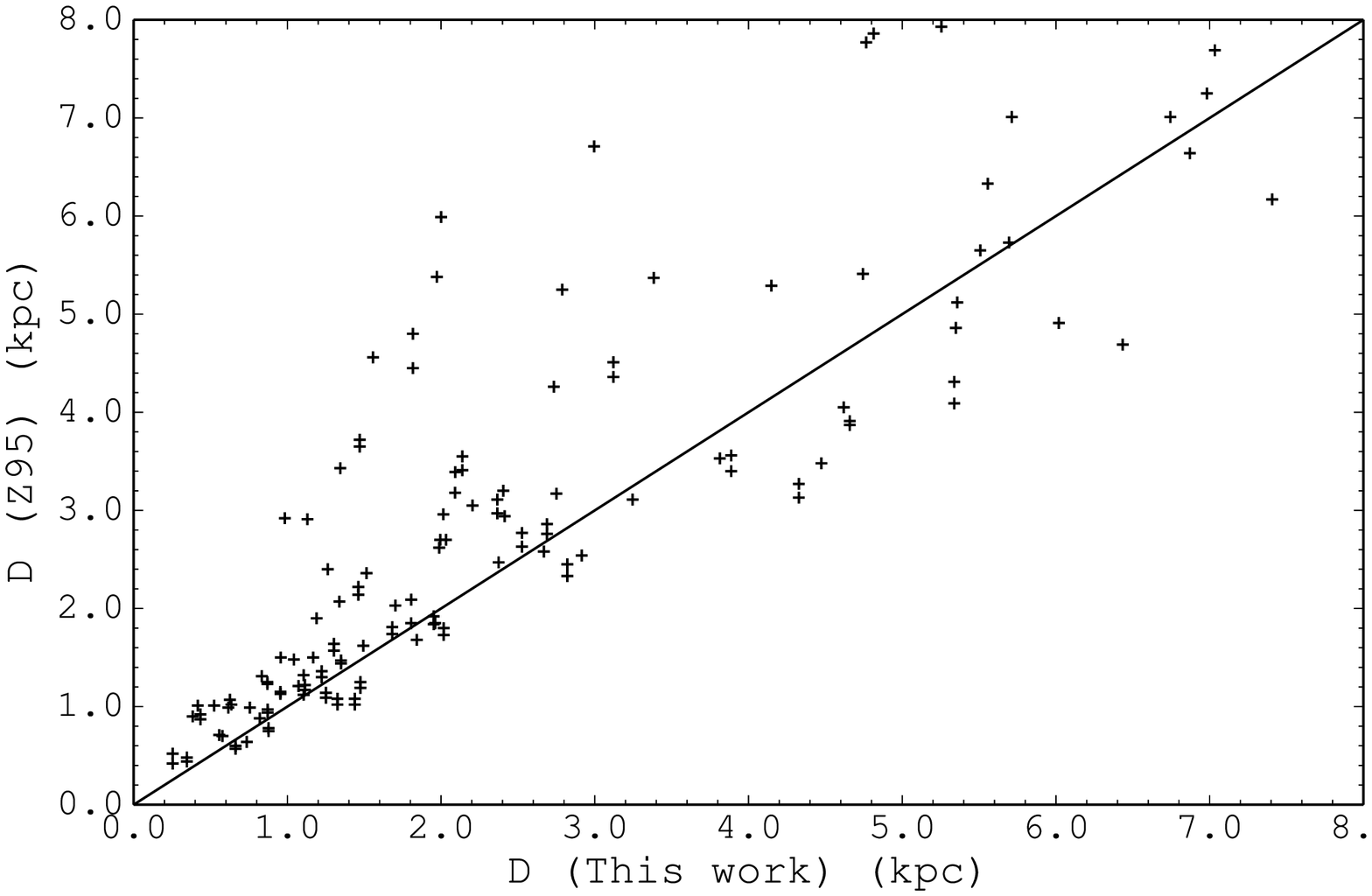}
\includegraphics[width=8.82cm]{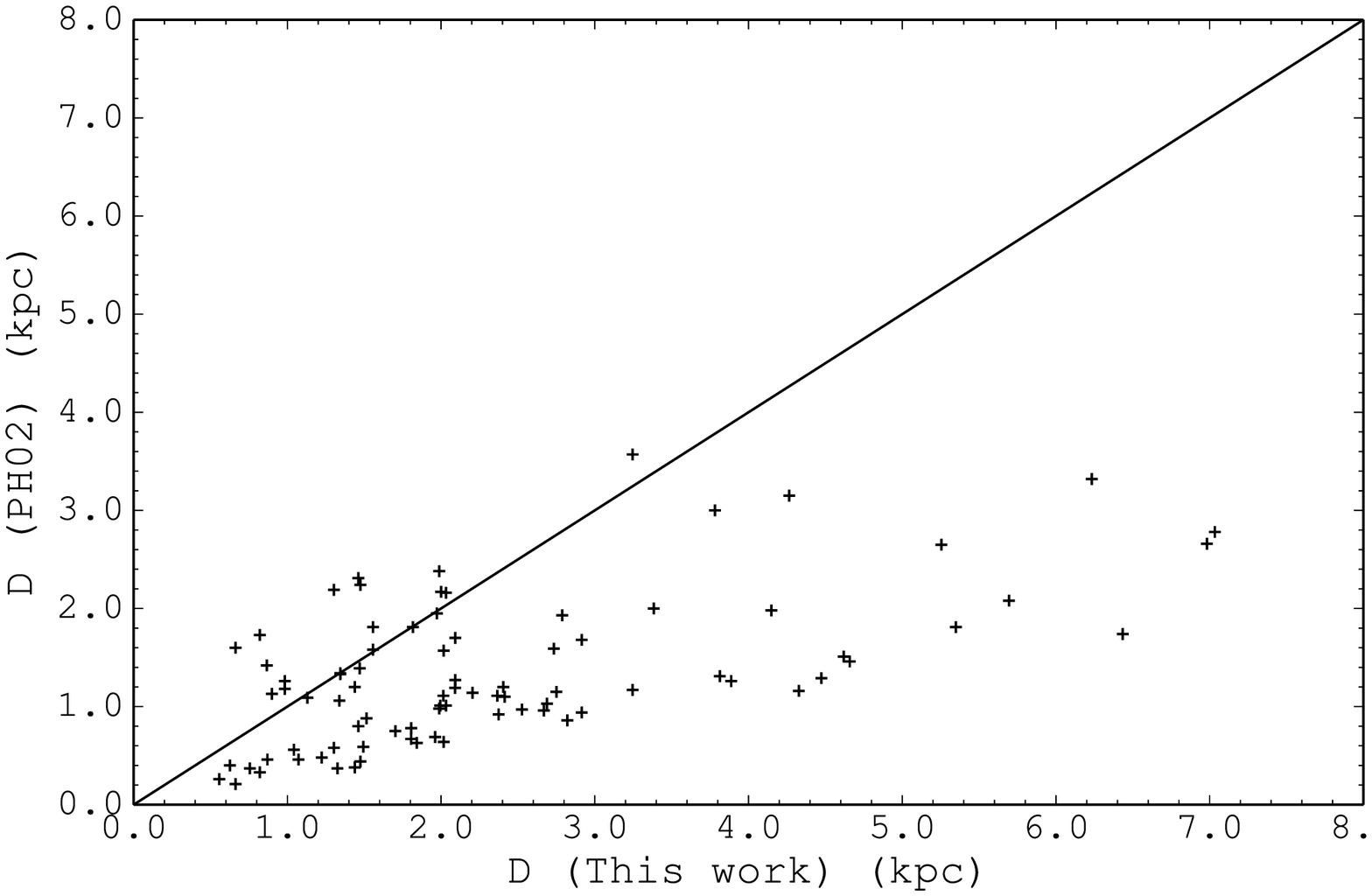}
\caption{Top row (L): The distances from SSV compared with our \SR\ distances, for PNe in common, with error bars omitted for clarity. (R): A comparison with data from Meatheringham et al. (1988), Kingsburgh \& Barlow (1992), and Kingsburgh \& English (1994).   Bottom row (L): A comparison with data from Zhang (1995). (R):  A comparison with data from Phillips (2002). 
}
\label{distance_scale_comparisons1}
\end{figure*}

From a review of the literature, it is seen that most published PN distance scales can be roughly divided into two camps, described as long and short (F08; Smith 2015) depending on whether they over- or underestimated the distances.  Clearly, the extant literature provides no consensus on the distance scale for evolved PNe, the most demographically abundant, with a factor of $\sim$3 discrepancy evident between the long and short scales, viz. Kingsburgh \& English (1992) and Phillips (2002) respectively.  In Figure\,\ref{distance_scale_comparisons1} we show a comparison of the distances from SSV, Meatheringham et al. (1988), Kingsburgh \& Barlow (1992), Kingsburgh \& English (1994), Zhang (1995), and Phiilips (2002) with the \SR\ distances from the present work.

To further compare the various published distance scales with one another, an index $\kappa_{\rm d}$ has been defined as $\overline{D}_{\rm Lit}$ / $\overline{D}_{\rm H\alpha}$ following Phillips (2002), where the mean distances for an ensemble of PNe using one of the distance scales from the literature are compared with distances for the same PNe using the \SR\ relation.   Table~\ref{rel_scales} shows a relative comparison of the most widely used recent distance scales discussed in the literature, expressed as approximate ratios relative to the present work (defined as $\kappa_{\rm d}$ = 1.00).  The distances derived here were directly related to the largest data sets of Zhang (1995), Kingsburgh \& English (1992), Van de Steene \& Zijlstra (1995), Mal'kov (1997, 1998), Bensby \& Lundstrom (2001), Phillips (2002), Phillips (2004b) and SSV.  For the older distance scales, we normalised the data summarised by Peimbert (1990) onto the distance scale of Daub (1982) for all PNe in common between the two studies, before linking that with the more recent data presented in the various papers of Phillips (2002, 2004b, 2005a) to get a fairly consistent set of ratios relative to the present work. 

Owing to the exact value of the $\kappa$-ratio being dependent on the subset of PNe used to make the comparison (i.e. whether the adopted distances of the calibrating PNe, or the statistical distances themselves were compared, or if subsets of compact or evolved PNe were used), statistical errors on the ratios are not formally given, but are estimated to be 20 per cent.  For example, for young, high-surface brightness PNe, the distance scale of Z95 agrees to within 10\% with the present work, but for the most evolved PNe which were not used as calibrators by Z95, his scale predicts distances roughly a factor of two too large (see figure~\ref{distance_scale_comparisons1}), and a factor of four larger than the SSV scale for evolved PNe.

Recall that the present mean scale is about three per cent longer than the mean scale of F08, in excellent agreement with the theoretical calibration of Jacob et al. (2013).   From Fig.\,\ref{distance_scale_comparisons1} and Table\,\ref{rel_scales} it can be seen that the Phillips (2002) scale is much too short, based primarily on a range of incorrect distances to his calibrating nebulae (non-PNe also contaminate his calibrating sample), with the Phillips (2004b) scale being a better match to the present scale.  Another of his distance scales (Phillips 2005b) could not be consistently normalised with respect to the present scale, but it is a short one, owing to the high number of PNe within 500\,pc in his sample.

Recently, Smith (2015) has analysed the Zhang (1995) and CKS/SSV scales in some depth.  For the mean Zhang scale, itself an average of two scales, one based on ionized mass versus radius, and the other a conventional $T_b$--$r$ relation, there is a considerable scale error in the distances for large PNe.  Smith finds only the $T_b$--$r$ relation should be used as a distance indicator.  For both the CKS and SSV scales, there is a substantial error dependence with PN radius at large radii, meaning that the distances for the demographically-common largest PNe are considerably underestimated, by a factor of two-or-so.

\begin{table}
\begin{center}
\caption{A selection of statistical distance scales from the literature, normalised to the present work.}
\label{rel_scales}
{\footnotesize
\begin{tabular}{llc}
\hline
Distance Scale 						&		Method					&  $\kappa_{\rm d}$	 		\\
\hline
O'Dell (1962)							&		Shklovsky method			&~~0.84~~					\\
Cahn \& Kaler (1971)					&		Shklovsky method			&  0.72	   					\\
Cudworth (1974)						&		Shklovsky method			&  0.95	  					\\
Milne \& Aller (1975)					&		Shklovsky method			&  0.72	  					\\	
Acker (1976, 1978)						&		synthetic					&  0.67	  					\\	
Maciel \& Pottasch (1980)				&		$M_{\rm ion}$--$r$ relation	&  0.83	  	 				\\	
Daub (1982)           						&		modified Shklovsky			&  0.56	  					\\	
Meatheringham et al. (1988)				&		nebular model				&  1.03						\\	
Cahn et al.  (1992; CKS)     				&		modified Shklovsky~~~		&  0.80						\\	
Kingsburgh \& Barlow (1992)  			&		nebular model				&  1.25	 					\\	
Zhang (1995)     						&		$T_b$--$r$ relation			&  1.02	 					\\	
Van de Steene \& Zijlstra (1995)~~		&		$T_b$--$r$ relation			&  0.93	  					\\	
Schneider \& Buckley (1996)				&		$T_b$--$r$ relation			&  0.91	  					\\	
Mal'kov (1997, 1998)					&		nebular model				&  1.05 	 					\\	
Bensby \& Lundstr\"om (2001)			&		$M_{\rm ion}$--$r$ relation	&  0.97	  					\\	
Phillips (2002)       						&		$T_b$--$r$ relation			&  0.37	 					\\	
Phillips (2004b)                  				&		$T_b$--$L_5$ relation		&  0.94	 					\\	
Phillips (2005a)       					&		standard candle			&  0.77	 					\\	
Stanghellini et al. (2008; SSV) 			&		modified Shklovsky			&  0.88						\\	
Frew (2008; F08)						&		\SR\ relation				&  0.97						\\ 
This work     							&		\SR\ relation				&  1.00	 	 				\\ 
\hline
\end{tabular}
}
\end{center}
\end{table}

\section{Summary and Future Work}\label{sec:conclusions}

We have critically compiled a catalogue of \ha\ fluxes, angular diameters, and distances for 207 Galactic and 126 extra-galactic PNe, to be used as primary calibrators for a newly established optical statistical distance indicator, the \ha\ surface brightness -- radius (\SR) relation.   Its application requires only an angular diameter, an integrated \ha\ flux, and the reddening to the PN.  From these quantities, an intrinsic radius is calculated, which when combined with the angular size, yields the distance.  
The \ha\ relation is also preferred to the equivalent \hb\ relation, as at a minimum, \ha\ fluxes are a factor of approximately three brighter. The \SR\ relation also has better utility than the equivalent [O{\sc\,iii}] and [N{\sc\,ii}] relations, as it includes both bright objects and the most senile PNe over a broad range of excitation, and best reflects the underlying ionised mass. The [N{\sc\,ii}] relation, especially, is strongly influenced by abundance variations between objects, and furthermore, there is negligible  [N{\sc\,ii}] emission in the PNe of highest excitation (F08).  

Furthermore, a number of recent and ongoing imaging surveys in \ha\ have become available which have allowed (and will continue to aid in) the determination of accurate integrated \ha\ fluxes for PNe and related nebulae.    We find that greater precision can be obtained by dividing PNe into two broad groups based on spectroscopic criteria.  Optically thick PNe populate the upper bound of the trend, while optically-thin (and generally high-excitation) PNe fall along the lower boundary in the \SR\ plane.  Using sub-trends has allowed more precision in the determination of distances, as good as $\pm$18 per cent in the case of optically-thin PNe.    The mean \SR\ relation of F08 has been independently validated by Jacob et al. (2013) and Smith (2015) as the most reliable statistical distance scale in the literature to date.  The present study improves this still further, and we complete this work by presenting an extensive catalogue of statistical distances obtained with our method, the largest such compilation in the literature.

In a follow-up paper (Frew et al., in preparation) we will present a further catalogue of distances for PNe that we are currently collecting new data for, including new objects discovered only recently (Kronberger et al. 2012, 2014; Sabin et al. 2014).  These catalogues of homogeneously derived distances will be a legacy to the community, and will be used to build the first accurate volume-limited PN census centred on the Sun (Frew 2008; Kastner et al. 2012; Frew et al., in prep.), as well as local PN luminosity functions in \ha\ and \OIII, to be presented in further papers in this series.  In the near future, new large-area radio surveys (e.g. Norris et al. 2013; Dickey et al. 2013) will allow distances to be obtained for PNe completely obscured at optical wavelengths, as will $S$--$r$ relations in the near-IR, now that integrated fluxes in the Paschen and Brackett hydrogen lines are becoming available (e.g. Wang et al. 2010; Dong et al. 2011).   New statistical calibrations in the radio and NIR domains will also be the subject of future work.

We expect our distance catalogues to remain useful even after the expected data avalanche from the Gaia satellite becomes available, as only a minority of the Galactic PN population will be able to have their distances determined.  Firstly, many compact, high-surface brightness PNe will have no astrometric data obtained, as they are larger than the Gaia's maximum angular size cutoff of 0.7\arcsec\ (Manteiga et al. 2014).  Only for more evolved PNe, where the central star is clearly visible against the surrounding nebular shell, can the immense resolving power of Gaia be utilized.  For any PN smaller than 0.7\arcsec\ across, and for more extended objects with bright central regions smaller than this limit, astrometric and spatial information will be recorded at a pixel scale of 59 mas\,pix$^{-1}$ and a point spread function of 180\,mas, brighter than a limiting magnitude of $\sim$20.  However, this size limit is smaller than the majority of known Galactic PNe, including most of those at the distance of the Bulge.  Second, more evolved bipolar PNe with bright, dense nebular cores can hide the central stars, even if they are formally brighter than the Gaia detection limit. Third, even at a relatively close distance 1.0\,kpc from the Sun, some PNe have central stars already below the  detection limit, so no parallax data will be obtained.   
Of course the new Gaia data will allow the refinement of our proposed sub-trends in S--$r$ space, enhancing its ability both as a diagnostic tool, and as a robust distance indicator for the many PNe which will not have Gaia distance estimates.

\section*{ACKNOWLEDGEMENTS}
D.J.F. thanks Macquarie University for a MQ Research Fellowship and I.S.B. is the recipient of an Australian Research Council Super Science Fellowship (project ID FS100100019).  Q.A.P.  thanks the Australian Astronomical Observatory for additional support.  We especially thank Arsen Hajian for providing his unpublished data, Ralf Jacob for providing his evolutionary tracks in machine-readable form, and the referee, Romano Corradi, whose valuable and insightful comments improved the content and layout of this paper.  We further thank our colleagues who have provided comments and advice, from the initial germination of this project to the present, in particular Martin Cohen, Romani Corradi, Hugh Harris, George Jacoby, Greg Madsen, Warren Reid, Detlef Sch\"onberner, Dick Shaw, Haywood Smith, and Albert Zijlstra. This research has made use of the SIMBAD database and the VizieR service, operated at CDS, Strasbourg, France, and also utilised data from the Southern \ha\ Sky Survey Atlas (SHASSA) and the Virginia-Tech Spectral Line Survey (VTSS), which were produced with support from the National Science Foundation (USA).  Additional data were used from the AAO/UKST \ha\ Survey, produced with the support of the Anglo-Australian Telescope Board and the Particle Physics and Astronomy Research Council (UK), and the Wisconsin H-Alpha Mapper (WHAM), produced with support from the National Science Foundation.


\section*{SUPPORTING INFORMATION}
Additional Supporting Information may be found in the online version of this article:\\
\\
Table A3. Final calibrating nebulae for the \SR\ relation.\\
Table A4. A catalogue of \SR\ distances for Galactic PNe.

\appendix

\section{The \SR\ Plane as a Diagnostic Tool}\label{sec:diagnostic_tool}

\subsection{Background}\label{sec:background}

Besides the ability of the \SR\ diagram to discriminate between optically-thick and optically-thin PNe, we are also interested in its ability to discriminate between bona fide PNe, transitional (and pre-) PNe, and the zoo of PN-like nebulae and outright mimics that are often confused with them (see Frew \& Parker 2010 for a review), both in the Milky Way and in the nearest external systems.  For instance,  the similarities and differences between bipolar PNe and symbiotic outflows have been discussed several times in the literature (e.g. Lutz et al. 1989; Corradi 1995; Schmeja \& Kimeswenger 2001; Frew \& Parker 2010), while compact \HII\ regions and the ejecta around massive stars were a contaminant in the earlier PN catalogues (Perek \& Kohoutek 1967; Kohoutek 2001).  Before discussing these in more detail, we briefly describe here four nebulae with accurate distances that have genuine affinities with bona fide PNe.\\

\noindent{\it Bode 1}:  ~This is the putative bipolar PN (Bode et al. 1987; Seaquist et al. 1989; Scott et al. 1994), around the classical nova GK\,Per.  The nebula has a distorted `bowtie' shape, consistent with shaping by an ISM interaction (Tweedy 1995; Bode et al. 2004; Shara et al. 2012).  We can derive an approximate \ha\ flux from the surface brightness data presented by Tweedy (1995).   Adopting dimensions of 780\arcsec\ $\times$ 450\arcsec for the outer nebula, and a mean \ha\ surface brightness of $2.5\pm1.3$ \ergcms, we determine log\,$F$(\ha) $\simeq$ $-11.15\pm0.30$.  The distance (477\,pc) is accurately known from an HST trigonometric parallax (Harrison et al. 2013).   Tweedy (1995) argued on evolutionary grounds that the nebula was unlikely to be a PN, but the lack of \NII\ emission shows it is not a reflection nebula around the current nova ejecta. It is likely to be a fossil nebula that was flash ionized by the 1901 eruption, analogous to the PN around V458\,Vul (Wesson et al. 2006).  From the observed \ha\ flux, reddening, diameter, and distance, the ionized mass is $\sim$0.1\,\msun\ and the mean electron density, $n_{e}$ = $\sim$10\,cm$^{-3}$, adopting a canonical filling factor of 0.3.  These numbers appear to rule out the bipolar nebula being an old nova shell from an earlier eruption, being more typical of an old PN (Frew \& Parker 2010).  While Bode\,1 is plots close to the optically-thin PN trend, we decline to use this as a calibrator due to lingering doubts over its nature.\\

\
\noindent{\it KjPn 8}:  ~This is a highly unusual nebula, with large, fast-expanding bipolar lobes extending over an angular size of 14\arcmin\ $\times$ 4\arcmin.  At the distance of $1.8\pm0.3$\,kpc (Boumis \& Meaburn 2013), the lobes extend over 7\,pc in length, and it may be the product of an Intermediate Luminosity Optical Transient (ILOT) event, powered by a binary interaction (Soker \& Kashi 2012).  The small, low-excitation core is only $\sim$6\arcsec\ $\times$ 4\arcsec\ across, is nitrogen enriched (V\'azquez et al. 1998), and has an integrated flux, $F$(\ha) = 2.4 $\times$ 10$^{-13}$\,\ergcms\ (L\'opez et al. 2000).   In addition, the compact core (but not the giant outflow) is detected in the radio at 6\,cm (Boji\v{c}i\'c et al. 2011a).   
We plot the nebular core  in Fig.\,\ref{fig:mimics} for illustrative purposes only.  KjPn\,8 has a number of properties in common with the southern nebula Hen\,2-111 (Webster 1978; Meaburn \& Walsh 1989; Cohen et al. 2011).  In the latter case however, the inner PN has a more normal ionized mass. \\

\noindent{\it PHR J1735-3333}:  ~This is the faint circular nebula around the OH/IR star V1018\,Sco, which may be a peculiar PN or an object more akin to the symbiotic outflows.  Two distance estimates are available: a maser phase-lag distance of $3.2\pm0.6$\,kpc from Cohen, Parker \& Chapman (2005a) and an SED distance of $3.76\pm0.66$\,kpc (Vickers et al. 2015).  These are consistent so we combine them to obtain $D$ = $3.5\pm0.5$\,kpc.  We obtain an integrated flux from the SHS following the recipe of Frew et al. (2014a), in order to plot this nebula in \SR\ space.  \\    

\noindent{\it SB\,17}:  ~The nebula around the unusual H-deficient star V348\,Sgr was discovered by Herbig (1958) and later catalogued by Beaulieu, Dopita \& Freeman (1999).  Since the distance is based on a model-dependent assumed luminosity (De Marco et al. 2002; Clayton et al. 2011) we do not use this object as a calibrator. \\  

\noindent{\it SMP LMC\,83}:  ~This unusual polypolar nebula (Shaw et al. 2006) surrounds a likely accreting binary system with a variable, H-deficient spectrum (Hamann et al. 2003).  This fast-expanding (Dopita, Ford \& Webster 1985), nitrogen-enriched nebula appears unusually massive for a PN, plotting $\sim$2$\sigma$ above the optically-thick \SR\ relation.   Owing to its suite of peculiarities, it is not included as a primary calibrator, but is shown in Fig.\,\ref{fig:mimics}.

\subsection{Pre-Planetary Nebulae and Related Objects}\label{sec:SED}

Some dusty pre-PNe are transition objects (e.g. Su\'arez et al. 2006) emitting in \ha, so can be plotted in the \SR\ plane.  For pre-PNe, as well as for the very youngest PNe, most of the luminosity is radiated in the thermal infrared (van de Veen, Habing \& Geballe 1989; Kwok, Hrivnak \& Langill 1993).  Thus comparing the observed bolometric flux from the spectral energy distribution (SED) with an assumed luminosity gives the distance (van de Veen et al. 1989; Goodrich 1991; Kwok et al. 1993; De Marco, Barlow \& Storey 1997).   The SED method is discussed in full in Vickers et al. (2015). 
For a few young PNe and transition objects, SED distances have been adopted from Vickers et al. (2015) if no other primary distance is available, in order to better populate and delineate the compact end of the \SR\ relation, but as these are statistical distances, they have been excluded as primary calibrators. These distances are presented in Table\,\ref{table:SED}, and plotted in Figure\,\ref{fig:mimics}.

\begin{table}
\begin{center}
\caption{SED distances to pre-PNe and very young PNe.} \label{table:SED}
{\footnotesize
\begin{tabular}{lll}
\hline
Name           				&  	~~~$D$ (kpc)					&     References~~~~~~~~  		\\
\hline
CRL 618					&     $1.22\pm0.16$				&	 VF15					\\   	
Hen 2-113				&	$1.48\pm0.30$~~~~~~~~		&	 DM97, VF15				\\   	
Hen 3-1333				&	$1.26\pm0.27$				&	 KH93, DM97, VF15		\\   	
IC 5117					&	$5.02\pm0.69$ 				&	 VF15 					\\
IRAS 21282+5050~~~~		&	$2.44\pm0.50$				&   	KH93, VF15				\\   	
M 2-56					&	$2.21\pm0.36$				&	G91	, VF15				\\   	
M 4-18					&	$6.89\pm1.45$				&	KH93, VF15				\\   	
PM 1-188				&	$4.43\pm1.05$				&	KH93, VF15 				\\   	
SwSt 1					&	$2.50\pm0.60$				&	DM97, VF15				\\   	
Vo 1						&	$2.91\pm0.58$				&    	VF15					\\   	
\hline
\end{tabular}
}
\end{center}
{\scriptsize
\begin{flushleft}
References:  DM97 -- De Marco et al. (1997); G91 -- Goodrich (1991);  KH93 -- Kwok et al.  (1993); VF15 -- Vickers et al. (2015).
\end{flushleft}
}
\end{table}

\subsection{\HII\ Regions in the ISM around White Dwarfs and Subdwarfs}

\HII\ regions around hot, low-mass stars have been repeatedly been confused with PNe in the literature (Frew \& Parker 2006, 2010).   As bona fide PNe at moderate to low electron density can be either optically thick (e.g. NGC\,2899, RCW\,69) or optically thin (e.g. NGC\,246, Abell\,39), we would expect to see Str\"omgren zones of similar or larger diameter around hot white dwarfs whose own PNe have dissipated into the ISM.  Likely examples are DeHt\,5 around WD\,2218+706 (F08; De Marco et al. 2013) and Sh\,2-174 around GD\,561 (F08; Frew \& Parker 2010), though Ransom et al. (2010, 2015) have argued that these two nebulae are fossil PNe.  The hot pre-WD KPD\,0005+5106 (Wassermann et al. 2010) is also surrounded by a large, low-density, high-excitation nebula (Chu et al. 2004; Sankrit \& Dixon 2009).   

Other \HII\ regions around low-mass stars are Abell~35, associated with BD\,$-$22\arcdeg3467\,B (F08), the nebulae around the DO stars PG\,0108+101 (Re\,1; Reynolds 1987), PG\,1034+001 (Hewett\,1; Hewett et al. 2003; Chu et al. 2004), and the nebulae around the subdwarf\,B stars PHL\,932 and EGB\,5 (Frew et al. 2010).  These latter objects are smaller and fainter than PNe, owing to the lower ionizing fluxes of these stars.   The integrated \ha\ fluxes, reddening values, and diameters for these objects have been taken primarily from F08, FBP13, Frew et al. (2014a), Madsen et al. (2006) and Parker et al. (in preparation), while the adopted distances are taken from the various literature sources given in the footnotes to Table\,\ref{table:mimics}, where the relevant data on these objects are presented.

\subsection{Compact \HII\ regions}
Discrete compact \HII\ regions have also been misclassified as PNe in the past (Frew \& Parker 2010).  We selected a representative sample of compact star-forming regions visible in the optical, especially those that are relatively symmetrical and which have detectable \OIII\ emission.  The adopted data for these objects presented in Table\,\ref{table:mimics}.

\subsection{Ejecta from Massive Stars}

Ejecta from massive stars have also been confused with bona fide PNe (Frew \& Parker 2010; Frew et al. 2014b).   In order to cover the widest parameter space possible, we plot several ejecta shells on the \SR\ plane surrounding WR and LBV stars.  As before, the adopted data for these objects are presented in Table\,\ref{table:mimics}. with the sources of the fluxes and distances given in the table footnotes.

\subsection{Bowshock nebulae}
We also investigate the ionized bowshock nebulae around a pair of nova-like cataclysmic variables: EGB\,4 around BZ\,Cam (Hollis et al. 1992) and Fr~2-11 around V341\,Ara (Frew, Madsen \& Parker 2006; F08).  The integrated \ha\ fluxes, reddening values, and diameters for these objects have been taken primarily from Greiner F08, FBP13, and Madsen et al. (2006), while the adopted distances are taken from the various literature sources given in the footnotes to Table\,\ref{table:mimics}, where the relevant data on these objects are presented.

\subsection{Discussion}
\begin{figure*}
\centering  
\includegraphics[width=16cm]{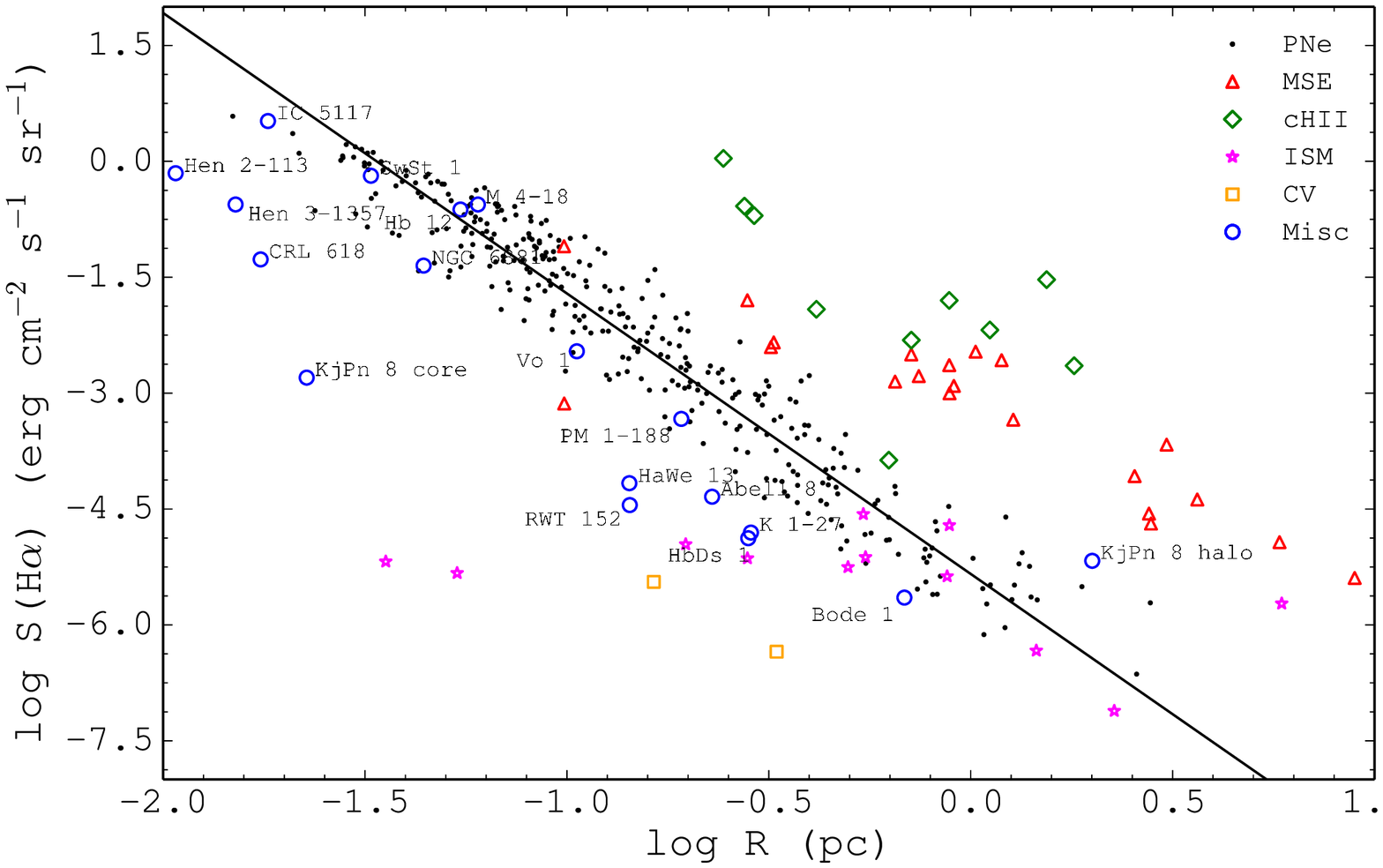}
\caption{PNe and mimics plotted in the \SR\ plane. Massive star ejecta (MSE), compact \HII\ regions, low-mass \HII\ regions in the ISM, and CV-bowshock nebulae have been plotted separately to bona fide PNe (small black points).  Several miscellaneous young PNe and PN-like nebulae discussed in the text are plotted as open blue circles with labels.}
\label{fig:mimics}
\end{figure*}

Resolved symbiotic outflows and their kin, many of which are morphologically similar to bipolar PNe, will be the subject of a separate investigation.  As expected, Fig.\,\ref{fig:mimics} shows that compact \HII\ regions and massive star ejecta (MSE) generally plot above the main PN locus, reflecting their larger ionized masses in the mean.  One c\HII\ region, We\,1-12 (Kimeswenger 1998), surrounds an early B star with an ionizing luminosity comparable to many CSPNe, thus it falls near the PN locus.  On the other hand, the \HII\ regions in the ISM ionized by low-mass stars are generally of low to very-low surface brightness and plot on and around the PN locus at medium to large radii.  The two known CV bowshock nebulae (EGB\,4 and Fr\,2-11) are clearly seen to be of substantially lower ionized mass than PNe, though apparently unrelated to classical nova shells (Frew \& Parker 2010). 

For the massive star ejecta, a surprisingly tight relation is shown in Fig.\,\ref{fig:mimics} if we exclude the young,low-mass nebula around the historical LBV, P Cygni.   The points fit a relation with a power law slope of $-2.3$, markedly shallower than the PN locus, or alternatively by two power laws with a break radius of $\sim$2\,pc.   Recalling equation\,\ref{eq:S_alpha}, we determine $\beta$ = 1.36, which indicates that an approximate distance scale can be developed for the ejecta around massive stars, at least for those examples that have not swept up large amounts of interstellar matter.  The distinct trend shown by massive stellar ejecta, separate to PNe, indicates that  \SR\ plane will be a useful adjunct to deep hydrogen-line surveys of the nearest galaxies with the next generation of telescopes.   We will explore these results in more detail in a companion paper.

\begin{table*}
\begin{center}
\caption{~Mimics plotted in the \SR\ plane.  Refer to the text for details. }
\label{table:mimics}
\begin{tabular}{lccclll}
\hline
Name				 &$\theta$~(\arcsec) 	&   log\,$S_{0}$(\ha)	&   ~\EBV~ 	&      ~~$D$~(pc) 		&        ~~~Type 			&    References  		\\
\hline
EGB 4				&  	82			&   	  $-6.35$				&   	0.05			& 	 $830\pm160$		&         CV bowshock			& 	RN98, GT01	 	\\
Fr~2-11				&  	208			&  	  $-5.40$				&   	0.05			& 	 $163^{+231}_{-37}$	&         CV bowshock			&	FM06, F08, vL07	 \\	
Abell 35				&  	419			&   	  $-5.12$				&    0.04			& 	 $220\pm100$		&	    Ionized ISM			& 	F08, Z12		\\
DeHt 5				&  	297			&   	  $-5.25$				&    0.10			& 	 $345^{+19}_{-17}$	&	    Ionized ISM			& 	F08, B09		\\
EGB 1  				&  	277	 		&  	  $-5.37$				&    0.23			& 	 $470\pm140$		&	    Ionized ISM			& 	F08, U		 	\\	
EGB 5  				&  	 90			&  	  $-5.12$				&    0.30			& 	 $550\pm140$		&	    Ionized ISM			& 	F08, U		 	\\	
HaWe 5				&  	17.5 		&  	  $-5.18$				&    0.20			& 	 $420\pm100$		&	    Ionized ISM			& 	N99, U		 	\\	%
HaWe 6				&  	53	 		&  	  $-5.33$				&    0.08			& 	 $209^{+19}_{-16}$	&	    Ionized ISM			& 	H07, U		 	\\	
Hewett 1				&  	1470		&  	  $-6.33$				&   	0.01			& 	 $211^{+67}_{-47}$	&	    Ionized ISM		& 	C04, H07, F08	\\
K 2-2				&  	 312			&  	  $-5.06$				&    0.03			& 	 $620\pm220$		&	     Ionized ISM			& 	F08, DD14, U	\\	
KPD\,0005+5106~~	&  	4500		&   	  $-5.74$				&   	0.05			& 	 $390\pm90$			&	    Ionized ISM			& 	C04, F08, U, W10	\\
PHL 932				&  	136			&   	  $-4.96$				&    0.02			& 	 $298^{+67}_{-47}$	&	     Ionized ISM		& 	H07, FM10		\\
Re 1					&  	1540		&   	  $-7.10$				&   	0.01			& 	 $300\pm100$		&	     Ionized ISM		& 	R87, FM10		\\	
Sh 2-174				&  	452			&   	  $-4.96$				&   	0.09			& 	 $410\pm120$		&	     Ionized ISM			& 	F08, U	 		\\
TK 2					&  	1040		&   	  $-6.87$				&   	0.03			& 	 $169^{+13}_{-11}$	&	     Ionized ISM			& 	H97, F08, U		\\
ESO 370-9			&  	25			&   	  $-1.94$				&   	1.27			& 	 $7600\pm900$		&	    c\HII\ region			& 	C07				\\
Hen 2-77				&  	10			&   	  $-0.58$				&   	2.55			& 	 $10000\pm2000$	&	    c\HII\ region			& 	CH87, CP11	\\
IC 1470				&  	38			&   	  $-1.55$				&   	1.31			& 	 $3000\pm600$		&	    c\HII\ region			& 	CD00			\\	 
K 2-15				&  	84			&   	  $-3.07$				&   	1.25			& 	 $3050\pm1450$		&	    c\HII\ region			& 	PC10				\\	 
M 2-62				&  	21			&   	  $-1.80$				&   	1.88			& 	 $8600\pm1000$			&	    c\HII\ region			& 	BR11, U				\\
NGC 2579			&  	35			&   	  $-1.47$				&   	1.14			& 	 $7600\pm900$			&	    c\HII\ region			& 	C07					\\
NGC 7538			&  224			&   	  $-2.25$				&   	1.46			& 	 $2650\pm900$			&	    c\HII\ region			& 	G68, MR09			\\
RCW 64				&  	93			&   	  $-2.65$				&   	1.38			& 	 $5400\pm1400^{\dagger}$ &	    c\HII\ region			& 	B86, R97			\\
RCW 71				&  	30			&   	  $-2.11$				&   	0.99 		& 	 $4900\pm1000$			&	    c\HII\ region			& 	WG89, U			\\
RCW 117				&  	22			&   	  $-0.09$				&   	2.76			& 	 $2600\pm500$			&	    c\HII\ region			& 	RF06, FBP13		\\
Sh 2-128				&  	33			&   	  $-2.15$				&   	1.80			& 	 $9400\pm400$			&	    c\HII\ region			& 	BT03, U				\\	
We 1-12				&  	56			&   	  $-3.87$				&   	0.69			& 	 $2300\pm1000$			&	    c\HII\ region			& 	K98					\\
HD 168625			&    27			&      $-1.34$				&     1.38			&     $2800\pm200^{\dagger}$	&	    LBV ejecta			& 	N96, U				\\	
Hen 2-58				&  	22			&   	  $-2.37$				&   	0.55			&     $6000\pm1000$			&	    LBV ejecta			& 	HL92, FB14	 		\\	
Hf 39				&  	33			&   	  $-3.42$				&   	1.14			& 	 $8000\pm1000$			&	    LBV ejecta			& 	SC94, FB14	 	\\	
HR Car				&   13.4			&   	  $-2.35$				&   	0.96			& 	 $5000\pm500^{\dagger}$	&	    LBV ejecta			&     vG91, CS95			\\	%
P Cyg				&  	11.3			&   	  $-3.13$				&   	0.60			& 	 $1800\pm200^{\dagger}$	&	    LBV ejecta			& 	BD94				\\	%
R 127				&  	4.2			&   	  $-2.42$				&   	0.14			& 	 $50000\pm1100$~~~~~	&	    LBV ejecta			& 	N97		  			\\	%
S 61					&  	3.7			&   	  $-2.64$				&   	0.21			& 	 $50000\pm1100$		&	    LBV ejecta			& 	PNC99 				\\	%
Wray 15-751			&  	11.0			&   	  $-2.43$				&   	1.80			& 	 $6000\pm1000$			&	    LBV ejecta		& 	P06, VN14 		\\	
NGC 6164-5			&    131			&      $-2.99$				&     0.55			&     $1380\pm120^{\dagger}$	&	    Of ejecta			& 	H78, N08, FB14	 \\	
Anon WR 8			&    173			&      $-4.82$				&     0.71			&     $3470\pm350^{\dagger}$	&	    WR ejecta				& 	vdH01, U		 	\\	%
Anon WR 16			&    240			&      $-4.57$				&     0.64			&     $2300\pm230^{\dagger}$	&	    WR ejecta				& 	vdH01, U			\\	%
Anon WR 71			&    292			&      $-5.39$				&     0.30			&     $6300\pm630^{\dagger}$	&	    WR ejecta				& 	IM83, vdH01, U		\\	%
BAT99 16			&    15.0			&      $-3.13$				&     0.24			&     $50000\pm1100$		&	    WR ejecta				& 	GC94, C99, U		\\	%
DuRe 1				&    22			&      $-4.81$				&     2.20			&     $11000\pm1100$		&	    WR ejecta				& 	FBP13, FB14, U	 	\\	%
M 1-67				&  	 40			&   	  $-2.24$				&   	1.31			& 	 $3350\pm670$			&	    WR ejecta				&    GM98, MM10, FBP13 \\
NGC 6888			&    441			&      $-3.81$				&     0.65			&     $1260\pm130^{\dagger}$	&	    WR ejecta				& 	WS75, vdH01, U 	\\	%
PCG 11				&  	 37			&   	  $-2.78$				&   	2.17			& 	 $4100\pm400$			&	    WR ejecta				& 	CP05, FB14~~~		\\
PMR 5				&  	16.7			&   	  $-1.81$ 				&   	3.25 		& 	 $3500\pm400$			&	    WR ejecta				& 	FB14, U				\\
RCW 58				&  	465			&   	  $-4.07$				&   	0.43			& 	 $2300\pm300^{\dagger}$	&	    WR ejecta			& 	FB14, U 	\\	
Sh 2-308				&  	1150			&   	  $-4.99$				&   	0.10			& 	 $970\pm100^{\dagger}$	&	    WR ejecta				& 	vdH01, U  			\\	%
\hline
\end{tabular}
\end{center}
{\scriptsize
\begin{flushleft}
References: ~B86 -- Brand (1986);  B09 -- Benedict et al. (2009); BD94 -- Barlow et al. (1994); BR11 -- Balser et al. (2011); BT03 -- Bohigas \& Tapia (2003); C99 -- Chu et al. (1999); C04 -- Chu et al. (2004); C07 -- Copetti et al. (2007);  CD00 -- Caplan et al. (2000);  CH87 -- Caswell \& Haynes (1987); CP05 -- Cohen et al. (2005b); CP11 -- Cohen et al. (2011); F08 -- Frew (2008); FBP13 -- Frew et al. (2013);  FB14 -- Frew et al. (2014a);  FM06 -- Frew et al. (2006a); FM10 -- Frew et al. (2010); G68 -- Gebel (1968); GC94 -- Garnett \& Chu (1994); GM98 -- Grosdidier et al. (1998); GT01 -- Greiner et al., 2001;  H78 -- Humphreys (1978);  H03 -- Hewett et al. (2003); H07 -- Harris et al. (2007); HH01 -- Herald et al. (2001);  HL92 -- Hoekzema et al. (1992); IM83 -- Isserstedt et al. (1983);  K98 -- Kimeswenger (1998); KB10 -- Kamohara et al. (2010); MM10 -- Marchenko et al. (2010); MR09 -- Moscadelli et al. (2009); N96 -- Nota et al. (1996); N97 -- Nota (1997);  N08 -- Naz\'e et al. (2008); P06 -- Pasquali et al. (2006); PC10 -- Pinheiro et al. (2010); PNC99 -- Pasquali et al. (1999); R87 -- Reynolds (1987); R97 -- Russeil (1997); RF06 -- Rudolph et al. (2006);  RN98 -- Ringwald \& Naylor (1998); SC94 -- Smith et al. (1994); U -- unpublished data;  vdH01 -- van der Hucht (2001);  vG91 -- van Genderen et al. (1991);  vL07 -- van Leeuwen (2007);  VN14 -- Vamvatira-Nakou et al. (2014);  W10 -- Wassermann et al. (2010); WG89 -- Westerlund \& Garnier (1989); WS75 -- Wendker et al. (1975); Z12 -- Ziegler et al. (2012b).   ~Note:~$^{\dagger}$Adopted uncertainty.   
\end{flushleft}
} 
\end{table*}

\clearpage

\onecolumn

\begin{center}
{\scriptsize
\begin{longtable}{llccccccccc}
\caption*{Table A3: Final calibrating nebulae for the $S$--$r$ relation.} 
\\
\hline
\multicolumn{1}{l}{PN G}	&	 {Name} 	 &   {$D$ (pc)}		& {Meth}	& {Trend}	& {Morph}     & {$a$ (\arcsec)}  & {$b$ (\arcsec)}	&  $E(B-V)$ 	&  {$S_0$(H$\alpha$)}	&  {log $r$ (pc)} 		\\ 
\hline 
\endfirsthead
																					
\multicolumn{11}{l}
{{   }} \\

\hline
\multicolumn{1}{l}{PN G}	&	 {Name} 	 &   {$D$ (pc)}		& {Meth}	& {Trend}	& {Morph}     & {$a$ (\arcsec)}  & {$b$ (\arcsec)}	&  $E(B-V)$ 	&  {$S_0$(H$\alpha$)}	&  {log $r$ (pc)} 		\\ 
\hline 
\endhead
\hline 
\endfoot
\hline
\endlastfoot	
002.1$+$01.7		&	JaFu~1		&	7200 $\pm$ 700		&	C	&	Inter		&	Eb	&	8.0	&	8.0	&	$1.93 \pm 0.21$	&	$-2.20 \pm 0.26$	&	$-0.86$	\\
002.4$+$05.8 		&	NGC~6369	&	1550	$\pm$ 300 		&	M 	&	Inter 		&	Eb	&	30.0	&	29.0	&	$1.31 \pm 0.16$	&	$-1.01 \pm 0.17$ 	&	$-0.96$	\\
003.5$-$04.6		&	NGC~6565	&	2000 $\pm$ 500		&	X	&	Inter		&	E	&	18.0	&	13.0	&	$0.31 \pm 0.10$ 	&	$-1.95 \pm 0.12$	&	$-1.13$	\\
004.0$-$03.0		&	M~2-29		&	7100 $\pm$ 2200		&	G	&	Thin		&	E	&	4.8	&	3.6	&	$0.72 \pm 0.14$	&	$-1.25 \pm 0.15$	&	$-1.16$	\\
010.4$+$04.4		&	DPV~1		&	3400 $\pm$ 500		&    M;Z	&	Thin		&	R	&	44.0	&	44.0	&	$0.71 \pm 0.08$	&	$-4.35 \pm 0.15$	&	$-0.51$	\\
010.8$-$01.8		&	NGC~6578	&	2900	$\pm$ 800		&	E	&	Inter		&	E	&	12.1	&	11.8	&	$0.93 \pm 0.10$	&	$-1.18 \pm 0.12$	&	$-1.08$	\\
011.7$-$00.6		&	NGC~6567	&	1680	$\pm$ 170		&    E;H	&	Thin		&	E	&	8.1	&	6.4	&	$0.48 \pm 0.10$	&	$-0.79 \pm 0.11$	&	$-1.52$	\\
013.8$-$02.8		&	SaWe~3		&	2100	$\pm$ 300		&	X	&	Thick	&	B	& 110.0	&	80.0	&	$0.72 \pm 0.27$	&	$-3.82 \pm 0.27$	&	$-0.32$	\\
\end{longtable}
}
\end{center}


\begin{center}
{\scriptsize
\begin{longtable}{llccccccccccc}
\caption*{Table A4:  A catalogue of \SR\ distances to Galactic PNe}\\
\hline \noalign{\smallskip}
PN\,G & 
Name & 
$a$ &
$b$ &
$E(B-V)$ &
method &
$\log$\,$S_0$(H$\alpha$) &
~~$\log$\,$r$~~  &
$D_{\rm mean}$ &
$D_{\rm thin}$&
$D_{\rm thick}$&
Notes \\
		&
		&
(\arcsec)	& 
(\arcsec)	& 
(mag)	&
		&
(cgs\,sr$^{-1}$)	&
(pc)		&
(kpc)		&
(kpc)		&
(kpc)		&
		&
\\
\hline \noalign{\smallskip}
\endfirsthead
\multicolumn{10}{c}{{}} \\
\hline \noalign{\smallskip}
PN\,G & 
Name & 
$a$&
$b$ &
$E(B-V)$ &
method &
$\log$\,$S_0$(H$\alpha$) &
~~$\log$\,$r$~~  &
$D_{\rm mean}$ &
$D_{\rm thin}$&
$D_{\rm thick}$&
Notes \\
		&
		&
(\arcsec)	& 
(\arcsec)	& 
(mag)	&
		&
(cgs\,sr$^{-1}$)	&
(pc)		&
(kpc)		&
(kpc)		&
(kpc)		&
		&
\\
\hline \noalign{\smallskip}

\endhead
\hline \noalign{\smallskip}
\endfoot
\hline \noalign{\smallskip}
\endlastfoot

000.0$-$06.8&H~1-62&$       5.0$&$       4.0$&$0.49 \pm 0.29$&$1,3$&$-1.25 \pm 0.29$&$     -1.12$&$6.97 \pm 2.43$&$...$&$...$&...\\
000.1$+$17.2&PC~12&$       2.3$&$       2.2$&$0.54 \pm 0.31$&$1,3$&$-0.65 \pm 0.32$&$     -1.29$&$9.46 \pm 3.39$&$...$&$...$&...\\
000.1$-$01.7&PHR~J1752-2941&$      16.7$&$      12.2$&$0.99 \pm 0.31$&$1$&$-3.07 \pm 0.33$&$     -0.62$&$6.95 \pm 2.54$&$...$&$...$&...\\
000.1$-$02.3&Bl~3-10&$       7.2$&$       6.9$&$0.64 \pm 0.25$&$3$&$-2.41 \pm 0.30$&$     -0.80$&$9.24 \pm 3.24$&$7.62 \pm 2.12$&$...$&...\\
000.1$-$05.6&H~2-40&$      18.3$&$      16.9$&$0.50 \pm 0.22$&$1$&$-3.22 \pm 0.23$&$     -0.58$&$6.19 \pm 1.99$&$...$&$...$&...\\
000.2$+$01.7&JaSt~19&$       7.2$&$       6.4$&$1.59 \pm 0.07$&$1,3$&$-2.22 \pm 0.13$&$     -0.85$&$8.50 \pm 2.49$&$...$&$...$&...\\
000.2$+$06.1&Terz~N~67&$      16.0$&$      12.0$&$0.76 \pm 0.13$&$1,3$&$-3.57 \pm 0.26$&$     -0.48$&$9.79 \pm 3.27$&$...$&$...$&...\\
000.2$-$01.9&M~2-19&$       9.4$&$       8.5$&$0.83 \pm 0.21$&$1,3$&$-1.78 \pm 0.22$&$     -0.97$&$4.89 \pm 1.55$&$...$&$...$&...\\
\end{longtable}
}
\end{center}

\end{document}